\def\isarxiv{1} %%% for icml submission version, we comment this line

\ifdefined\isarxiv
\documentclass[11pt]{article}

\usepackage[numbers]{natbib}

\else
\documentclass{article}
\usepackage{neurips_2022}
\fi

\usepackage{amsmath}
\usepackage{amsthm}
\usepackage{amssymb}
\usepackage{algorithm}
\usepackage{subfig}
\usepackage{algpseudocode}
\usepackage{graphicx}
\usepackage{grffile}
\usepackage{wrapfig,epsfig}
\usepackage{url}
\usepackage{xcolor}
\usepackage{epstopdf}

\usepackage{bbm}
\usepackage{dsfont}

\allowdisplaybreaks

\ifdefined\isarxiv

\usepackage{tikz}
\usepackage{hyperref}  %%% arxiv don't allow this.
\hypersetup{colorlinks=true,citecolor=blue,linkcolor=blue} %%% Zhao : maybe we should comment this in submission.
\usetikzlibrary{arrows}
\usepackage[margin=1in]{geometry}

\else

\usepackage{microtype}
\usepackage{hyperref}
\definecolor{mydarkblue}{rgb}{0,0.08,0.45}
\hypersetup{colorlinks=true, citecolor=mydarkblue,linkcolor=mydarkblue}

\fi
\graphicspath{{./figs/}}

\newtheorem{theorem}{Theorem}[section]
\newtheorem{lemma}[theorem]{Lemma}
\newtheorem{definition}[theorem]{Definition}

\newtheorem{fact}[theorem]{Fact}

\newcommand{\wt}{\widetilde}
\newcommand{\ov}{\overline}

\newcommand{\R}{\mathbb{R}}

\renewcommand{\d}{\mathrm{d}}

\newcommand{\Tmat}{{\cal T}_{\mathrm{mat}}}

\DeclareMathOperator{\poly}{poly}

\DeclareMathOperator{\nnz}{nnz}

\DeclareMathOperator{\rank}{rank}
\DeclareMathOperator{\diag}{diag}

\DeclareMathOperator{\vect}{vec}

\DeclareMathOperator{\reg}{reg}
\DeclareMathOperator{\A}{\mathsf{A}}

\makeatletter
\newcommand*{\RN}[1]{\expandafter\@slowromancap\romannumeral #1@}
\makeatother

\usepackage{lineno}

\begin{document}

\ifdefined\isarxiv

\date{}

\title{A Fast Optimization View: Reformulating Single Layer Attention \\in LLM Based on Tensor and SVM Trick, and Solving It in Matrix Multiplication Time}

\author{
Yeqi Gao\thanks{\texttt{a916755226@gmail.com}. The University of Washington.}
\and 
Zhao Song\thanks{\texttt{zsong@adobe.com}. Adobe Research.}
\and 
Weixin Wang\thanks{\texttt{wwang176@jh.edu}. Johns Hopkins University.}
\and 
Junze Yin\thanks{\texttt{junze@bu.edu}. Boston University.}
}

\else

\title{Intern Project} 
\maketitle 
\fi

\ifdefined\isarxiv
\begin{titlepage}
  \maketitle
  \begin{abstract}

Large language models have played a pivotal role in revolutionizing various facets of our daily existence. Serving as the cornerstone of virtual assistants, they have seamlessly streamlined information retrieval and task automation. Spanning domains from healthcare to education, these models have made an enduring impact, elevating productivity, decision-making processes, and accessibility, thereby influencing and, to a certain extent, reshaping the lifestyles of people.

Solving attention regression is a fundamental task in optimizing LLMs. In this work, we focus on giving a provable guarantee for the one-layer attention network objective function
\begin{align*}
   L(X,Y) = \sum_{j_0=1}^n \sum_{i_0=1}^d ( \langle \langle \exp( \A_{j_0} x ) , {\bf 1}_n \rangle^{-1} \exp( \A_{j_0} x ), A_{3} Y_{*,i_0} \rangle - b_{j_0,i_0} )^2
\end{align*}
Here $\mathsf{A} \in \mathbb{R}^{n^2 \times d^2}$ is Kronecker product between $A_1 \in \mathbb{R}^{n \times d}$ and $A_2 \in \mathbb{R}^{n \times d}$. $A_3$ is a matrix in $\mathbb{R}^{n \times d}$, $\mathsf{A}_{j_0} \in \mathbb{R}^{n \times d^2}$ is the $j_0$-th block of $\mathsf{A}$. The $X, Y \in \mathbb{R}^{d \times d}$ are variables we want to learn. $B \in \mathbb{R}^{n \times d}$ and $b_{j_0,i_0} \in \mathbb{R}$ is one entry at $j_0$-th row and $i_0$-th column of $B$, $Y_{*,i_0} \in \mathbb{R}^d$ is the $i_0$-column vector of $Y$, and $x \in \mathbb{R}^{d^2}$ is the vectorization of $X$.

In a multi-layer LLM network, the matrix $B \in \R^{n \times d}$ can be viewed as the output of a layer, and $A_1=  A_2 = A_3 \in \R^{n \times d}$ can be viewed as the input of a layer. The matrix version of $x$ can be viewed as $QK^\top$ and $Y$ can be viewed as $V$.  We provide an iterative greedy algorithm to train loss function $L(X,Y)$ up $\epsilon$ that runs in $\wt{O}( ({\cal T}_{\mathrm{mat}}(n,n,d) + {\cal T}_{\mathrm{mat}}(n,d,d) + d^{2\omega}) \log(1/\epsilon) )$ time. Here $\Tmat(a,b,c)$ denotes the time of multiplying $a \times b$ matrix another $b \times c$ matrix, and $\omega\approx 2.37$ denotes the exponent of matrix multiplication.

  \end{abstract}
  \thispagestyle{empty}
\end{titlepage}

{%\hypersetup{linkcolor=black}
%\tableofcontents
}
\newpage

\else

\begin{abstract}

\end{abstract}

\fi

\section{Introduction}

Large language models (LLMs) like GPT-1 \cite{rns+18}, BERT \cite{dclt18}, GPT-2 \cite{rwc+19}, GPT-3 \cite{bmr+20}, ChatGPT \cite{cha22}, GPT-4 \cite{o23}, OPT \cite{zrg+22}, Llama \cite{tli+23}, and Llama 2 \cite{tms+23} have demonstrated impressive capabilities in natural language processing (NLP). These models understand and generate complex language, enabling a wide range of applications such as sentiment analysis \cite{zdl+23}, language translation \cite{aaa+23}, question answering \cite{bhs+23}, and text summarization \cite{pd23}. Despite their high-quality performance, there remains untapped potential in optimizing and training these massive models, making it a challenging endeavor in the present day.

The primary technical foundation supporting the capabilities of LLMs is the attention matrix \cite{rns+18,vsp+17,bmr+20,dclt18}. The central concept of attention is to learn representations that emphasize the most relevant parts of the input. To be more specific, the attention mechanism compares the query vectors (the output tokens) with the key vectors (the input tokens). The attention weights are then determined based on the similarity of this comparison, indicating the relative importance of each input token. These attention weights are used to compute weighted averages of the value vectors, resulting in the output representation. By leveraging attention, LLMs acquire the ability to focus on the crucial aspects of the input, allowing them to gather pertinent information more efficiently and precisely. This capability enables LLMs to process longer texts effectively and comprehend intricate semantic relationships. Notably, the self-attention mechanism enables LLMs to establish connections between various segments of the input sequence, enhancing their contextual understanding.

We start with defining the general Attention forward layer,

\begin{definition}[$\ell$-th layer forward computation]
Let ${\bf 1}_n$ be the $n$-dimensional vector whose entries are all $1$. Let $\diag : \R^n \to \R^{n \times n}$ be a function: each entry of the vector in $\R^n$ is mapped to the diagonal entry of the matrix in $\R^{n \times n}$ and other entries of this matrix are all $0$'s.
Given weights $Q,K, V \in \R^{d \times d}$, let $X_{\ell} \in \R^{n \times d}$ denote the $\ell$-th layer input and $X_{\ell+1} \in \R^{n \times d}$
\begin{align*}
 X_{\ell+1} \gets D^{-1} \exp(X_{\ell} Q K^\top X_{\ell}^\top) X_{\ell} V
\end{align*} 
where $D:= \diag( \exp( X_{\ell} Q K^\top X_{\ell}^\top ){\bf 1}_n )$
\end{definition}

Mathematically, a general optimization with respect to attention computation is defined as:
\begin{definition}[Attention optimization]\label{def:attention}
    Let $A_1, A_2, A_3, B \in \R^{n \times d}$ and $X, Y \in \R^{d \times d}$. The attention computation is defined as:
    \begin{align*}
        \min_{X,Y \in \R^{d \times d}}  \| D(X)^{-1} \exp(A_1 X A_2^\top) A_3 Y - B \|_F^2,
    \end{align*}
    where $D(X) \in \R^{n \times n}$ is $D(X) := \diag( \exp(A_1 X A_2^\top ) {\bf 1}_n )$.
\end{definition}

\begin{figure}[!ht]
    \centering
    \includegraphics[width = \linewidth]{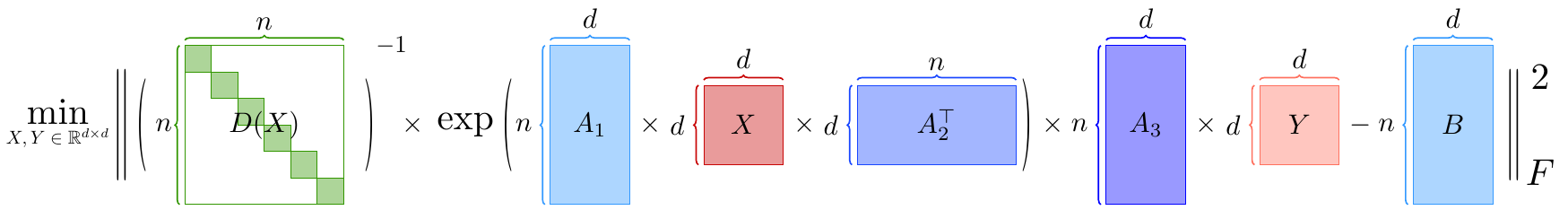}
    \includegraphics[width = 0.7\linewidth]{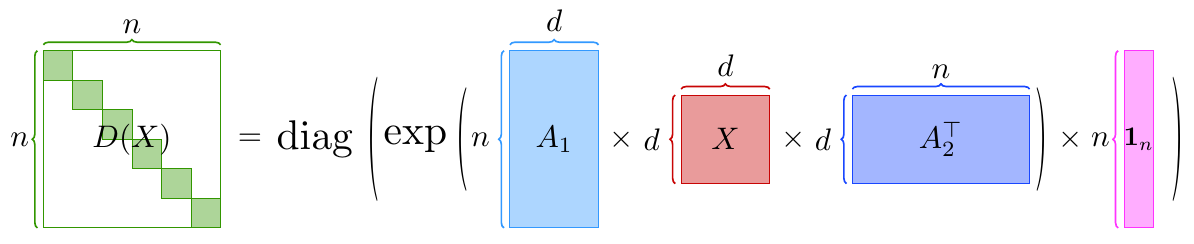}
    \caption{The visualization of the attention optimization (see Definition~\ref{def:attention}). Let $A_1, A_2, A_3, B \in \R^{n \times d}$ and $X, Y \in \R^{d \times d}$. We first get $\exp(A_1 X A_2^\top) \in \R^{n \times n}$ by multiplying $A_1$, $X$, and $A_2^\top$. Then, we have $D(X) \in \R^{n \times n}$ by computing $\diag( \exp(A_1 X A_2^\top ) {\bf 1}_n )$. After that, we multiply $D(X)^{-1}$, $\exp(A_1 X A_2^\top )$, $A_3$, and $Y$ and subtract $B$ from their product. Finally, we compute the minimum of the Frobenius norm of their difference. The blue rectangles represent the $n \times d$ matrices, the purple rectangle represents the $n$-dimensioal vector, the red squares represent the $d \times d$ matrices, and the green squares represent the $n \times n$ diagonal matrices.}
    \label{fig:attention_optimization}
\end{figure}

\begin{figure}[!ht]
    \centering
    \includegraphics[width = \linewidth]{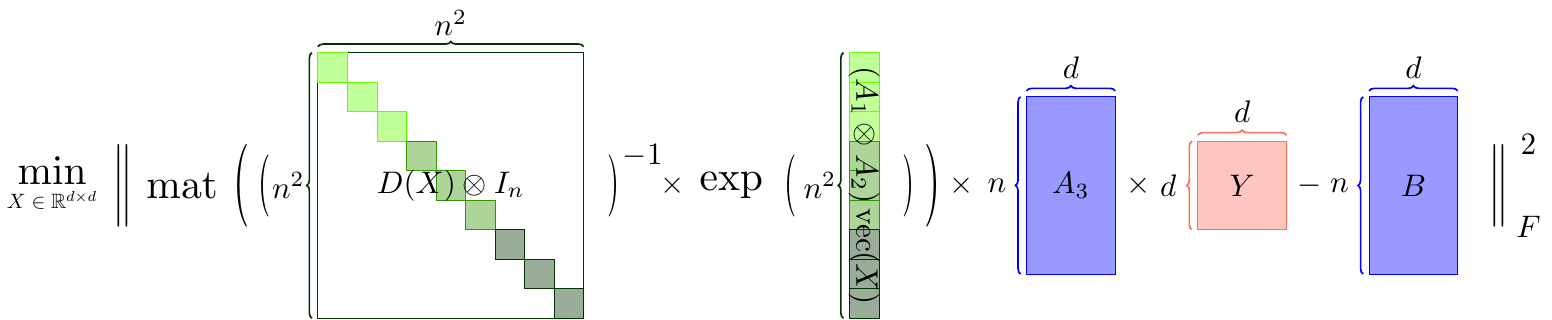}
    \caption{The visualization of a variation of Definition~\ref{def:attention}. Let $A_1, A_2, A_3, B \in \R^{n \times d}$, $X \in \R^{d \times d}$, $D(X) \in \R^{n \times n}$ (see Figure~\ref{fig:attention_optimization} and Definition~\ref{def:attention}), and $\A = A_1 \otimes A_2 \in \R^{n^2 \times d^2}$. $\mathrm{mat} : \R^{n^2} \to \R^{n \times n}$ is defined by $X_{i, j} = \mathrm{mat}(x)_{i, j} := x_{(i - 1) \cdot n + j}$, and $\vect = \mathrm{mat}^{-1}$. We first get that $(D(X) \otimes I_n)^{-1} \in \R^{n^2 \times n^2}$ and multiply $\A$ with $\vect(X)$. Then, we multiply $(D(X) \otimes I_n)^{-1} \in \R^{n^2 \times n^2}$ with $\A \cdot \vect(X) \in \R^{n^2}$, which gives us a vector in $\R^{n^2}$. We use $\mathrm{mat}$ to transform that into a matrix in $\R^{n \times n}$. After that, we multiply this matrix with $A_3 Y \in \R^{n \times d}$. Finally, we compute the minimum of the Frobenius norm of $\mathrm{mat}((D(X) \otimes I_n)^{-1} \cdot \exp(\A \vect(X))) A_3 Y - B$. In this figure, we give an example when $n = 3$: in the matrix $D(X) \otimes I_n$, the three light green squares (and their nearby white area) make up the first chunk, the three middle green squares (and their nearby white area) make up the second chunk, and the three dark green squares (and their nearby white area) make up the third chunk. The blue rectangles represent the matrices in $\R^{n \times d}$. The red rectangle represents the matrix in $\R^{d \times d}$.}
    \label{fig:softmax}
\end{figure}

Here $X = QK^\top, Y = V$ are the weights we want to learn, and $A_1,A_2,A_3$ are the input of a layer $X_{\ell}$, and the $B$ are the output layer $X_{\ell+1}$.
Attention computation has been analyzed in many recent works \cite{zhdk23,as23,bsz23,gsyz23_quantum,dms23,syz23,tlto23,pmxa23,mgn+23,zpga23,psza23,sht23}, 
but none of them give a complete analysis of the full version of the attention computation problem. They all simplify this problem by different strategies (see details in Table~\ref{tab:previous_attention}). 

However, simplifying this problem may lead to a significant decrease in the model performance, which may require extra model training or fine-tuning. This results in deployment obstacles.

In this paper, our focus is on optimizing the attention mechanism. Our goal is to present a complete, un-simplified analysis of the attention problem defined in Definition~\ref{def:attention}, a task that, to the best of our knowledge, has not been done before. We provide a provable guarantee for optimizing the attention function in the case of a single-layer attention network. Our motivation stems from the critical role of the attention optimization problem in the functionality of LLMs, and we firmly believe that our theoretical analysis will significantly influence the development of LLMs.

As \cite{as23}, they show that one step forward computation of attention can be done in $o(n^2)$ time without formulating the $n \times n$ matrix. However, it is still an open problem about how fast we optimize the loss function via the iterative method.
\begin{center}
    {\it How fast can we optimize the training process of attention matrix (See Definition~\ref{def:attention})?} 
\end{center}
In this study, we make progress towards this fundamental question. 

To establish the correctness of our algorithm, we conduct a comprehensive analysis of the positive semi-definite (PSD) property and the Lipschitz continuity of the Hessian matrix constructed from the attention matrix. These two properties provide the necessary assurance for employing TensorSRHT and Newton's method, ensuring both fast computation and convergence, respectively. 

Now, we will present our main result as follows.

\begin{theorem}[Informal version of our main theorem]\label{thm:main_informal}
Let $A_1, A_2, A_2 \in \R^{n \times d}$. 
    There is an algorithm that runs in $\wt{O}( \Tmat(n,d,n) + \Tmat(n,d,d) + d^{2\omega}) \log(1/\epsilon)  $ solves to the attention problem up to $\epsilon$ accuracy with probability $1-1/\poly(n)$. Here $\omega \approx 2.37$.
\end{theorem}
Here $\omega$ denotes the exponent of matrix multiplication \cite{w12,lg14,aw21,dwz23,lg23,wxxz23}, $\Tmat(a,b,c)$ denotes the time of multiplying an $a \times b$ size matrix with another $b \times c$ size matrix, and $\Tmat(n,n,n) = n^{\omega}$. See more details of matrix multiplication notation in Section~\ref{sub:preli:fast_matrix_multi}.

\paragraph{Relationship with the Softmax Regression Problem}

Moreover, the attention weight can be viewed as the output of a softmax regression model, which is defined as follows:
\begin{definition}[Single softmax regression \cite{dls23} and multiple softmax regression \cite{gsx23_incontext}]\label{def:softmax}
    Given a matrix $A \in \R^{n \times d}$ and a vector $c \in \R^n$, the single softmax regression problem is defined as
    \begin{align*}
     {\bf Part~1.}   \min_{x \in \R^d} \| \langle \exp(Ax) , {\bf 1}_n \rangle^{-1} \exp(Ax) - c \|_2^2 .
    \end{align*}

    Let $D(X) \in \R^{n \times n}$ be defined as in Definition~\ref{def:attention}. Given $A_1, A_2 \in \R^{n \times d}$ and $X \in \R^{d \times d}$
    \begin{align*}
    {\bf Part~2.} \min_{X \in \R^{d \times d}} \| D(X)^{-1} \exp(A_1 X A_2^\top) - C \|_F^2
\end{align*}
\end{definition}
On the one hand, due to the observation in \cite{gsx23_incontext,gsy23_coin}, the equation in Part 1 of Definition~\ref{def:softmax} can be viewed as one row of the equation in Part 2 of Definition~\ref{def:softmax}.

On the other hand, due to the well-known tensor trick\footnote{Given matrices $A_1, A_2 \in \R^{n \times d}$ and $X \in \R^{d \times d}$, the well-known tensor-trick suggests that $\vect(A_1 X A_2^\top) = (A_1 \otimes A_2) \vect(X) \in \R^{n^2}$.} (see \cite{dssw18,djs+19} as an example), the Part 2 equation Definition~\ref{def:softmax} is equivalent to 
\begin{align}\label{eq:softmax}
    \min_{X \in \R^{d \times d}}\| (D(X) \otimes I_n)^{-1} \exp(\A \vect(X) ) - \vect(C) \|_2^2,
\end{align}
which can be a slightly more complicated version of the Part 1 equation in Definition~\ref{def:softmax}. In particular, instead of one re-scaling factor, we will have $n$ rescaling factor. We split $\exp(\A \vect(X)) \in \R^{n^2}$ into $n$ chunks, and each chunk has size $n$. For each chunk, we use the same rescaling factor.

\begin{figure}[!ht]
    \centering
    \includegraphics[width = \linewidth]{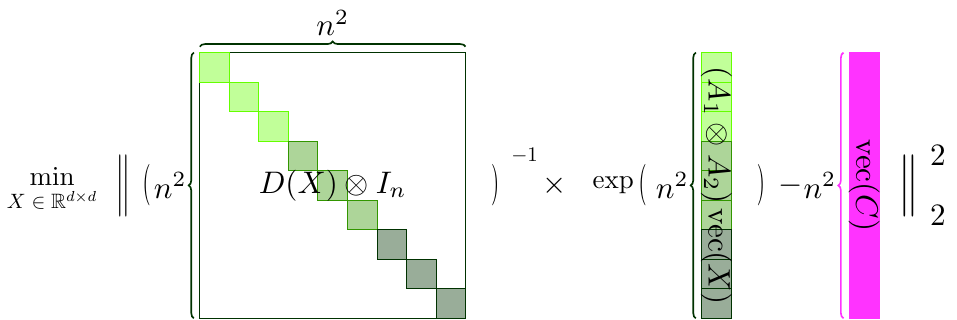}
    \caption{The visualization of Eq.~\eqref{eq:softmax}. Let $A_1, A_2 \in \R^{n \times d}$, $X \in \R^{d \times d}$, $C, D(X) \in \R^{n \times n}$ (see Figure~\ref{fig:attention_optimization} and Definition~\ref{def:attention}), and $\A = A_1 \otimes A_2 \in \R^{n^2 \times d^2}$. We first get that $(D(X) \otimes I_n)^{-1} \in \R^{n^2 \times n^2}$ and multiply $\A$ with $\vect(X)$. Then, we multiply $(D(X) \otimes I_n)^{-1} \in \R^{n^2 \times n^2}$ with $\A \cdot \vect(X) \in \R^{n^2}$. After that, we compute $c = \vect(C) \in \R^{n^2}$ and subtract it from $(D(X) \otimes I_n)^{-1} \exp(\A \cdot \vect(X))$. Finally, we compute the minimum of the $\ell_2$ norm of $(D(X) \otimes I_n)^{-1} \exp(\A \cdot \vect(X)) - c$. In this figure, we give an example when $n = 3$: in the matrix $D(X) \otimes I_n$, the three light green squares (and their nearby white area) make up the first chunk, the three middle green squares (and their nearby white area) make up the second chunk, and the three dark green squares (and their nearby white area) make up the third chunk. The purple rectangle represents $c = \vect(C) \in \R^{n^2}$.}
    \label{fig:eq1}
\end{figure}

Note that the multiple softmax regression problem is a simplified version of what we study in Definition~\ref{def:attention}. We believe that our work can also support the study of softmax regression.

\paragraph{Relatinship with Support Vector Machines (SVM)} 
The usual SVM \cite{j06,cl01,gsz23,tlto23} objective function in optimization can be viewed as a product of a summation of a batch of inner product. Inspired by that, we can define $n$ functions $f(x)_{j_0 }\in \R^n$ for each $j_0 \in [n]$ and $d$ functions $h(Y)_{i_0} \in \R^n$. Here $x$ is the vectorization of $X$ and $y$ is the vectorization of $Y$. Then the objective function in Definition~\ref{def:attention} $\| D(X)^{-1} \exp(A_1 X A_2^\top) A_3 Y - B \|_F^2$ can be turned into 
\begin{align}\label{eq:svm}
    \sum_{j_0=1}^n \sum_{i_0=1}^d ( \langle f(x)_{j_0}, h(Y)_{i_0} \rangle - b_{j_0,i_0} )^2
\end{align}
where $b_{j_0,i_0}$ is the entry of matrix $B \in \R^{n \times d}$. 
 We call this formulation SVM-inspired formulation.

\begin{figure}[!ht]
    \centering
    \includegraphics[width = 0.8\linewidth]{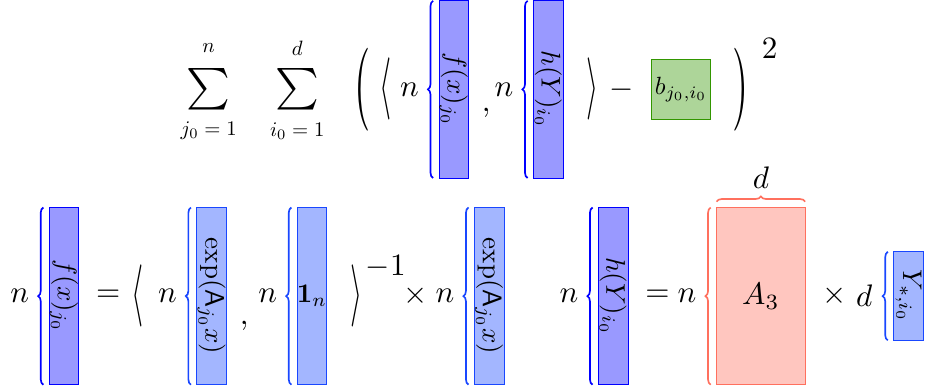}
    \caption{The visualization of Eq.~\eqref{eq:svm}. Let $A_1, A_2, A_3, B \in \R^{n \times d}$ and $X, Y \in \R^{d \times d}$. We have $\A = A_1 \otimes A_2 \in \R^{n^2 \times d^2}$ and $\A_{j_0} \in \R^{n \times d^2}$ is the $j_0$-th block of $\A$. $x = \vect(X) \in \R^{d^2}$. First, we use the definition of $f(x)_{j_0} \in \R^n$ (see Definition~\ref{def:f}) and $h(Y)_{i_0} \in \R^n$ (see Definition~\ref{def:h}) to compute them. Then, we find their inner produce and subtract the entry of $B$ at $j_0$-th row and $i_0$-column from the inner produce. Finally, we compute the square of this difference and add all of them from $i_0 = 1$ to $i_0 = d$ and from $j_0 = 1$ to $j_0 = n$. In this figure, we use blue rectangles to represent vectors, where the dark blue represents $f(x)_{j_0}$ and $h(Y)_{i_0}$, and the light blue represents the terms used to compute $f(x)_{j_0}$ and $h(Y)_{i_0}$. The green square represents the scalar. The red rectangle represents the matrix.}
    \label{fig:svm}
\end{figure}

\paragraph{Roadmap}

In Section~\ref{sec:related_work}, we introduce related research work. In Section~\ref{sec:tech_overview}, we provide an overview of the techniques we will use throughout the rest of the paper. In Section~\ref{sec:preli}, we present the basic notations we use, some mathematical facts, and helpful definitions that support the following proof. In Section~\ref{sec:gradient}, we compute the gradients of the helpful functions defined earlier. In Section~\ref{sec:hessian}, we define the Hessian for further discussion. In Section~\ref{sec:hessian_X}, we compute the Hessian matrix with respect to $X$. In Section~\ref{sec:lips_H_xx}, we demonstrate that the Hessian for $X$ is Lipschitz. In Section~\ref{sec:psd_H_xx}, we show that the Hessian matrix with respect to $X$ is positive semidefinite (PSD). In Section~\ref{sec:hessian_Y}, we compute the Hessian matrix with respect to $Y$ and show that it is Lipschitz and positive semidefinite (PSD). In Section~\ref{sec:hessian_XY}, we compute the Hessian matrix with respect to both $X$ and $Y$. In Section~\ref{sec:lips_H_xy}, we demonstrate that the Hessian matrix with respect to both $X$ and $Y$ is Lipschitz. In Section~\ref{sec:tensorsketch}, we introduce some tensor sketch techniques to obtain fast approximations of the Hessian. In Section~\ref{sec:newton}, we introduce the Newton step.

 %%% Section 1. Introduction

\section{Related Work}\label{sec:related_work}

\paragraph{Attention}

\cite{bcb14} represents one of the earliest works that employed attention in NLP. They assumed that a fixed-length vector could enhance the performance of the encoder-decoder design by incorporating an attention mechanism. This mechanism allows the decoder to focus on relevant words in the source sentence while generating translations. Consequently, this approach significantly improves the performance of machine translation models compared to those without an attention mechanism. Subsequently, \cite{lpm15} explained two variants of attention: local attention, which considers a subset of source words at a time, and global attention, which attends to all source words.

Attention finds extensive applications across various domains. In image captioning, \cite{xbk+15} utilizes attention matrices to align specific parts of an image with words in a caption. In the context of the Transformer model \cite{vsp+17}, attention matrices capture differences between words in a sentence. In the realm of graph neural networks, \cite{vcc+17} investigates these neural network architectures designed for graph-structured data, computing attention matrices between each node and its neighbors.

On the theoretical side, after the emergence of LLMs, there has been a substantial body of work dedicated to studying attention computation \cite{dms23, as23, zhdk23, clp+21, lsz23, bsz23, kkl20}. Notably, recent research by \cite{zhdk23, clp+21, kkl20} employs Locality Sensitive Hashing (LSH) techniques to approximate attention mechanisms. In particular, \cite{zhdk23} introduces $\mathsf{KDEformer}$, an efficient algorithm for approximating dot-product attention. This algorithm provides provable spectral norm bounds and outperforms various pre-trained models. Additionally, current research explores both static and dynamic approaches to calculating attention, as evidenced by the works of \cite{bsz23} and \cite{as23}. Furthermore, \cite{lsz23} delves into the regularization of hyperbolic regression problems, which involve functions like $\exp$, $\sinh$, and $\cosh$. Lastly, \cite{dms23} proposes randomized and deterministic algorithms for reducing the dimensionality of attention matrices in LLMs, achieving high accuracy while significantly reducing feature dimensions.

Additionally, numerous studies have attempted to analyze theoretical attention from the perspectives of optimization and convergence \cite{llr23, gms23, szks21, zkv+20}. \cite{llr23} investigated how transformers acquire knowledge about word co-occurrence patterns. \cite{gms23} focused on studying regression problems inspired by neural networks that employ exponential activation functions. \cite{szks21} analyzed why models occasionally prioritize significant words and explained how the attention mechanism evolves during the training process. \cite{zkv+20} demonstrated that the presence of a heavy-tailed noise distribution contributes to the bad performance of stochastic gradient descent (SGD) compared to adaptive methods.

\paragraph{Theoretical LLMs}

There are numerous amount of works focusing on the theoretical aspects of LLMs. In \cite{ryw+19}, the syntactic representations of the attention matrix and the individual word embeddings are presented, together with the mathematical justification of elucidating the geometrical properties of these representations. \cite{hm19} introduces a structural probe that analyzes, under the linear transformation of a word representation space of a neural network, whether or not syntax trees are embedded. 

\cite{clmy21,llh+23,rsm+23,kmh+20} study the optimization of LLMs. \cite{clmy21} proposes a new algorithm called ZO-BCD. It has favorable overall query complexity and a smaller computational complexity in each iteration. \cite{llh+23} creates a simple scalable second-order optimizer, called Sophia. In different parts of the parameter, Sophia adapts to the curvature. This may be strongly heterogeneous for language modeling tasks. The bound of the running time does not rely on the condition number of the loss.

Other theoretical LLM papers study the knowledge and skills of LLMs. \cite{wwz+22} analyzes distinct ``skill" neurons, which are regarded as robust indicators of downstream tasks when employing the process of soft prompt-tuning, as discussed in \cite{ll21}, for language models. \cite{ddh+21} find a positive relationship between the activation of these neurons and the expression of their corresponding facts, through analyzing BERT. Simultaneously, \cite{byks22} employs a fully unsupervised approach to extract latent knowledge from a language model's internal activations. In addition, \cite{hbkg23} and \cite{mbab22} show that in the feed-forward layers of pre-trained models, language models localize knowledge. \cite{xqp+22} explores the feasibility of selecting a specific subset of layers for modification and determining the optimal location for integrating a classifier. \cite{lyb+22} demonstrate that large trained transformers exhibit sparsity in their feedforward activations. Zero-th order algorithm for training LLM has been analyzed \cite{mgn+23,dlms23,zhl+23}.

\paragraph{LLMs Application and Evaluation}

Recently, there has been much interest in developing LLM-based systems for conversational AI and task-oriented dialogue, like Google's Meena chatbot \cite{r20}, Microsoft 365 Copilot \cite{s23}, Adobe firefly, Adobe Photoshop, GPT series \cite{rns+18,rwc+19,bmr+20,cha22,o23}, and BERT \cite{dclt18}.

Moreover, LLM evaluation is also a popular research area. Within the field of NLP, LLMs are evaluated based on natural language understanding \cite{bcl+23,lbl+22,lbr+23,cpk+23}, reasoning \cite{bhs+23,wqr+23,xlh+23}, natural language generation \cite{wlj+23,qzz+23,pd23,chbp23,cwj+23}, and multilingual tasks \cite{amc+23,aho+23,lnv+23,zag+23}. Robustness \cite{llgb23,whh+23,zpd+23}, ethics \cite{czl+23}, biases \cite{f23}, and trustworthiness \cite{hf23} are also important aspects. More specifically, the abilities of LLMs in social science \cite{dgg23,fr23,nkl+23}, mathematics \cite{as+23,dl23,wll+23,bce+23}, science \cite{cp23,ggl+23}, engineering \cite{bce+23,lxwz23,pmm+23,sm+23}, and medical applications \cite{clbj23,jgp+23} are evaluated.

\paragraph{Sketching}

Sketching is a powerful tool that is used to accelerate the performance of machine learning algorithms and optimization processes. The fundamental concept of sketching is to partition a large input matrix into a significantly smaller sketching matrix but still preserve the main characteristics of the original matrix. Therefore, the algorithms may work with the smaller matrix instead of the huge original, which leads to a substantial reduction in processing time. Many previous works have studied sketching, proposed sketching algorithms, and supported these algorithms with robust theoretical guarantees. For example, the Johnson-Lindenstrauss lemma is proposed by \cite{jl84}: it shows that under a certain high-dimensional space, projecting points to a lower-dimensional subspace may preserve the pairwise distances between these points. This mathematical property becomes the foundation of the development of faster algorithms for tasks such as nearest neighbor search. In addition, as explained in \cite{ac06}, the Fast Johnson-Lindenstrauss Transform (FJLT) introduces a specific family of structured random projections that can be applied to a matrix in input sparsity time.

More recently, sketching has been applied to many numerical linear algebra tasks, such as linear regression \cite{cw13,nn13}, dynamic kernel estimation \cite{qrs+22}, submodular maximization \cite{qsw23}, matrix sensing \cite{qsz23}, gradient-based algorithm \cite{xss21}, clustering \cite{dswy22,emz21}, convex programming \cite{sy21,qszz23,jswz21,jlsw20,lsz19}, online optimization problems \cite{rrs+21}, training neural networks \cite{szz21,xzz18,syz21,gqsw22,bpsw20}, reinforcement learning \cite{wzd+20,ssx23}, tensor decomposition \cite{swz19_tensor,dsy23}, relational database \cite{qjs+22}, low-rank approximation \cite{bw14,mrs20,mm13,als+18,swz17}, distributed problems \cite{bwz16,wz16}, weighted low rank approximation \cite{rsw16,gsyz23,syyz23_weight}, CP decomposition \cite{ms21}, regression inspired by softmax \cite{lsz23,gsy23_hyper,ssz23,dls23}, matrix sensing \cite{qsz23}, and Kronecker product regression \cite{rsz22}.

\paragraph{Second-order Method}

Second-order method have been used for solving many convex optimization and non-convex optimization problems, such as linear programming \cite{cls19,b20,jswz21,sy21,gs22,hlz23}, empirical risk minimization \cite{lsz19,qszz23}, support vector machines \cite{gsz23}, cutting plan method \cite{lsw15,jlsw20}, semi-definite programming \cite{jkl+20,hjs+22,gs22,syz23_sdp}, hyperbolic programming/polynomials \cite{dszz23,zz23}, streaming algorithm \cite{lsz+23,bs23,syz23_sdp}, federated learning \cite{bsy23}.

\paragraph{Convergence and Deep Neural Network Optimization}

Many works focus on analyzing optimization, convergence guarantees, and training improvement. \cite{ll18} shows that stochastic gradient descent optimizes over-parameterized neural networks on structured data, while \cite{dzps18} demonstrates that gradient descent optimizes over-parameterized neural networks. In \cite{azls19a}, a convergence theory for over-parameterized deep neural networks via gradient descent is developed. \cite{azls19b} analyzes the convergence rate of training recurrent neural networks. \cite{adh+19a} provides a fine-grained analysis of optimization and generalization for over-parameterized two-layer neural networks. \cite{adh+19b} studies exact computation with an infinitely wide neural network. \cite{cgh+19} proposes a Gram-Gauss-Newton method for optimizing over-parameterized neural networks. \cite{zg19} improves the analysis of the global convergence of stochastic gradient descent when training deep neural networks, requiring a milder over-parameterization compared to prior research. Other research, such as \cite{os20,jt19,zpd+20}, focuses on optimization and generalization, while \cite{gms23,lsz23} emphasize the convergence rate and stability. Works like \cite{bpsw20,szz21,als+22,mosw22,z22} concentrate on specialized optimization algorithms and techniques for training neural networks, and \cite{lss+20,hlsy21} concentrate on leveraging neural network structure.

\paragraph{Algorithmic Regularization}

There is a significant body of research exploring the latent bias inherent in gradient descent when applied to separable classification tasks. This research typically employs logistic or exponentially-tailed loss functions to maximize margins, as demonstrated in previous studies \cite{jt20, glss18, kpot21, jt21, shn+18, mwg+20, nlg+19}. These novel findings have also been applied to non-separable data through the utilization of gradient-based techniques \cite{jdst20, jt19_alg, jt18}. Analysis of implicit bias in regression problems and associated loss functions is carried out using methods such as mirror descent \cite{ykm20, aw20a, aw20b, vkr19, sata22, wgl+20, alh21, glss18} and stochastic gradient descent \cite{hwlm21, lwa21, lr20, zwb+21, dml21, lwm19, bgvv20}. These findings extend to the implicit bias of adaptive and momentum-based optimization methods \cite{jst21, wmcl21, wmz+21, qq19}.

\section{Technique Overview}\label{sec:tech_overview}
In this section, we will introduce the primary technique employed in this paper. The notations used in this section are presented in Preliminary (Section~\ref{sec:preli}). 

\subsection{Analysis}

\paragraph{Split Hessian into blocks ($X,Y$)}
In the fast approximation and convergence guarantee of the training process for the attention matrix, the positive semi-definite property is a key focus in Section~\ref{sec:hessian}. In comparison to single/multiple softmax regression, both the weights $X$ and $Y$ (refer to Definition~\ref{def:attention}) need to be considered. Therefore, our Hessian matrix discussed in Section~\ref{sec:hessian} has the following format
\begin{align*}
H = \begin{bmatrix}
H_{x,x} & H_{x,y} \\
H_{y,x} & H_{y,y}
\end{bmatrix}
\end{align*}
To establish the positive semi-definite property, we will examine the properties of the matrix above individually.

\paragraph{Positive Semi-Definite For Hessian $H_{x,x}$, $H_{y,y}$}
The positive semi-definite of the Hessian, denoted as ${ H_{x,x}, H_{y,y} }$, constitutes a crucial initial step in the proof outlined in Lemma~\ref{lem:hessian_lower_bound}. These Hessian are discussed in detail in Section~\ref{sec:psd_H_xx} and Section~\ref{sec:hessian_Y}.

Leveraging Lemma~\ref{lem:hessian_property:y} and Lemma~\ref{lem:hessian_property:x}, we can establish the following results if the regularization weight sufficiently large (see Section~\ref{sec:psd_H_xx} in details), then 
\begin{align*}
H(x) \succeq l\cdot I_{d^2} ~ ~\text{and}~~H(y) \succeq l\cdot I_{d^2}
\end{align*}
\paragraph{Spectral upper bound for $H_{x,y}$, $H_{y,x}$}
To establish the spectral upper bound of $H_{x,y}$, we can decompose $H_{x,y}$ into $\{ {G_i} \}_{i=1}^4$ as described in Lemma~\ref{lem:summary_Gi_xy_psd}. Building upon the results from Lemma~\ref{lem:summary_Gi_xy_psd}, we obtain:
$\max_{i \in [n]} \| G_i \| \leq R^2$.
The spectral upper bound for $H_{x,y}$ is then established in Lemma~\ref{lem:lipschitz_xy_G4} as follows:
$\| H(x,y) \| \leq nd \cdot 10 R^2$.

Given this upper bound, our final focus in the proof of the positive semi-definite property (PSD) will be as follows.

\paragraph{PSD for Hessian $H$}

 The Hessian matrix $H$ can be regarded as a combination of four matrices. The norm of the diagonal elements ($H_{x,x}$ in Section~\ref{sec:psd_H_xx} and $H_{y,y}$ in Section~\ref{sec:hessian_Y}) can be guaranteed to have a higher lower bound than $H_{x,y}$  and $H_{y,x}$ which are discussed in Section~\ref{sec:hessian_XY}. Consequently, based on the positive semi-definite property of the diagonal matrix, the computation of the off-diagonal part of the matrix does not affect the positivity of the entire matrix, thereby establishing a positive semi-definite. With $a_1$, $a_2$, $a_3$ as the bound of the matrix above respectively in Lemma~\ref{lem:hessian_lower_bound}, we have the following result 
\begin{align*}
    H \succeq \{ \alpha_1 - \alpha_3, \alpha_2 - \alpha_3 \} \cdot I_{2d^2}
\end{align*}
Given the relationship of $\{ a_i\}_{i = 1}^3$ as discussed above, the positive semi-definite property of the Hessian matrix is established.

\paragraph{Lipschitz property for Hessian}

The Lipschitz property of the Hessian is determined by the upper bound and Lipschitz property of the basic functions that constitute the Hessian matrix $H$. Since $H$ has three parts $H_{x,x}$, $H_{x,y}$ and $H_{y,y}$. In Section~\ref{sec:hessian_Y}, due to $H(y)$ is independent of $y$, the Lipschitz property can be easily established.
For details of others, we refer the readers to read Section~\ref{sec:lips_H_xy}. 

To compute the Lipschitz continuity of $H_{x,x}$, we will begin by providing a brief explanation. In our proof, we first establish upper bounds for the functions $u(x)$, $c(x)$, and $f(x)$ in Lemma~\ref{lem:upper_bound}, which together form the matrix $H_{x,x}$ (as detailed in Section~\ref{sub:preli:help_def_x}). Importantly, we ensure that these basic functions possess the Lipschitz property in Lemma~\ref{lem:basic_lips}.
Using the foundational components mentioned above, we can decompose $H_{x,x}$ into four distinct parts denoted as $\{ G_k \}_{k=1}^4$. We will leverage the Lipschitz property of the basic functions above and a method introduced below, 
The following task is extensively involved in the Lipschitz proof (for each $G_k$), we want to bound 
\begin{align*}
| \prod_{i=1}^t \beta_i(x) - \prod_{i=1}^t \beta_i(\wt{x}) |,
\end{align*}
which can be upper bounded by
\begin{align*}
    \sum_{j=0}^{t-1} | \prod_{i=0}^j \beta_i( \wt{x} ) \prod_{i'=j+1}^t \beta_{i'} ( x ) - \prod_{i=1}^{j+1} \beta_i( \wt{x} ) \prod_{i'=j+2}^{t+1} \beta_{i'} ( x )  |
\end{align*}
where assume that $\beta_0(x)= 1$ and $\beta_{t+1}(x) = 1$ for convenient.  
We will then proceed to establish the Lipschitz continuity of $H_{x,x}$

\begin{align*}
        \sum_{k=1}^K \|G_{k }(x,y) - G_{k}(\wt{x}, \wt{y})\| \leq n^{1.5} \exp(20R^2) ( \|x -\wt{x} \|_2 +  \| y -\wt{y} \|_2 )
    \end{align*}

\subsection{Algorithm}

\paragraph{Forward Computation}

To simplify the computation of the attention matrix, we can decompose the computation process into three components: $f$, $c$, and $h$ as defined in Section~\ref{sub:preli:general_def}. The forward computation can then be completed in $O(\Tmat(n,d,d) + \Tmat(n,n,d))$ time, as stated in Lemma~\ref{lem:forward_computation}.

\paragraph{Gradient Computation}
We can compute the gradient in Section~\ref{sec:gradient} as follows:
\begin{align*}
\frac{\d L(x,y)}{\d x} &= \vect ( A_1^\top p(x,y) A_2 ),
\end{align*}
for some matrix $p(x,y) \in \R^{n \times n}$.
Here $A_1^\top p(x,y) A_2$ can be computed in $\Tmat(n,d,n) + \Tmat(d,n,d)$ time. 
Similarly,
\begin{align*}
\frac{\d L(x,y)}{\d y} &= \vect( A_3^\top \widetilde{q}(x,y) ),
\end{align*}
which also takes $\Tmat(n,n,d) + \Tmat(n,d,d)$ time. We will now establish the overall running time for gradient computation.
By utilizing the results from Lemma~\ref{sub:gradient:reform_x} and Lemma~\ref{sub:gradient:reform_y}, we can efficiently compute the gradients of $g(x(t))$ and $g(y(t))$ in $\Tmat(n,d,n) + \Tmat(n,d,d)$ time.

\paragraph{Straightforward Hessian Computation}

Computing the Hessian in straightforward way would take $\Tmat(d^2, n^2, d^2)$ time, because we need to explicitly write down $\A^\top \A \in \R^{d^2 \times d^2}$ where $\A \in \R^{n^2 \times d^2}$. This is too slow, we will use sketching ideas to speed up this running time. Using sketching matrices to speed up the Hessian computation has been extensively studied in convex and non-convex optimization \cite{jswz21,lsz19,sy21,gs22,gsz23,qszz23}.

\paragraph{TensorSRHT Fast Approximation for Hessian}
Building upon the aforementioned properties, we can apply the Newton Method in Section~\ref{sec:newton} to establish convergence for the regression problem. Now, let's delve into the primary contribution of this paper. Given that $\A \in \R^{n^2 \times d^2}$ (Refer to Definition~\ref{def:u}), the time complexity of regression becomes prohibitively expensive. Our contribution aims to execute a fast approximation to significantly reduce the time complexity when using the Newton Method. Employing a matrix sketching approach, we can construct a sparse Hessian. This reduces the time from $\Tmat(d^2,n^2,d^2)$ down to $\wt{O}(nd) + \Tmat(d^2,d^2,d^2)$ (we consider the regime $n \gg d$ in the paper which is the most common setting in practice because $n$ is the length of the document, and $d$ is feature dimension).

\paragraph{Overall Time}

In Summary, we know that 
\begin{itemize}
    \item Computing forward function $\Tmat(n,n,d) + \Tmat(n,d,d)$ time (Lemma~\ref{lem:forward_computation})
    \item Computing gradient takes $\Tmat(n,n,d) + \Tmat(n,d,d)$ time (Lemma~\ref{lem:compute_gradient_x} and Lemma~\ref{lem:compute_gradient_y})
    \item Compute Hessian takes $\wt{O}(nd) + \Tmat(d^2,d^2,d^2)$ (Lemma~\ref{lem:compute_hessian_approximate})
    \item Compute $g$ times inverse of approximate hessian, this can be done in $\Tmat(d^2,d^2,d^2) = d^{2\omega}$
\end{itemize}

The total time can be expressed as $\widetilde{O}(\Tmat(n,d,n) + \Tmat(n,d,d) + d^{2\omega}) \log(1/\epsilon)$. Here $\omega$ is the exponent of matrix multiplication.

%\newpage
\section{Preliminary}\label{sec:preli}

\begin{table}[!ht]
    \centering
    \begin{tabular}{|c|c|p{6cm}|}
    \hline
       Previous works  & Simplified version of Def.~\ref{def:attention} & How Def.~\ref{def:attention} is simplified \\
    \hline
       \cite{zhdk23,as23,bsz23}  & $D^{-1} \exp(A_1 X A_2^\top) A_3 Y$ & $Q= A_1 X$, $K = A_2$, $V = A_3 Y$, both $X,Y$ are not considered \\
    \hline
    \cite{dls23} & $( D^{-1} \exp(A_1 X A_2^\top) )_{i,*}$ & one row, $A_3 Y$ are not considered  \\ \hline
    \cite{gsx23_incontext,gsy23_coin} & $D^{-1} \exp(A_1 X A_2^\top)$ & $A_3 Y$ are not considered \\ \hline
       \cite{gsyz23_quantum}  & $D^{-1} \exp(A_1 X A_2^\top)$ & $A_3 Y$ is not considered and need the symmetric assumption for matrix \\
    \hline
       \cite{dms23}  & $D^{-1} \exp(A_2 A_2^\top)$ & symmetric, $A_3Y$ is not considered \\
    \hline
       \cite{syz23}  & $A_1 X A_2^\top A_3$ & $D$ and $\exp$ are both removed, $Y$ is not considered\\
    \hline
    \end{tabular}
    \caption{Here $D := \diag( \exp( A_1 X A_2^\top ) {\bf 1}_n )$.}
    \label{tab:previous_attention}
\end{table}

In Section~\ref{sub:preli:basic_facts}, we present the basic mathematical properties of vectors, norms and matrices. In section~\ref{sub:preli:general_def}, we provide a definition of $L(X,Y)$. In Section~\ref{sub:preli:help_def_x}, we define a series of helpful functions with respect to $X$. In section~\ref{sub:preli:help_def_y}, we define a series of helpful functions with respect to $Y$. In Section~\ref{sub:preli:help_def_xy}, we define a series of helpful functions with respect to both $X$ and $Y$. In Section~\ref{sub:preli:regularization}, we define the regularization function. In Section~\ref{sub:preli:fast_matrix_multi}, we introduce facts related to fast matrix multiplication.

\paragraph{Notation}

Now we define the basic notations we use in this paper. 

First, we define the notations related to the sets. We use $\mathbb{N}$ to denote the set of positive integers, namely $\mathbb{N} := \{1, 2, 3, \dots\}$. Let $n$ and $d$ be in $\mathbb{N}$. We define $[n] := \{1, 2, \dots, n\}$. We use $\R, \R^n, \R^{n \times d}$ to denote the set containing all real numbers, all $n$-dimensional vectors, and $n \times d$ matrices, whose entries are all in $\R$. We use $\R_+$ to denote the set containing all positive real numbers.

Then, we define the notations related to vectors. Let $x,y \in \R^d$. For all $i \in [d]$, we define $x_i \in \R$ as the $i$-th entry of $x$. We define $\langle \cdot, \cdot \rangle : \R^d \times \R^d \to \R$ as $\langle x, y \rangle := \sum_{i = 1}^d x_i y_i$, which is called the inner product between $x$ and $y$. We define $x \circ y \in \R^d$ as $(x \circ y)_i := x_i \cdot y_i$, for all $i \in [d]$. For all $p \in \{1, 2, \infty\}$, we define $\|x\|_p : = (\sum_{i \in [d]} |x_i|^p)^{1/p}$, which is the $\ell_p$ norm of $x$. We use ${\bf 1}_d$ and ${\bf 0}_d$ to denote the $d$-dimensional vectors whose entries are all $1$'s and $0$'s, respectively.

After that, we define the notations related to matrices. Let $A \in \R^{n \times d}$. For all $i \in [n]$ and $j \in [d]$, we use $A_{i, j} \in \R$ to denote the entry of $A$ at $i$-th row and $j$-th column, use $A_{i, *} \in \R^d$ and $A_{*, j} \in \R^n$ to denote vectors, where $(A_{i, *})_j = A_{i, j} = (A_{*, j})_i$. We use $A^\top \in \R^{d \times n}$ to denote the transpose of the matrix $A$, where $A_{i, j}^\top = A_{j, i}$. For $X \in \R^{d \times d}$, we define $x = \vect(X) \in \R^{d^2}$ as $X_{i,j} = \vect(X)_{(i - 1) \times d + j}$. For $x \in \R^d$, we define $\diag(x) \in \R^{d \times d}$ as $\diag(x)_{i, i} = x_i$, for all $i \in [d]$ and other entries of $\diag(x)$ are all $0$'s. $\| A \|_F \in \R$ and $\|A\| \in \R$ denote the Frobenius norm and the spectral norm of $A \in \R^{n \times d}$, respectively, where $\|A\|_F := \sqrt{\sum_{i \in [n]} \sum_{j \in [d]} |A_{i, j}|^2}$ and $\| A \| := \max_{x \in \R^d} \| A x \|_2 / \| x \|_2$. Let $\A \in \R^{n^2 \times d^2}$. For each $j_1 \in [n]$, we use $\A_{j_1} \in \R^{n \times d^2}$ to denote one $n \times d^2$ block from $\A \in \R^{n^2 \times d^2}$. Let $C, D \in \R^{d \times d}$ be symmetric matrices, $C \succeq D$ if for all $y \in \R^{d}$, $y^\top C y \geq y^\top D y$. $C$ is said to be a positive semidefinite (PSD) matrix if $y^\top C y \geq 0$. We use $I_d$ to denote the $d \times d$ identity matrix. $\nnz(A)$ represents the number of entries in the matrix $A$ that are not equal to zero. ${\bf 0}_{n \times n} \in \R^{n \times n}$ is a matrix, where for all $i,j \in [n]$, $({\bf 0}_{n \times n})_{i, j} = 0$.

Let $n_1, n_2, d_1, d_2$ be positive integers. Let $A \in \R^{n_1 \times d_1}$ and $B \in \R^{n_2 \times d_2}$. We define the Kronecker product between matrices $A$ and $B$, denoted $A \otimes B \in \R^{n_1 n_2 \times d_1 d_2}$, as $(A \otimes B)_{(i_1 - 1) n_2 + i_2, (j_1-1)d_2+j_2}$ 
is equal to $A_{i_1,j_1} B_{i_2,j_2}$, where $i_1 \in [n_1], j_1 \in [d_1], i_2 \in [n_2], j_2 \in [d_2]$. $\mathrm{mat} : \R^{n^2} \to \R^{n \times n}$ is defined by $X_{i, j} = \mathrm{mat}(x)_{i, j} := x_{(i - 1) \cdot n + j}$, and $\vect = \mathrm{mat}^{-1}$.

\begin{figure}[!ht]
    \centering
    \includegraphics[width = 0.8\linewidth]{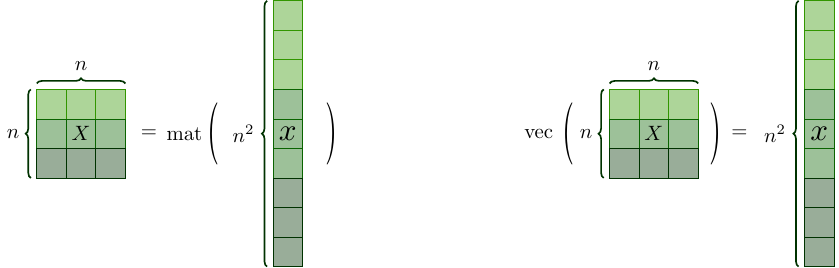}
    \caption{The visualization of the functions $\mathrm{mat} : \R^{n^2} \to \R^{n \times n}$ and $\vect = \mathrm{mat}^{-1} : \R^{n \times n} \to \R^{n^2}$. We have $x \in \R^{n^2}$ and $X \in \R^{n \times n}$. In this figure, we give an example of $n = 3$. In the left figure, by the function $\mathrm{mat}$, the first three entries of the vector $x$ are mapped to $X_{1, 1}$, $X_{1, 2}$, and $X_{1, 3}$ respectively, the second three entries of the vector $x$ are mapped to $X_{2, 1}$, $X_{2, 2}$, and $X_{2, 3}$ respectively, and the third three entries of the vector $x$ are mapped to $X_{3, 1}$, $X_{3, 2}$, and $X_{3, 3}$ respectively. For the right figure, every entry in $X$ is mapped to $x$ by $\vect$ in the reverse pattern of $\mathrm{mat}$.}
    \label{fig:vec_mat}
\end{figure}

\subsection{Basic Facts}\label{sub:preli:basic_facts}

In this section, we will introduce the basic mathematical facts. 

\begin{fact}\label{fac:circ_rules}
    Let $a, b \in \R$.
    
    For all vectors $u,v,w \in \R^{n}$, we have 
    \begin{itemize}
        \item $\langle u,v \rangle = \langle u \circ v, {\bf 1}_n \rangle =  u^\top \mathrm{diag}(v)  {\bf 1}_n $
        \item $\langle u \circ v, w\rangle = \langle u \circ w, v\rangle$
        \item $\langle u \circ v, w \rangle =  \langle u \circ v \circ w, {\bf 1}_n  \rangle = u^\top \diag(v) w$
        \item $\langle u \circ v \circ w \circ z , {\bf 1}_n \rangle = u^\top \diag(v \circ w) z$
        \item $u \circ  v = v \circ u = \diag (u) \cdot v = \diag (v) \cdot u$ 
        \item $u^{\top}(v \circ w) = v^{\top}(u \circ w) = w^{\top}(u \circ v)= u^{\top}\diag(v) w = v^{\top}\diag(u) w = w^{\top}\diag(u) v$
        \item $ \diag (u)^{\top} = \diag (u)$
        \item $\diag (u) \cdot \diag (v) \cdot {\bf 1}_n = \diag(u) v$
        \item $\diag (u \circ v) = \diag (u) \diag (v)$
        \item $\diag (u) + \diag (v) = \diag (u +v)$
        \item $\langle u,v \rangle = \langle v,u \rangle$
        \item $\langle u,v \rangle = u^\top v = v^\top u$
        \item $a\langle w, v \rangle + b\langle u, v \rangle = \langle aw + bu, v \rangle = \langle v, aw + bu \rangle = a\langle v, w \rangle + b\langle v, u \rangle$.
    \end{itemize}
\end{fact}
\begin{fact}\label{fac:vector_norm}
Let $R > 0$.

For vectors $x,y \in \R^n$, and a constant $\alpha \in \R$ we have
\begin{itemize}
    \item $\| x \circ y \|_2 \leq \| x \|_{\infty} \cdot \| y \|_2$
    \item $\| x \|_{\infty} \leq \| x \|_2 \leq \sqrt{n} \| x \|_{\infty}$
    \item $\| \exp(x) \|_{\infty} \leq \exp(\| x \|_2)$ 
    \item $\|x + y \|_2 \leq \| x\|_2 + \| y\|_2$
    \item $ \| \alpha x \|_2 \leq |\alpha| \cdot \| x\|_2$
    \item For any $\| x \|_2, \|y\|_2 \leq R$, we have $\| \exp(x) - \exp(y) \|_2 \leq \exp(R) \cdot \| x - y \|_2$
\end{itemize}
\end{fact}

\begin{fact}\label{fac:matrix_norm}
For matrices $X ,Y \in \R^{n \times n}$, and for any vector $x \in \R^n$, we have 
\begin{itemize}
    \item $\| X^\top \| = \| X \|$ 
    \item $\| X \| \geq \| Y \| - \| X - Y \|$
    \item $\| X + Y \| \leq \| X \| + \| Y \|$
    \item $\| X \cdot Y \| \leq \| X \| \cdot \| Y \|$ 
    \item If $X \preceq \alpha \cdot Y$, then $\| X \| \leq \alpha \cdot \| Y \|$
    \item $\|Yx \|_2 \leq \| Y\| \cdot \|x \|_2$
\end{itemize}
\end{fact}
\begin{fact}\label{fac:psd_rule}
    For any vectors $u,v \in \R^n$, we have
    \begin{itemize}
        \item Part 1. $u u^{\top} \preceq \| u\|_2^2 \cdot I_n $
        \item Part 2. $\diag(u) \preceq \|u\|_2 \cdot I_n$
        \item Part 3. $\diag(u \circ u) \preceq \|u\|_2^2 \cdot I_n$
        \item Part 4. $uv^{\top} + vu^{\top} \preceq uu^
        \top + vv^{\top}$
        \item Part 5. $uv^{\top} + vu^{\top} \succeq -( uu^
        \top + vv^{\top})$
        \item Part 6. $(v \circ u) (v \circ u)^{\top} \preceq \| v\|^2_{\infty} u u^{\top}$
        \item  Part 7. $\diag(u \circ v) \preceq \|u\|_2\|v\|_2 \cdot I_n$
    \end{itemize}
\end{fact}

\begin{fact} \label{fac:exponential_der_rule}
Let $g, f: \R^d \to \R^n$ and $q: \R^d \to \R$. 

Let $x \in \R^d$ be an arbitrary vector.

Let $a \in \R$ be an arbitrary real number.

Then, we have
    \begin{itemize}
        \item $\frac{\d q(x)^a}{\d x} =  a\cdot q(x)^{a-1} \cdot \frac{\d q(x)}{\d x}$
        \item $\frac {\d \|f(x) \|^2_2}{\d t} = 2 \langle f(x) , \frac{\d f(x)}{\d t} \rangle $
        \item $\frac{\d \langle f(x), g(x) \rangle}{\d t} = \langle \frac{\d f(x)}{\d t} , g(x) \rangle + \langle f(x), \frac{\d g(x)}{\d t} \rangle$
        \item $\frac{\d (g(x) \circ f(x))}{\d t} = \frac{\d g(x)}{\d t} \circ f(x) + g(x) \circ \frac{\d f(x)}{\d t}$ (product rule for Hadamard product)
    \end{itemize}
\end{fact}

\subsection{General Definitions}\label{sub:preli:general_def}

In this section, we introduce some general definitions.

\begin{definition}[Index summary]
We use $i$ to denote indices in $[d^2]$ range, and $j$ to denote indices in $[n^2]$ range.

We use $i_0, i_1, i_2$ to denote indices in $[d]$, and $j_0, j_1, j_2$ to denote indices in $[n]$.
\end{definition}

\begin{definition}\label{def:L}
If the following conditions hold
\begin{itemize}
    \item $A_1 \in \R^{n \times d}$
    \item $A_2 \in \R^{n \times d}$
    \item Let $\mathsf{A} \in \R^{n^2 \times d^2}$ denote the Kronecker product between $A_1, A_2$
    \begin{itemize}
        \item For each $j_0 \in [n]$, we use $\A_{j_0} \in \R^{n \times d^2}$ to be one $n \times d^2$ block from $\A \in \R^{n^2 \times d^2}$
    \end{itemize}
    \item $A_3 \in \R^{n \times d}$
    \item $B \in \R^{n \times d}$, let $b_{j_0,i_0}$ denote the $(j_0,i_0)$-th entry in $B \in \R^{n \times d}$ for each $j_0 \in [n]$ and $i_0 \in [d]$  
    \item $X \in \R^{d \times d}$
\end{itemize}
 Our final goal is to study the loss function, defined as:
\begin{align*}
   L(X,Y) := 0.5 \cdot \| \underbrace{ D(X)^{-1} }_{n \times n} \underbrace{ \exp(A_1 X A_2^\top) }_{n \times n} \underbrace{ A_3 }_{n \times d} \underbrace{Y}_{d \times d} - \underbrace{ B  }_{n \times d} \|_F^2
\end{align*}
where
\begin{itemize}
    \item we define $D(X) \in \R^{n \times n}$ as $D(X) := \diag( \exp(A_1 X A_2^\top ) {\bf 1}_n )$  
    \item For each $j_0 \in [n]$, we define $D(X)_{j_0,j_0} \in \R$ to be $\langle \exp( \A_{j_0} x ), {\bf 1}_n \rangle$ where $\A_{j_0} \in \R^{n \times d^2}$ is the $j_0$-th block of $\A \in \R^{n^2 \times d^2}$ and $x \in \R^{d^2}$ is the vectorization of $X \in \R^{d \times d}$  
\end{itemize}
Further, for each $j_0 \in [n], i_0 \in [d]$, we define $L(X,Y)_{j_0,i_0}$ as follows: 
\begin{align*}
    L(X,Y)_{j_0,i_0} := 0.5 ( \langle \langle \exp(\A_{j_0} x ) , {\bf 1}_n \rangle^{-1} \exp (\A_{j_0} x ) , A_3 Y_{*,i_0} \rangle - b_{j_0,i_0})^2 
\end{align*}

Using tensor-trick in \cite{gsx23_incontext,gsy23_coin}, we can see that
\begin{align*}
 L(X,Y) = \sum_{j_0=1}^n \sum_{i_0=1}^d L(X,Y)_{j_0,i_0}.
\end{align*}
\end{definition}

\subsection{Helpful Definitions With Respect to \texorpdfstring{$X$}{}}\label{sub:preli:help_def_x}

Now, we introduce a few helpful definitions related to $X \in \R^{d \times d}$.

\begin{definition}\label{def:u}
Let $\A = A_1 \otimes A_2 \in \R^{n^2 \times d^2}$, where $A_1, A_2 \in \R^{n \times d}$, and $\A_{j_0} \in \R^{n \times d^2}$ be one $n \times d^2$ block from $\A$.

We define $u(x)_{j_0}: \R^{d^2} \rightarrow \R^n$ as follows:
\begin{align*}
    u(x)_{j_0} := \underbrace{ \exp( \A_{j_0} x ) }_{n \times 1}.
\end{align*}
\end{definition}

\begin{definition}\label{def:alpha}
Let $\A = A_1 \otimes A_2 \in \R^{n^2 \times d^2}$, where $A_1, A_2 \in \R^{n \times d}$, and $\A_{j_0} \in \R^{n \times d^2}$ be one $n \times d^2$ block from $\A$.

We define $\alpha(x)_{j_0}: \R^{d^2} \rightarrow \R$ as:
\begin{align*}
  \alpha(x)_{j_0}:= \langle \underbrace{ \exp( \A_{j_0} x ) }_{n \times 1} , \underbrace{ {\bf 1}_n }_{n \times 1} \rangle.
\end{align*}
\end{definition}

\begin{definition}\label{def:f}

Let $\alpha(x)_{j_0} \in \R$ be defined as in Definition~\ref{def:alpha}.

Let $u(x)_{j_0} \in \R^n$ be defined as in Definition~\ref{def:u}.

We define $f(x)_{j_0} : \R^{d^2} \rightarrow \R^n$
\begin{align*}
    f(x)_{j_0} := \underbrace{ \alpha(x)_{j_0}^{-1} }_{ \mathrm{scalar} } \underbrace{ u(x)_{j_0} }_{ n \times 1 } .
\end{align*}
\end{definition}

\subsection{A Helpful Definition With Respect to \texorpdfstring{$Y$}{}}\label{sub:preli:help_def_y}

In this section, we introduce a helpful definition related to $Y \in \R^{d \times d}$.

\begin{definition}\label{def:h}
For each $i_0 \in [d]$, we define $h()_{i_0} : \R^{d \times d} \rightarrow \R^n$ as:
\begin{align*}
    h(Y)_{i_0}:= \underbrace{ A_3 }_{n \times d} \underbrace{ Y_{*,i_0} }_{d \times 1}.
\end{align*}

\end{definition}

\subsection{Helpful Definitions With Respect to Both \texorpdfstring{$X$}{} and \texorpdfstring{$Y$}{}} \label{sub:preli:help_def_xy}

In this section, we introduce some helpful definitions related to both $X \in \R^{d \times d}$ and $Y \in \R^{d \times d}$.

\begin{definition}\label{def:c}
We define $c(x,y)_{j_0,i_0}: \R^{d^2} \times \R^{d^2} \rightarrow \R$ as follows:
\begin{align*}
    c(x,y)_{j_0,i_0}:= \langle f(x)_{j_0}, h(y)_{i_0} \rangle - b_{j_0,i_0}.
\end{align*}
Furthermore, we define $c(x, :)_{j_0,i_0}$ as follows
\begin{align*}
    c(x,:)_{j_0,i_0} := \langle f(x)_{j_0}, v \rangle - b_{j_0,i_0}
\end{align*}
for some fixed vector $v \in \R^n$ which doesn't depend on $x$ and also doesn't depend on $y$. 

Similarly, we also define $c(:,y)_{j_0,i_0}$ as follows
\begin{align*}
    c(:,y)_{j_0,i_0} := \langle v, h(y)_{i_0} \rangle - b_{j_0,i_0}
\end{align*}
for some fixed vector $v \in \R^n$ which doesn't depend on $x$ and also doesn't depend on $y$.
\end{definition}

\begin{definition}\label{def:l}
We define
\begin{align*}
 L(x,:)_{j_0,i_0} := 0.5 c(x,:)_{j_0,i_0}^2
\end{align*}
and 
\begin{align*}
     L(:,y)_{j_0,i_0} := 0.5 c(:,y)_{j_0,i_0}^2
\end{align*}
and 
\begin{align*}
     L(x,y)_{j_0,i_0} := 0.5 c(x,y)_{j_0,i_0}^2
\end{align*}
\end{definition}

\subsection{Regularization}\label{sub:preli:regularization}

In this section, we define the regularization loss we use.

\begin{definition}
Let $W \in \R^{n \times n}$ denote a positive diagonal matrix. 
We use the following regularization loss
\begin{align*}
   \| (W \otimes I) (A_1 \otimes A_2)x \|_2^2 + \| W A_3 y \|_F^2 
\end{align*}
Note that $\| W A_3 y \|_F^2 = \sum_{i_0=1}^d \| W A_3 y_{i_0} \|_2^2$.
\end{definition}

\subsection{Fast Matrix Multiplication}\label{sub:preli:fast_matrix_multi}

We use $\Tmat(a,b,c)$ to denote the time of multiplying an $a \times b$ matrix with another $b \times c$ matrix. Fast  matrix multiplication \cite{c82,w12,lg14,gu18,cglz20,aw21,dwz23,lg23,wxxz23} is a fundamental tool in theoretical computer science.

\begin{fact}\label{fac:Tmat}
$\Tmat(a,b,c) = O( \Tmat(b,a,c) ) = O( \Tmat(a,c,b) )$.
\end{fact}

For $k \in \R_+$, we define $\omega(k) \in \R_+$ to be the value such that $\forall n \in \mathbb{N}$, $\Tmat(n,n,n^k) = O(n^{\omega(k)})$.

For convenience, we define three special values of $\omega(k)$. We define $\omega$ to be the fast matrix multiplication exponent, i.e., $\omega := \omega(1)$. We define $\alpha \in \R_+$ to be the dual exponent of matrix multiplication, i.e., $\omega(\alpha) = 2$. We define $\beta := \omega(2)$.

The following fact can be found in Lemma 3.6 of \cite{jkl+20}, also see \cite{bcs97}.
\begin{fact}[Convexity of $\omega(k)$]\label{fact:omega_k_convex}
The function $\omega(k)$ is convex.
\end{fact}

%\newpage
\section{Gradient}\label{sec:gradient}

In Section~\ref{sec:gradient:x}, we show the gradient with respect to variables $x$. In Section~\ref{sec:gradient:y}, we prove the gradient with respect to variables $y$. In Section~\ref{sub:gradient:compute_c_f_h}, we compute running time of $c,f,h$. In Section~\ref{sub:gradient:reform_x}, we reformulate the gradient with respect to $X$ to compute time complexity. In Section~\ref{sub:gradient:reform_y}, we reformulate the gradient with respect to $Y$ to compute time complexity.

\subsection{Gradient for \texorpdfstring{$x$}{}}\label{sec:gradient:x}

In this section, we compute the gradient for $x$. Most of the following gradient computations can be found in \cite{gsx23_incontext,gsy23_coin}. 
\begin{lemma}[Gradient with respect to $x$]\label{lem:gradient_x}
If the following conditions hold
\begin{itemize}
    \item For each $i \in [d^2]$, let $\A_{j_0,i} \in \R^n$ denote the $i$-th column for $\A_{j_0} \in \R^{n \times d}$  
     \item Let $u(x)_{j_0} \in \R^n$ be defined as Definition~\ref{def:u}
    \item Let $\alpha(x)_{j_0} \in \R$ be defined as Definition~\ref{def:alpha}
    \item Let $f(x)_{j_0} \in \R^n$ be defined as Definition~\ref{def:f}
    \item Let $c(x,:)_{j_0,i_0} \in \R$ be defined as Definition~\ref{def:c}
    \item Let $L(x,:)_{j_0,i_0} \in \R$ be defined as Definition~\ref{def:l}
\end{itemize}
Then, for each $i \in [d^2]$, for each $j_0 \in [n]$, we have
\begin{itemize}
    \item {\bf Part 1.}
    \begin{align*}
        \frac{ \d u(x)_{j_0} }{ \d x_i } = u(x)_{j_0} \circ \A_{j_0,i} 
    \end{align*}
    \item {\bf Part 2.} 
    \begin{align*}
        \frac{\d \alpha(x)_{j_0}}{\d x_i} = \langle u(x)_{j_0} \circ \A_{i_0,i} , {\bf 1}_n \rangle
    \end{align*}
    \item {\bf Part 3.}
    \begin{align*}
        \frac{ \d f(x)_{j_0} }{ \d x_i } = f(x)_{j_0} \circ \A_{j_0,i} - f(x)_{j_0} \cdot \langle f(x)_{j_0} , \A_{j_0,i}\rangle
    \end{align*}
    \item {\bf Part 4.} For a fixed vector $v \in \R^n$ (which doesn't depend on $x$), we have
    \begin{align*}
        \frac{ \d \langle f(x)_{j_0} , v \rangle }{ \d x_i } = \langle  f(x)_{j_0} \circ \A_{j_0,i}, v \rangle - \langle  f(x)_{j_0} , v \rangle \cdot \langle f(x)_{j_0}, \A_{j_0,i} \rangle
    \end{align*}
    \begin{figure}[!ht]
    \centering
    \includegraphics[width = \linewidth]{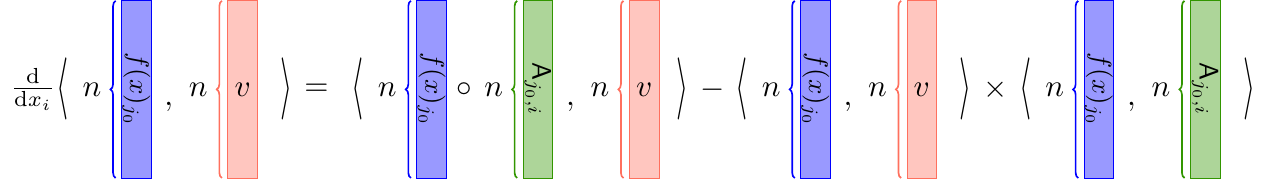}
    \caption{The visualization of Part 4 of Lemma~\ref{lem:gradient_x}. We are given $f(x)_{j_0} , v, \A_{j_0,i} \in \R^n$. The left-hand side of the equation is the derivative of the inner product of $f(x)_{j_0}$ and $v$ with respect to $x_i \in \R$. For the right-hand side, we have three steps. Step 1: we compute the Hadamard product of $f(x)_{j_0}$ and $\A_{j_0,i}$. Step 2: We find the inner product of this Hadamard product and $v$. Step 3: We subtract the product of two inner products, one is of $f(x)_{j_0}$ and $v$ and the other is of $f(x)_{j_0}$ and $\A_{j_0,i}$, from the result of step 2. The purple rectangles represent the vector $f(x)_{j_0}$. The red rectangles represent the vector $v$. The green rectangles represent the vector $\A_{j_0, i}$.}
    \label{fig:gradient_part_4}
\end{figure}
    \item {\bf Part 5.} For each $i_0 \in [d]$
    \begin{align*}
        \frac{ \d c(x,:)_{j_0,i_0} }{ \d x_i } = \langle  f(x)_{j_0} \circ \A_{j_0,i}, v \rangle - \langle  f(x)_{j_0} , v \rangle \cdot \langle f(x)_{j_0}, \A_{j_0,i} \rangle
    \end{align*}
    \item {\bf Part 6.}
    \begin{align*}
        \frac{\d L(x,:)_{j_0,i_0}}{\d x_i} = c(x,:)_{j_0, i_0} \cdot ( \langle  f(x)_{j_0} \circ \A_{j_0,i}, v \rangle - \langle  f(x)_{j_0} , v \rangle \cdot \langle f(x)_{j_0}, \A_{j_0,i} \rangle )
    \end{align*}
    \item {\bf Part 7.} (for hessian diagonal term)
    \begin{align*}
        \frac{ \d  \langle  f(x)_{j_0} \circ \A_{j_0,i}, v \rangle }{\d x_i} = \langle f(x)_{j_0} \circ \A_{j_0,i} \circ \A_{j_0,i}, v \rangle - \langle f(x)_{j_0} \circ \A_{j_0,i} , v \rangle \cdot \langle f(x)_{j_0} , \A_{j_0,i} \rangle
    \end{align*}
    \begin{figure}[!ht]
    \centering
    \includegraphics[width = \linewidth]{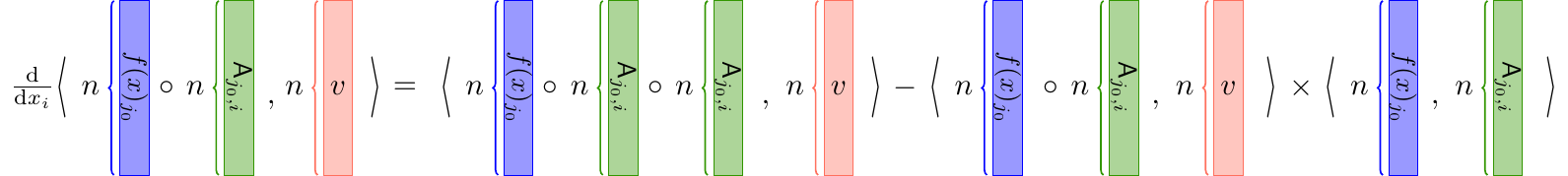}
    \caption{The visualization of Part 7 of Lemma~\ref{lem:gradient_x}. We are given $f(x)_{j_0} , v, \A_{j_0,i} \in \R^n$. First, we compute the Hadamard product between $f(x)_{j_0}$ and $\A_{j_0,i}$. The left-hand side of the equation is the derivative of the inner product of this Hadamard product and $v$ with respect to $x_i \in \R$. For the right-hand side, we have four steps. Step 1: We compute the inner product of the Hadamard product of $f(x)_{j_0}, \A_{j_0,i}, \A_{j_0,i}$ and $v$. Step 2: We compute the inner product of the Hadamard product of $f(x)_{j_0}, \A_{j_0,i}$ and $v$. Step 3: We compute the inner product between $f(x)_{j_0}$ and $\A_{j_0,i}$. Step 4: We subtract the product of steps 2 and 3 from step 1. The purple rectangles represent the vector $f(x)_{j_0}$. The red rectangles represent the vector $v$. The green rectangles represent the vector $\A_{j_0, i}$.}
    \label{fig:gradient_part_7}
\end{figure}
    \item {\bf Part 8.} (for hessian off-diagonal term)
    \begin{align*}
        \frac{ \d  \langle  f(x)_{j_0} \circ \A_{j_0,i}, v \rangle }{\d x_l} = \langle f(x)_{j_0} \circ \A_{j_0,i} \circ \A_{j_0,l}, v \rangle - \langle f(x)_{j_0} \circ \A_{j_0,l} , v \rangle \cdot \langle f(x)_{j_0} , \A_{j_0,i} \rangle
    \end{align*}
    \item {\bf Part 9} (for hessian diagonal term, this can be obtained by using Part 4 as a black-box)
    \begin{align*}
       \frac{ \d \langle f(x)_{j_0}, \A_{j_0,i} \rangle }{\d x_i} = \langle f(x)_{j_0} , \A_{j_0,i} \circ \A_{j_0,i} \rangle - \langle f(x)_{j_0}, \A_{j_0,i} \rangle \cdot \langle f(x)_{j_0}, \A_{j_0,i} \rangle
    \end{align*}
    \begin{figure}[!ht]
    \centering
    \includegraphics[width = \linewidth]{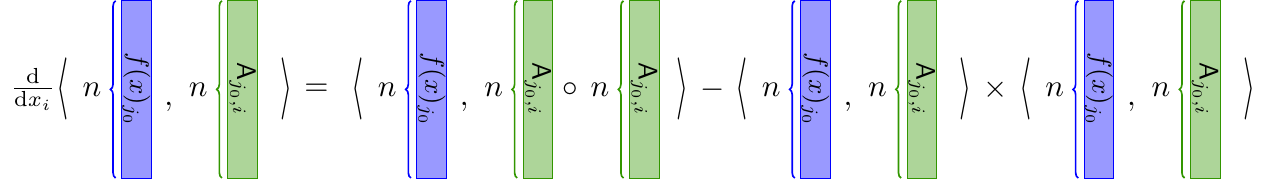}
    \caption{The visualization of Part 9 of Lemma~\ref{lem:gradient_x}. We are given $f(x)_{j_0}, \A_{j_0,i} \in \R^n$. The left-hand side of the equation is the derivative of the inner product of $f(x)_{j_0}$ and $\A_{j_0,i}$ with respect to $x_i \in \R$. For the right-hand side, we have three steps. Step 1: we compute the Hadamard product of $\A_{j_0,i}$ and $\A_{j_0,i}$. Step 2: We find the inner product of $f(x)_{j_0}$ and this Hadamard product. Step 3: We subtract the square of inner product of $f(x)_{j_0}$ and $\A_{j_0,i}$ from the result of step 2. The purple rectangles represent the vector $f(x)_{j_0}$. The green rectangles represent the vector $\A_{j_0, i}$.}
    \label{fig:gradient_part_9}
\end{figure}
    \item {\bf Part 10} (for hessian off-diagonal term, this can be obtained by using Part 4 as a black-box)
    \begin{align*}
       \frac{ \d \langle f(x)_{j_0}, \A_{j_0,i} \rangle }{\d x_l} = \langle f(x)_{j_0} , \A_{j_0,i} \circ \A_{j_0,l} \rangle - \langle f(x)_{j_0}, \A_{j_0,i} \rangle \cdot \langle f(x)_{j_0}, \A_{j_0,l} \rangle
    \end{align*}
\end{itemize}
\end{lemma}
\begin{proof}

{\bf Proof of Part 1.}
See Part 4 of Proof of Lemma 5.18 in \cite{gsx23_incontext} (Page 14).

{\bf Proof of Part 2.}
See Part 5 of Proof of Lemma 5.18 in \cite{gsx23_incontext} (Page 14).

{\bf Proof of Part 3.}
See Part 9 of Proof of Lemma 5.18 in \cite{gsx23_incontext} (page 15).

{\bf Proof of Part 4.}
See Part 14 of Proof of Lemma 5.18 in \cite{gsx23_incontext} (page 15).

{\bf Proof of Part 5.}

Note that by Definition~\ref{def:c}, we have
\begin{align}\label{eq:c_x:}
    c(x,:)_{j_0,i_0} := \langle f(x)_{j_0}, v \rangle - b_{j_0,i_0}
\end{align}

Therefore, we have
\begin{align*}
        \frac{ \d c(x,:)_{j_0,i_0} }{ \d x_i } 
        = & ~ \frac{ \d (\langle f(x)_{j_0}, v \rangle - b_{j_0,i_0}) }{ \d x_i } \\
        = & ~ \frac{ \d \langle f(x)_{j_0}, v \rangle }{ \d x_i } \\
        = & ~ \langle  f(x)_{j_0} \circ \A_{j_0,i}, v \rangle - \langle  f(x)_{j_0} , v \rangle \cdot \langle f(x)_{j_0}, \A_{j_0,i} \rangle,
    \end{align*}
    where the first step comes from Eq.~\eqref{eq:c_x:}, the second step follows from $\frac{\d b_{j_0,i_0}}{\d x_i} = 0$, and the third step is due to {\bf Part 4}.

{\bf Proof of Part 6.}
Noted that by Definition~\ref{def:l}, we have
\begin{align}\label{eq:l_x:}
    L(x,:)_{j_0,i_0} = 0.5 c(x,:)_{j_0,i_0}^2 
\end{align}

Therefore, we have
\begin{align*}
    \frac{\d L(x,:)_{j_0,i_0}}{\d x_i} 
    = & ~ \frac{\d (0.5 c(x,:)_{j_0,i_0}^2)}{\d x_i} \\
    = & ~ c(x,:)_{j_0,i_0} \frac{\d c(x,:)}{\d x_i} \\
    = & ~ c(x,:)_{j_0,i_0} \cdot (\langle  f(x)_{j_0} \circ \A_{j_0,i}, v \rangle - \langle  f(x)_{j_0} , v \rangle \cdot \langle f(x)_{j_0}, \A_{j_0,i} \rangle),
\end{align*}
where the first step is due to Eq.~\eqref{eq:l_x:}, the second step is because of chain rule of derivative, the last step comes from {\bf Part 5}.

{\bf Proof of Part 7.}

We have
\begin{align*}
    \frac{ \d  \langle  f(x)_{j_0} \circ \A_{j_0,i}, v \rangle }{\d x_i} 
    = & ~   \langle  \frac{ \d (f(x)_{j_0} \circ \A_{j_0,i}) }{\d x_i} , v \rangle \\
    = & ~   \langle  \frac{ \d f(x)_{j_0}}{\d x_i} \circ \A_{j_0,i}  , v \rangle \\
    = & ~   \langle  (f(x)_{j_0} \circ \A_{j_0,i} - f(x)_{j_0} \cdot \langle f(x)_{j_0} , \A_{j_0,i}\rangle) \circ \A_{j_0,i}  , v \rangle \\
    = & ~   \langle  f(x)_{j_0} \circ \A_{j_0,i} \circ \A_{j_0,i} - f(x)_{j_0} \cdot \langle f(x)_{j_0} , \A_{j_0,i}\rangle \circ \A_{j_0,i}  , v \rangle \\
    = & ~   \langle  f(x)_{j_0} \circ \A_{j_0,i} \circ \A_{j_0,i} , v \rangle - \langle f(x)_{j_0} \cdot \langle f(x)_{j_0} , \A_{j_0,i}\rangle \circ \A_{j_0,i}  , v \rangle \\
    = & ~   \langle  f(x)_{j_0} \circ \A_{j_0,i} \circ \A_{j_0,i} , v \rangle - \langle f(x)_{j_0} , \A_{j_0,i}\rangle \cdot \langle f(x)_{j_0} \circ \A_{j_0,i}  , v \rangle
\end{align*}
where the first step is due to Fact~\ref{fac:exponential_der_rule}, the second step comes from Fact~\ref{fac:exponential_der_rule}, the third step is because of  {\bf Part 4}, the fourth step is owing to simple algebra, the fifth step follows from Fact~\ref{fac:circ_rules}, and the last step comes from Fact~\ref{fac:circ_rules}.

{\bf Proof of Part 8.}

We have 
\begin{align*}
    \frac{ \d  \langle  f(x)_{j_0} \circ \A_{j_0,i}, v \rangle }{\d x_l} 
    = & ~ \langle  \frac{ \d (f(x)_{j_0} \circ \A_{j_0,i}) }{\d x_l} , v \rangle \\
    = & ~ \langle  \frac{ \d f(x)_{j_0}}{\d x_l} \circ \A_{j_0,i}  , v \rangle \\
    = & ~ \langle (f(x)_{j_0} \circ \A_{j_0, l} - f(x)_{j_0} \cdot \langle f(x)_{j_0}, \A_{j_0,l} \rangle) \circ \A_{j_0,i}, v \rangle \\
    = & ~ \langle  f(x)_{j_0} \circ \A_{j_0,i} \circ \A_{j_0,l} - f(x)_{j_0} \cdot \langle f(x)_{j_0} , \A_{j_0,l}\rangle \circ \A_{j_0,i}  , v \rangle \\
    = & ~ \langle  f(x)_{j_0} \circ \A_{j_0,i} \circ \A_{j_0,l} , v \rangle - \langle f(x)_{j_0} \cdot \langle f(x)_{j_0} , \A_{j_0,l}\rangle \circ \A_{j_0,i}  , v \rangle \\
    = & ~ \langle  f(x)_{j_0} \circ \A_{j_0,i} \circ \A_{j_0,l} , v \rangle - \langle f(x)_{j_0} , \A_{j_0,l}\rangle \cdot \langle f(x)_{j_0} \circ \A_{j_0,i}  , v \rangle
\end{align*}
where the first step comes from Fact~\ref{fac:exponential_der_rule}, the second step is because of Fact~\ref{fac:exponential_der_rule}, the third step follows from {\bf Part 4}, the fourth step is due to simple algebra, the fifth step is owing to Fact~\ref{fac:circ_rules}, and the last step comes from Fact~\ref{fac:circ_rules}.

{\bf Proof of Part 9.}

We have
\begin{align*}
    \frac{ \d \langle f(x)_{j_0}, \A_{j_0,i} \rangle }{\d x_i}
    = & ~  \langle \frac{ \d f(x)_{j_0} }{\d x_i}, \A_{j_0,i} \rangle \\
    = & ~  \langle f(x)_{j_0} \circ \A_{j_0,i} - f(x)_{j_0} \cdot \langle f(x)_{j_0} , \A_{j_0,i}\rangle, \A_{j_0,i} \rangle \\
    = & ~ \langle f(x)_{j_0} , \A_{j_0,i} \circ \A_{j_0,i} \rangle - \langle f(x)_{j_0}, \A_{j_0,i} \rangle \cdot \langle f(x)_{j_0}, \A_{j_0,i} \rangle
\end{align*}
where the first step is due to Fact~\ref{fac:exponential_der_rule}, the second step comes from {\bf Part 4}, and the last step is because of Fact~\ref{fac:circ_rules}.

{\bf Proof of Part 10.}
We have 
\begin{align*}
     \frac{ \d \langle f(x)_{j_0}, \A_{j_0,i} \rangle }{\d x_l}
     = & ~  \langle \frac{ \d f(x)_{j_0}}{\d x_l}, \A_{j_0,i}  \rangle \\
     = & ~ \langle f(x)_{j_0} \circ \A_{j_0,l} - f(x)_{j_0} \cdot \langle f(x)_{j_0} , \A_{j_0,l}\rangle, \A_{j_0,i} \rangle \\
     = & ~ \langle f(x)_{j_0} , \A_{j_0,i} \circ \A_{j_0,l} \rangle - \langle f(x)_{j_0}, \A_{j_0,i} \rangle \cdot \langle f(x)_{j_0}, \A_{j_0,l} \rangle
\end{align*}
where the first step comes from Fact~\ref{fac:exponential_der_rule}, the second step is owing to {\bf Part 4}, and the last step is due to Fact~\ref{fac:circ_rules}.
\end{proof}

\subsection{Gradient With Respect to \texorpdfstring{$y$}{}}\label{sec:gradient:y}

In this section, we compute the gradient with respect to $y$.

\begin{lemma}\label{lem:gradient_y}
If the following conditions hold
\begin{itemize}
\item  Let $v \in \R^n$ which doesn’t depend on x and also doesn’t depend on y.
\item Let $c(:,y)_{j_0,i_0} \in \R$ be defined as Definition~\ref{def:c}.
\item Let $L(:,y)_{j_0,i_0} \in \R$ be defined as Definition~\ref{def:l}.
\item Let 
 $h(y_{i_0}):= \underbrace{ A_3 }_{n \times d} \underbrace{ y_{i_0} }_{d \times 1}.$
 \item Let $h(y_{i_0}) = h(y)_{i_0}$ for convenient
 \item Let $A_{3,*,i_2} \in \R^n$ denote the $i_2$-th column of matrix $A_{3} \in \R^{n \times d}$ for each $i_2 \in [d]$
 \end{itemize}
 Then, we have
 \begin{itemize}
 \item {\bf Part 1.} If $i_1=i_0$
 \begin{align*}
    \frac{\d h(y_{i_0}) }{ \d y_{i_1,i_2} } = A_{3,*, i_2}
 \end{align*}
 \item {\bf Part 2.} If $i_1\neq i_0$
 \begin{align*}
    \frac{\d h(y_{i_0}) }{ \d y_{i_1,i_2} } = {\bf 0}_n
 \end{align*}
 \item {\bf Part 3.} If $i_1 = i_0$
 \begin{align*}
    \frac{\d \langle v, h(y)_{i_0} \rangle }{ \d y_{i_1,i_2}} = \langle v, A_{3,*,i_2} \rangle
 \end{align*}
 \item {\bf Part 4.} If $i_1 \neq i_0$
  \begin{align*}
    \frac{\d \langle v, h(y)_{i_0} \rangle }{ \d y_{i_1,i_2}}  = 0
 \end{align*}
\item {\bf Part 5.} If $i_1 = i_0$
 \begin{align*}
    \frac{\d c(:,y)_{j_0,i_0} }{ \d y_{i_1,i_2}} = \langle v, A_{3,*,i_2} \rangle
 \end{align*}
 \item {\bf Part 6.} If $i_1 \neq i_0$
  \begin{align*}
    \frac{\d c(:,y)_{j_0,i_0} }{ \d y_{i_1,i_2}}  = 0
 \end{align*}
 \item {\bf Part 7.} If $i_1 = i_0$
 \begin{align*}
    \frac{\d L(:,y)_{j_0,i_0} }{ \d y_{i_1,i_2}} = c(:,y)_{j_0,i_0} \langle v, A_{3,*,i_2} \rangle
 \end{align*}
 \item {\bf Part 8.} If $i_1 \neq i_0$
  \begin{align*}
    \frac{\d L(:,y)_{j_0,i_0} }{ \d y_{i_1,i_2}}  = 0
 \end{align*}
 \end{itemize}
 \end{lemma}
 \begin{proof}
{\bf Proof of Part 1.}
\begin{align*}
    \frac{\d h(y_{i_0})}{\d y_{i_1,i_2} }
    = & ~ \frac{\d A_{3} y_{i_0}}{\d y_{i_1,i_2}} \\
    = & ~ A_{3,*,i_2}
\end{align*}
where the first step is due to the definition of $h(y_{i_0})$ (see the Lemma statement), and the last step comes from that for $i \neq i_2,~\frac{\d}{\d y_{i_2}}f(y_{i}) = 0$.

{\bf Proof of Part 2.}
\begin{align*}
    \frac{\d h(y_{i_0})}{\d y_{i_1,i_2} } = & ~ {\bf 0}_n
\end{align*}
where the first step is due to $i_1 \neq i_2$.

{\bf Proof of Part 3.}
\begin{align*}
    \frac{\d \langle v, h(y)_{i_0} \rangle }{ \d y_{i_1,i_2}} = & ~ \langle v, \frac{\d h(y_{i_0})}{\d y_{i_1,i_2}} \rangle \\
    = & ~ \langle v, A_{3,*,i_2} \rangle
\end{align*}
where the first step comes from Fact~\ref{fac:exponential_der_rule}, the second step is due to the result of {\bf Part 1}.

{\bf Proof of Part 4.}
\begin{align*}
    \frac{\d \langle v, h(y)_{i_0} \rangle }{ \d y_{i_1,i_2}}  
    = & ~ \langle v, \frac{\d h(y_{i_0})}{\d y_{i_1,i_2}} \rangle \\
    = & ~ 0
\end{align*}
where the first step is becaues of Fact~\ref{fac:exponential_der_rule}, the second step comes from the result of {\bf Part 2}.

{\bf Proof of Part 5.}
\begin{align*}
    \frac{\d c(:,y)_{j_0,i_0} }{ \d y_{i_1,i_2}} 
    =  & ~ \frac{\d \langle v, h(y)_{i_0} \rangle - b_{j_0,i_0}}{\d y_{i_1,i_2}}\\\
    = & ~ \frac{\d \langle v, h(y)_{i_0} \rangle }{ \d y_{i_1,i_2}} \\
    = & ~ \langle v, A_{3,*,i_2} \rangle
\end{align*}
where the first step comes from the Definition~\ref{def:c}, the second step is because of $\frac{\d b_{j_0,i_0} }{ \d y_{i_1,i_2}} = 0$, and the last step is due to {\bf Part 3}.

{\bf Proof of Part 6.}
\begin{align*}
    \frac{\d c(:,y)_{j_0,i_0} }{ \d y_{i_1,i_2}} 
    =  & ~ \frac{\d \langle v, h(y)_{i_0} \rangle - b_{j_0,i_0}}{\d y_{i_1,i_2}}\\\
    = & ~ \frac{\d \langle v, h(y)_{i_0} \rangle }{ \d y_{i_1,i_2}} \\
    = & ~ 0
\end{align*}
where the first step is due to the Definition~\ref{def:c}, the second step comes from $\frac{\d b_{j_0,i_0} }{ \d y_{i_1,i_2}} = 0$, and the last step is owing to {\bf Part 4}.

{\bf Proof of Part 7.}
\begin{align*}
    \frac{\d L(:,y)_{j_0,i_0} }{ \d y_{i_1,i_2}} 
    = & ~ \frac{\d 0.5 c(:,y)_{j_0,i_0}^2}{\d y_{i_1,i_2}} \\
    = & ~ c(:,y)_{j_0,i_0} \cdot \frac{\d c(:,y)_{j_0,i_0}}{\d y_{i_1,i_2}} \\
    = & ~  c(:,y)_{j_0,i_0} \langle v, A_{3,*,i_2} \rangle
\end{align*}
where the first step is due to the Definition~\ref{def:l}, the second step comes from the chain rule of derivative, and the last step is owing to {\bf Part 5}.

{\bf Proof of Part 8.}
\begin{align*}
    \frac{\d L(:,y)_{j_0,i_0} }{ \d y_{i_1,i_2}} 
    = & ~ \frac{\d 0.5 c(:,y)_{j_0,i_0}^2}{\d y_{i_1,i_2}} \\
    = & ~ c(:,y)_{j_0,i_0} \cdot \frac{\d c(:,y)_{j_0,i_0}}{\d y_{i_1,i_2}} \\
    = & ~  0
\end{align*}
where the first step is because of the Definition~\ref{def:l}, the second step is due to the chain rule of derivative, and the last step comes from {\bf Part 6}.
 \end{proof}

 \subsection{Computation of \texorpdfstring{$c,f,h$}{}}\label{sub:gradient:compute_c_f_h}

 In this section, we explain how to compute $c(x,y), f(x), h(y)$.
\begin{lemma}\label{lem:forward_computation}
If the following conditions hold
\begin{itemize}
    \item For each $j_0 \in [n]$, $i_0 \in [d]$, let $c(x,y)_{j_0,i_0} \in \R$ be defined as Definition~\ref{def:c}. (We can view $c(x,y)$ as an $n \times d$ matrix)
    \item For each $j_0 \in [n]$, let $f(x)_{j_0} \in \R^n$ be defined as Definition~\ref{def:f}. (We can view $f(x)$ as an $n \times n$ matrix)
    \item For each $i_0 \in [d]$, let $h(y)_{i_0} \in \R^n$ be defined as Definition~\ref{def:h}. (We can view $h(y)$ as $n \times d$ matrix)
    \item Let $A_3 \in \R^{n \times d}$
    \item We can view $y$ as an $d \times d$ matrix
\end{itemize}
Then, we can compute $f,h, c$ in $O(\Tmat(n,d,d) + \Tmat(n,n,d))$ time.
\end{lemma}
\begin{proof}
By definition~\ref{def:h}, we have 
\begin{align}\label{eq:h_y}
      \underbrace{h(y) }_{n \times d} =  \underbrace{ A_3 }_{n \times d} \underbrace{ y }_{d \times d}.
\end{align}
First $h(y) \in \R^{n \times d}$ can be viewed as multiplying $n \times d$ matrix ($A_3$) and $d \times d$ matrix ($y$), this can be computed in $\Tmat(n,d,d)$.

\begin{figure}[!ht]
    \centering
    \includegraphics[width = 0.6\linewidth]{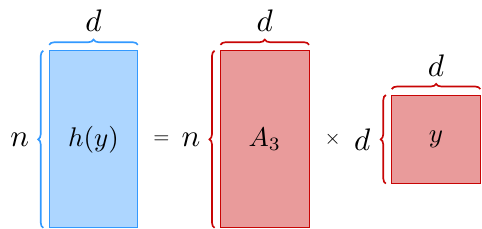}
    \caption{The visualization of Eq.~\eqref{eq:h_y}. We have $A_3 \in \R^{n \times d}$. $h: \R^{d \times d} \to \R^{n \times d}$ is a function, which maps the matrix $y \in \R^{d \times d}$ to $h(y)$ by multiplying $A_3$ and $y$. The red rectangles represent matrices which are the factors, and the blue rectangle represents the matrix which is the product.}
    \label{fig:h_y}
\end{figure}

We also have
\begin{align}\label{eq:f_x}
    \underbrace{ f(x) }_{n \times n} = \underbrace{ D(X)^{-1} }_{n \times n} \exp( \underbrace{ A_1 }_{n \times d} \underbrace{ X }_{d \times d} \underbrace{ A_2^\top }_{d \times n} ), \mathrm{~~~and~~~} D(X) = \diag( \exp(A_1XA_2^\top) {\bf 1}_n )
\end{align}
Then the computation of $f(x) \in \R^{n \times n}$ can be done in $\Tmat(n,n,d) + \Tmat(n,d,d)$.

\begin{figure}[!ht]
    \centering
    \includegraphics[width = \linewidth]{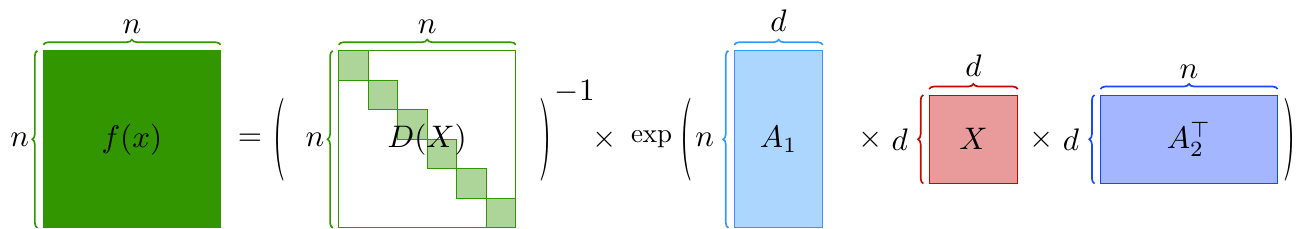}
    \caption{The visualization of Eq.~\eqref{eq:f_x}. We have $A_1, A_2 \in \R^{n \times d}$, $X \in \R^{d \times d}$, and $D(X) \in \R^{n \times n}$ (see Definition~\ref{def:attention} and Figure~\ref{fig:attention_optimization}). First, we find the inverse of the matrix $D(X)$ and compute $\exp(A_1 X A_2^\top) \in \R^{n \times n}$, as shown in Figure~\ref{fig:attention_optimization}. Then, we multiply $D(X)^{-1}$ and $\exp(A_1 X A_2^\top)$ to get $f(x) \in \R^{n \times n}$. The green squares represent the square matrices in $\R^{n \times n}$. The blue rectangles represent the matrices in $\R^{n \times d}$ (the dark blue denotes the transpose of the matrix in $\R^{n \times d}$). The red square represents the square matrices in $\R^{d \times d}$.}
    \label{fig:f_x}
\end{figure}

Given that 
\begin{align}\label{eq:c_xy}
    \underbrace{ c (x,y) }_{n \times d} = \underbrace{ f(x) }_{n \times n} \underbrace{ h(y) }_{n \times d} - \underbrace{ B }_{n \times d}
\end{align}

Then $c$ can be done in $\Tmat(n,n,d)$.

\begin{figure}[!ht]
    \centering
    \includegraphics[width = 0.7\linewidth]{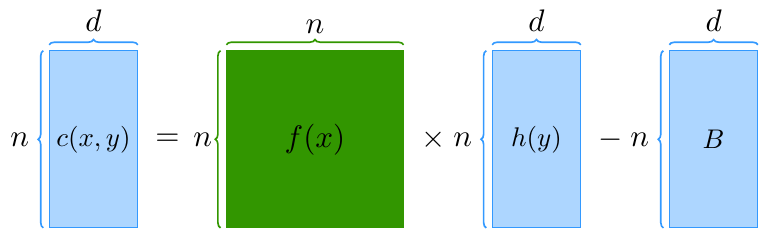}
    \caption{The visualization of Eq.~\eqref{eq:c_xy}. Let $f(x) \in \R^{n \times n}$ (see Figure~\ref{fig:f_x}) and $h(y) \in \R^{n \times d}$ (see Figure~\ref{fig:h_y}). We have $B \in \R^{n \times d}$. We multiply $f(x)$ with $h(y)$ and subtract $B$ from their product to get $c(x, y) \in \R^{n \times d}$. The green square represents the square matrices in $\R^{n \times n}$. The blue rectangles represent the matrix in $\R^{n \times d}$.}
    \label{fig:c_xy}
\end{figure}
\end{proof}

 \subsection{Reformulating Gradient (\texorpdfstring{$x$}{}) in Matrix View}\label{sub:gradient:reform_x}

 In this section, we reformulate the gradient $x$ in the matrix's view.

 \begin{lemma}\label{lem:compute_gradient_x}
If the following conditions hold
    \begin{itemize}
        \item  $\frac{\d L(x,y)_{j_0,i_0}}{\d x_i} 
    =  c(x,y)_{j_0,i_0} \cdot (\langle  f(x)_{j_0} \circ \A_{j_0,i}, h(y)_{i_0} \rangle - \langle  f(x)_{j_0} , h(y)_{i_0} \rangle \cdot \langle f(x)_{j_0}, \A_{j_0,i} \rangle)$
    \item Let $c(x,y) \in \R^{n \times d}$
    \item Let $f(x)_{j_0} \in \R^n$
    \item Let $h(y)_{i_0} \in \R^n$
    \item Let $\frac{\d L(x,y)}{\d x} = \sum_{j_0=1}^n \sum_{i_0=1}^d \frac{\d L(x,y)_{j_0,i_0} }{ \d x }$
    \item Let \begin{align*}
        q(x,y)_{j_0} = \sum_{i_0=1}^d c(x,y)_{j_0,i_0} h(y)_{i_0}
\end{align*}
    \end{itemize}
    then, we have
    \begin{itemize}
        \item {\bf Part 1.}
        \begin{align*}
            \frac{\d L(x,y)_{j_0,i_0} } { \d x } = \underbrace{ c(x,:)_{j_0,i_0} }_{ \mathrm{scalar} } \cdot \underbrace{ \A_{j_0}^\top }_{d^2 \times n} \underbrace{ ( f(x)_{j_0} - f(x)_{j_0} f(x)_{j_0}^\top ) }_{n \times n} \underbrace{ h(y)_{i_0} }_{n \times 1}  
        \end{align*}
        \item {\bf Part 2.} Suppose $c(x,y), \A, f(x), h(y)$ are given, then $\frac{\d L(x,y)_{j_0,i_0} } { \d x } $ can be computed in $O(n d^2 )$ time.
        \item {\bf Part 3.} 
        \begin{align*}
            \frac{\d L(x,y)}{\d x} = \sum_{j_0=1}^n \A_{j_0}^\top (f(x)_{j_0} - f(x)_{j_0} f(x)_{j_0}) q(x,y)_{j_0}
        \end{align*}
        \item {\bf Part 4.}
        Suppose $c(x,y), \A, f(x), h(y)$ are given, then 
        $
            \frac{\d L(x,y)}{\d x} \in \R^{d^2} 
       $
        can be computed in $\Tmat(n,d,n) + \Tmat(n,d,d)$ time
    \end{itemize}
 \end{lemma}
 \begin{proof}

{\bf Proof of Part 1.}

Note that by Fact~\ref{fac:circ_rules}, we have
\begin{align*}
    \langle  f(x)_{j_0} \circ \A_{j_0,i}, h(y)_{i_0} \rangle = \A_{j_0,i}^\top \diag(f(x)_{j_0}) h(y)_{i_0}
\end{align*}
and 
\begin{align*}
    \langle  f(x)_{j_0} , v \rangle \cdot \langle f(x)_{j_0}, \A_{j_0,i} \rangle
    = \A_{j_0,i}^\top f(x)_{j_0} f(x)_{j_0}^\top h(y)_{i_0}
\end{align*}
Thus, we complete the proof.

{\bf Proof of Part 2.}

We first compute 
$( \diag( f(x)_{j_0} ) - f(x)_{j_0} f(x)_{j_0}^\top ) h(y)_{i_0}$, this can be done in $O(n)$ time.

Then we can compute the rest, it takes $O(nd^2)$ time.

{\bf Proof of Part 3 and Part 4.}

Firstly, we can compute $q(x,y)_{j_0}  \in \R^n$. 

Recall from the Lemma statement, we have 
\begin{align}\label{eq:q_xy_j0}
q(x,y)_{j_0} = \sum_{i_0=1}^d c(x,y)_{j_0,i_0} h(y)_{i_0}.
\end{align}

Let $q(x,y)_{j_0} \in \R^n$ denote the $j_0$-th column of $q(x,y)$.

Then we have
\begin{align*}
    q(x,y) = \underbrace{ h(y) }_{n \times d} \underbrace{ c(x,y)^\top }_{d \times n}
\end{align*}

This takes $\Tmat(n,d,n)$ time.

Then, we compute 
\begin{align}\label{eq:p_xy_j0}
p(x,y)_{j_0} = ( \diag( f(x)_{j_0} ) - f(x)_{j_0} f(x)_{j_0}) q(x,y)_{j_0}.
\end{align}

This takes $O(n^2)$ time in total.

We can show that
\begin{align*}
    & ~ \frac{\d L(x,y)}{\d x} \\
    = & ~ \sum_{j_0=1}^n \sum_{i_0=1}^d \frac{\d L(x,y)_{j_0,i_0} }{ \d x } \\
    = & ~ \sum_{j_0=1}^n \sum_{i_0=1}^d \underbrace{ c(x,y)_{j_0,i_0} }_{ \mathrm{scalar} } \cdot \underbrace{ \A_{j_0}^\top }_{d^2 \times n} \underbrace{ ( \diag( f(x)_{j_0} ) - f(x)_{j_0} f(x)_{j_0}^\top ) }_{n \times n} \underbrace{ h(y)_{i_0} }_{n \times 1}  \\
    = & ~ \sum_{j_0=1}^n \A_{j_0}^\top ( \diag( f(x)_{j_0} ) - f(x)_{j_0} f(x)_{j_0}) q(x,y)_{j_0} \\
    = & ~ \sum_{j_0=1}^n \A_{j_0}^\top p(x,y)_{j_0} \\
    = & ~ \vect( A_1^\top p(x,y) A_2 )
\end{align*}
where the first step is based on Definition~\ref{def:L}, the second step is because of {\bf Part 1}, the third step is due to Eq.~\eqref{eq:q_xy_j0}, the fourth step follows from Eq.~\eqref{eq:p_xy_j0}, and the last step due to tensor-trick.

Note that $A_1^\top p(x,y) A_2$ can be computed in $\Tmat(n,d,n) + \Tmat(d,n,d)$ time.
 \end{proof}

\subsection{Reformulating Gradient (\texorpdfstring{$y$}{}) in Matrix View}\label{sub:gradient:reform_y}

In this section, we reformulate the gradient $y$ in the matrix's view.

\begin{lemma}\label{lem:compute_gradient_y}
If the following conditions hold
\begin{itemize}
    \item if $i_1=i_0$, $\frac{\d L(x,y)_{j_0,i_0} }{ \d y_{i_1,i_2}} = c(x,y)_{j_0,i_0} \langle f(x)_{j_0}, A_{3,*,i_2} \rangle$
    \item if $i_1\neq i_0$, $\frac{\d L(x,y)_{j_0,i_0} }{ \d y_{i_1,i_2}} =0$
    \item Let $\frac{\d L(x,y) }{\d y_{i_0,i_2}} = \sum_{j_0=1}^n  c(x,y)_{j_0,i_0} \langle f(x)_{j_0}, A_{3,*,i_2} \rangle $
    \item Let $\wt{q}(x,y)_{i_0} = \sum_{j_0=1}^n f(x)_{j_0} c(x,y)_{j_0,i_0}$
\end{itemize}
Then we have
\begin{itemize}
    \item {\bf Part 1.} 
    \begin{align*}
        \frac{\d L(x,y)_{j_0,i_0} }{ \d y_{i_0,i_2}} = A_{3,*,i_2}^\top f(x)_{j_0} c(x,y)_{j_0,i_0}
    \end{align*}    
    \item {\bf Part 2.}
    \begin{align*}
        \frac{\d L(x,y) }{\d y_{i_0,i_2}} = A_{3,*,i_2}^\top \wt{q}(x,y)_{i_0}
    \end{align*}
    \item {\bf Part 3.}
    \begin{align*}
        \frac{\d L(x,y) }{\d y} = \vect ( \underbrace{ A_3^\top}_{d \times n} \underbrace{ \wt{q}(x,y) }_{n \times d} )
    \end{align*}
    \item {\bf Part 4}. Computing $ \frac{\d L(x,y) }{\d y} $ takes $\Tmat(n,n,d) + \Tmat(n,d,d)$
\end{itemize}
\end{lemma}
\begin{proof}
{\bf Proof of Part 1.}
\begin{align*}
    \frac{\d L(x,y)_{j_0,i_0} }{ \d y_{i_0,i_2}}
    = & ~ c(x,y)_{j_0,i_0} \langle f(x)_{j_0}, A_{3,*,i_2} \rangle \\
    = & ~ A_{3,*,i_2}^{\top} f(x)_{j_0} c(x,y)_{j_0,i_0} 
\end{align*}
where the first step comes from the assumption from the Lemma statement and the second step is based on Fact~\ref{fac:circ_rules}.

{\bf Proof of Part 2.}
\begin{align*}
    \frac{\d L(x,y) }{\d y_{i_0,i_2}} = & ~ \sum_{j_0=1}^n  c(x,y)_{j_0,i_0} \langle f(x)_{j_0}, A_{3,*,i_2} \rangle \\
    = & ~ \sum_{j_0=1}^n A_{3,*,i_2}^{\top} f(x)_{j_0} c(x,y)_{j_0,i_0} \\
    = & ~ A_{3,*,i_2}^\top \wt{q}(x,y)_{i_0}
\end{align*}
where the first step is due to the assumption from the Lemma statement,
the second step is because of Fact~\ref{fac:circ_rules}, and the last step comes from the definition of $\wt{q}(x,y)_{i_0}$ (see from the Lemma statement).

{\bf Proof of Part 3.}
\begin{align*}
    \frac{\d L(x,y) }{\d y} = \vect (  A_3^\top \wt{q}(x,y)  )
\end{align*}
where the first step comes from tensor trick based on {\bf Part 2}.

{\bf Proof of Part 4.}
Computing $\wt{q}(x,y) \in \R^{n \times d}$ takes $\Tmat(n,n,d)$ time.

Computing $A_3^\top \wt{q}(x,y)$ takes $\Tmat(n,d,d)$ time. 

\end{proof}

%\newpage
\section{Hessian}\label{sec:hessian}
In this section, we provide more details related to Hessian.

Finally the hessian $H \in \R^{2d^2 \times 2 d^2}$ which can be written as 
\begin{align*}
H
= \begin{bmatrix}
H_{x,x} & H_{x,y} \\
H_{y,x} & H_{y,y}
\end{bmatrix}
\end{align*}
where
\begin{itemize}
    \item $H_{x,x} \in \R^{d^2 \times d^2}$ is $\frac{\d^2 L}{\d x \d x}$ (see details in Section~\ref{sec:hessian_X})
    \item $H_{x,y}$, $H_{y,x} \in \R^{d^2 \times d^2}$ is $\frac{\d^2 L}{\d x \d y}$ (see details in Section~\ref{sec:hessian_XY})
    \item $H_{y,y} \in \R^{d^2 \times d^2}$ is $\frac{\d^2 L}{\d y \d y}$ (see details in Section~\ref{sec:hessian_Y})
    \begin{itemize}
        \item We can view $H_{y,y} = 
        \begin{bmatrix}
        H_{y,y,1,1} & 0 & 0 & \cdots & 0\\
        0 & H_{y,y,2,2} & 0 & \cdots & 0 \\
        0 & 0 & H_{y,y,3,3} & \cdots & 0 \\
        \vdots & \vdots & \vdots & \ddots & \vdots \\
        0 & 0 & 0 & \cdots & H_{y,y,d,d}
        \end{bmatrix}$
        \item 
        where $H_{y,y,i_0,i_0} = \sum_{j_0=1}^n \frac{\d^2 L_{j_0,i_0} }{ \d y_{i_0,*} \d y_{i_0,*}} \in \R^{d \times d}$ for each $i_0 \in [d]$
    \end{itemize}
\end{itemize}

\begin{lemma}\label{lem:hessian_lower_bound}
If the following conditions hold
\begin{itemize}
    \item $H_{x,x} \succeq \alpha_1 I_{d^2}$
    \item $H_{y,y} \succeq \alpha_2 I_{d^2}$
    \item $\| H_{x,y} \| \leq \alpha_3$
    \item $\| H_{y,x} \| \leq \alpha_3$
    \item Let $\alpha_1 \geq \alpha_3 >0$, $\alpha_2 \geq \alpha_3 >0$
\end{itemize}
Then we have
\begin{align*}
    H \succeq \{ \alpha_1 - \alpha_3, \alpha_2 - \alpha_3 \} \cdot I_{2d^2}
\end{align*}
\end{lemma}
\begin{proof}
 
Let $u, v \in \R^{d^2}$, then we have
\begin{align*}
\begin{bmatrix}
u^\top & v^\top 
\end{bmatrix}
H
\begin{bmatrix}
u \\
v
\end{bmatrix}
= & ~ u^\top H_{x,x} u + v^\top H_{y,y} v + u^\top H_{x,y} v + v^\top H_{y,x} u \\
\geq & ~ \| u \|_2^2 \cdot \alpha_1 + \| v \|_2^2 \cdot \alpha_2 + u^\top H_{x,y} v + v^\top H_{y,x} u \\
\geq & ~  \| u \|_2^2 \cdot \alpha_1 + \| v \|_2^2 \cdot \alpha_2 - \| u \|_2 \| v \|_2 ( \| H_{x,y} \| + \| H_{y,x} \| ) \\
\geq & ~  \| u \|_2^2 \cdot \alpha_1 + \| v \|_2^2 \cdot \alpha_2 - \| u \|_2 \| v \|_2 2 \alpha_3 \\
\geq & ~ \| u \|_2^2 \cdot \alpha_1 + \| v \|_2^2 \cdot \alpha_2  - (\| u \|_2^2 + \| v \|_2^2 ) \alpha_3 \\
\geq & ~ (\| u \|_2^2 + \| v\|_2^2) \cdot \min \{ \alpha_1 - \alpha_3, \alpha_2 - \alpha_3 \}
\end{align*}
where the first step is based on the expansion of $H$, the second step is due to $H_{x,x} \succeq \alpha_1 I_{d^2}, H_{y,y} \succeq \alpha_2 I_{d^2}$, the third step comes from Fact~\ref{fac:vector_norm} and Fact~\ref{fac:matrix_norm} , the fourth step is because of $\| H_{x,y} \| \leq \alpha_3, \| H_{y,x} \| \leq \alpha_3$, the fifth step is owing to $2\|u \|_2\|v\|_2 \leq \| u\|_2^2+ \|v\|_2^2$, and the last step is based on the simple algebra.

Thus, it implies
\begin{align*}
    H \succeq  \{ \alpha_1 - \alpha_3, \alpha_2 - \alpha_3 \} \cdot I_{2d^2}
\end{align*}
\end{proof}

\begin{algorithm}[!ht]\caption{ Our Algorithm }\label{alg:main_result}
\begin{algorithmic}[1]
\Procedure{TrainingAlgorithm}{$A_1, A_2, A_3, B \in \R^{n \times d}$} \Comment{Theorem~\ref{thm:main_informal}}
   \State Let $x(0), y(0) \in \R^{d^2}$ denote initialization point
   \For{$t = 0 \to T-1$}
        \State {\color{blue} /*Forward*/ }
        \State Compute $h(y(t) ) \in \R^{n \times d}$ \Comment{$\Tmat(n,d,d)$ time}
        \State Compute $f(x(t)) \in \R^{n \times n}$ \Comment{$\Tmat(n,d,n)$ time}
        \State Compute $c(x(t),y(t)) \in \R^{n \times d}$ (based on $f(x(t))$, $h(y(t))$) \Comment{$\Tmat(n,d,d)$ time}
        \State {\color{blue} /*Gradient*/}
        \State Compute $g(x(t))$ based on Lemma~\ref{lem:compute_gradient_x} \Comment{$\Tmat(n,d,n) + \Tmat(n,d,d)$ time}
        \State Compute $g(y(t))$ based on Lemma~\ref{lem:compute_gradient_y} \Comment{$\Tmat(n,d,n) + \Tmat(n,d,d)$ time}
        \State {\color{blue} /*Hessian*/}
        \State Compute $\wt{H}$ via {\sf TensorSRHT} \Comment{$\wt{O}(nd + d^{2\omega})$}
        \State {\color{blue} /*Update*/}
        \State $\begin{bmatrix} x(t+1) \\ y(t+1) \end{bmatrix} \gets \begin{bmatrix} x(t) \\ y(t) \end{bmatrix} -  \begin{bmatrix} g( x(t)) \\ g(y(t)) \end{bmatrix} \wt{H}^{-1}$ \Comment{$O(d^{2\omega})$}
   \EndFor 
\EndProcedure
\end{algorithmic}
\end{algorithm}

%\newpage
\section{Hessian for \texorpdfstring{$X$}{}}\label{sec:hessian_X}
In Section~\ref{sub:hessian_x:hessian}, we compute the Hessian matrix with respect to $x$. In Section~\ref{sub:hessian_x:help_lem}, we present a helpful lemma to simplify the Hessian. In Section~\ref{sub:hessian_x:B}, we define $B(x)$, representing the Hessian.

\subsection{Hessian}\label{sub:hessian_x:hessian}

Now, we start to compute the Hessian matrix with respect to $x$.

\begin{lemma}\label{lem:hessian_l}
If the following conditions hold
\begin{itemize}
    \item Let $\gamma(x)_{j_0} := \langle f(x)_{j_0}, v \rangle$ 
    (We define this notation for easy of writing proofs.)
\end{itemize}
Then we have for each $i \in [d^2]$, $l \in [d^2]$
\begin{itemize}
    \item Part 1. $i = l$ Hessian diagonal term
    \begin{align*}
        \frac{\d^2 L_{j_0,i_0} }{\d x_i \d x_i} = & ~  ( \langle  f(x)_{j_0} \circ \A_{j_0,i}, v \rangle - \gamma_{j_0}(x) \cdot \langle f(x)_{j_0}, \A_{j_0,i} \rangle )^2 \\
        & ~ + c(x,:)_{j_0,i_0} \cdot \\
        & ~ ( \\
        & ~ + \langle f(x)_{j_0} \circ \A_{j_0,i} \circ \A_{j_0,i} , v \rangle (1-\gamma_{j_0}(x)) \\
        & ~ - 2 \langle f(x)_{j_0} \circ \A_{j_0,i}, v \rangle \cdot \langle f(x)_{j_0}, \A_{j_0,i} \rangle \\
        & ~ + 2 \langle f(x)_{j_0}, \A_{j_0,i} \rangle^2 \cdot \gamma_{j_0}(x) \\
        & ~ ) 
    \end{align*}
    \item Part 2. $i \neq l$ Hessian off-diagonal term
    \begin{align*}
        \frac{ \d^2 L_{j_0,i_0} }{ \d x_i \d x_l } = & ~  ( \langle  f(x)_{j_0} \circ \A_{j_0,i}, v \rangle - \gamma_{j_0}(x) \cdot \langle f(x)_{j_0}, \A_{j_0,i} \rangle )  \\
        & ~ \cdot ( \langle  f(x)_{j_0} \circ \A_{j_0,l}, v \rangle - \gamma_{j_0}(x) \cdot \langle f(x)_{j_0}, \A_{j_0,l} \rangle )  \\
        & ~ + c(x,:)_{j_0,i_0} \cdot \\
        & ~ ( \\
        & ~ + \langle f(x)_{j_0} \circ \A_{j_0,i} \circ \A_{j_0,l} , v \rangle (1-\langle f(x)_{j_0}, v\rangle)) \\
        & ~ - \langle f(x)_{j_0} \circ \A_{j_0,i}, v \rangle \cdot \langle f(x)_{j_0}, \A_{j_0,l} \rangle  - \langle f(x)_{j_0} \circ \A_{j_0,l}, v \rangle \cdot \langle f(x)_{j_0}, \A_{j_0,i} \rangle \\
        & ~ + 2\langle f(x)_{j_0}, \A_{j_0,i} \rangle \langle f(x)_{j_0}, \A_{j_0,l} \rangle \cdot \gamma_{j_0}(x) \\
& ~ ) 
    \end{align*}
\end{itemize}
\end{lemma}
\begin{proof}

{\bf Proof of Part 1.}

At first, we have
\begin{align*}
& ~ \frac{\d}{\d x_i}  ( \langle  f(x)_{j_0} \circ \A_{j_0,i}, v \rangle - \langle  f(x)_{j_0} , v \rangle \cdot \langle f(x)_{j_0}, \A_{j_0,i} \rangle ) \\
= & ~ \underbrace{ \frac{\d}{\d x_i}  \langle  f(x)_{j_0} \circ \A_{j_0,i}, v \rangle }_{\mathrm{Part~7~of~Lemma~\ref{lem:gradient_x}}}  \\
& ~ - \underbrace{ ( \frac{\d }{\d x_i} \langle f(x)_{j_0}, v \rangle ) }_{ {\mathrm{Part~4~of~Lemma~\ref{lem:gradient_x}}} } \cdot \langle f(x)_{j_0}, \A_{j_0,i} \rangle \\
& ~ - \underbrace{ ( \frac{\d }{\d x_i}  \langle f(x)_{j_0}, \A_{j_0,i} \rangle ) }_{ {\mathrm{Part~9~of~Lemma~\ref{lem:gradient_x}}} } \cdot \langle f(x)_{j_0}, v \rangle \\
= & ~ \langle f(x)_{j_0} \circ \A_{j_0,i} \circ \A_{j_0,i}, v \rangle - \langle f(x)_{j_0} \circ \A_{j_0,i} , v \rangle \cdot \langle f(x)_{j_0} , \A_{j_0, i} \rangle \\
& ~ -  ( \langle f(x)_{j_0} \circ \A_{j_0,i}, v \rangle - \langle f(x)_{j_0}, v \rangle \cdot \langle f(x)_{j_0}, \A_{j_0, i} \rangle ) \cdot \langle f(x)_{j_0} , \A_{j_0,i} \rangle \\
& ~ - ( \langle f(x)_{j_0} \circ \A_{j_0,i} , \A_{j_0,i} \rangle - \langle f(x)_{j_0}, \A_{j_0,i} \rangle \langle f(x)_{j_0}, \A_{j_0,i} \rangle ) \cdot \langle f(x)_{j_0} , v \rangle \\
= & ~ \langle f(x)_{j_0} \circ \A_{j_0,i} \circ \A_{j_0,i} , v \rangle \\
& ~ - 2 \langle f(x)_{j_0} \circ \A_{j_0,i}, v \rangle \cdot \langle f(x)_{j_0}, \A_{j_0,i} \rangle \\
& ~ + 2 \langle f(x)_{j_0}, \A_{j_0,i} \rangle^2 \cdot \langle f(x)_{j_0}, v \rangle \\
& ~ - \langle f(x)_{j_0} \circ \A_{j_0,i} \circ \A_{j_0,i}, v \rangle \cdot \langle f(x)_{j_0}, v\rangle
\end{align*}
where the first step is based on the product rule of derivative, the second step comes from {\bf Part 4, Part 7, and Part 9} of Lemma~\ref{lem:gradient_x}, and the last step is due to simple algebra.

Then we can show that
\begin{align*}
& ~ \frac{\d }{\d x_i} ( \frac{\d }{\d x_i} L_{j_0,i_0} ) \\
= & ~ \frac{\d }{\d x_i} (  c(x,:)_{j_0, i_0} \cdot ( \langle  f(x)_{j_0} \circ \A_{j_0,i}, v \rangle - \langle  f(x)_{j_0} , v \rangle \cdot \langle f(x)_{j_0}, \A_{j_0,i} \rangle ) ) \\
= & ~ ( \langle  f(x)_{j_0} \circ \A_{j_0,i}, v \rangle - \langle  f(x)_{j_0} , v \rangle \cdot \langle f(x)_{j_0}, \A_{j_0,i} \rangle ) ^2 \\
& ~ + c(x,:)_{j_0, i_0} \cdot \frac{\d}{\d x_i}  ( \langle  f(x)_{j_0} \circ \A_{j_0,i}, v \rangle - \langle  f(x)_{j_0} , v \rangle \cdot \langle f(x)_{j_0}, \A_{j_0,i} \rangle ),
\end{align*}
where the first step comes from {\bf Part 6} of Lemma~\ref{lem:gradient_x} and the second step is due to {\bf Part 5} of Lemma~\ref{lem:gradient_x}.

Combining the above two equations, we complete the proof.

{\bf Proof of Part 2.}

Firstly, we can show that
\begin{align*}
& ~ \frac{\d}{\d x_l}  ( \langle  f(x)_{j_0} \circ \A_{j_0,i}, v \rangle - \langle  f(x)_{j_0} , v \rangle \cdot \langle f(x)_{j_0}, \A_{j_0,i} \rangle ) \\
= & ~ \underbrace{ \frac{\d}{\d x_l}  \langle  f(x)_{j_0} \circ \A_{j_0,i}, v \rangle }_{\mathrm{Part~8~of~Lemma~\ref{lem:gradient_x}}}  \\
& ~ - \underbrace{ ( \frac{\d }{\d x_l} \langle f(x)_{j_0}, v \rangle ) }_{ {\mathrm{Part~4~of~Lemma~\ref{lem:gradient_x}}} } \cdot \langle f(x)_{j_0}, \A_{j_0,i} \rangle \\
& ~ - \underbrace{ ( \frac{\d }{\d x_l}  \langle f(x)_{j_0}, \A_{j_0,i} \rangle ) }_{ {\mathrm{Part~10~of~Lemma~\ref{lem:gradient_x}}} } \cdot \langle f(x)_{j_0}, v \rangle \\
= & ~ \langle f(x)_{j_0} \circ \A_{j_0,i} \circ \A_{j_0,l}, v \rangle - \langle f(x)_{j_0} \circ \A_{j_0,l} , v \rangle \cdot \langle f(x)_{j_0} , \A_{j_0, i} \rangle \\
& ~ -  ( \langle f(x)_{j_0} \circ \A_{j_0,l}, v \rangle - \langle f(x)_{j_0}, v \rangle \cdot \langle f(x)_{j_0}, \A_{j_0, l} \rangle ) \cdot \langle f(x)_{j_0} , \A_{j_0,i} \rangle \\
& ~ - ( \langle f(x)_{j_0} \circ \A_{j_0,i} , \A_{j_0,l} \rangle - \langle f(x)_{j_0}, \A_{j_0,i} \rangle \langle f(x)_{j_0}, \A_{j_0,l} \rangle ) \cdot \langle f(x)_{j_0} , v \rangle \\
= & ~ \langle f(x)_{j_0} \circ \A_{j_0,i} \circ \A_{j_0,l} , v \rangle \\
& ~ - \langle f(x)_{j_0} \circ \A_{j_0,i}, v \rangle \cdot \langle f(x)_{j_0}, \A_{j_0,l} \rangle - \langle f(x)_{j_0} \circ \A_{j_0,l}, v \rangle \cdot \langle f(x)_{j_0}, \A_{j_0,i} \rangle \\
& ~ + 2\langle f(x)_{j_0}, \A_{j_0,i} \rangle \langle f(x)_{j_0}, \A_{j_0,l} \rangle \cdot \langle f(x)_{j_0}, v \rangle \\
& ~ - \langle f(x)_{j_0} \circ \A_{j_0,i} \circ \A_{j_0,l}, v \rangle \cdot \langle f(x)_{j_0}, v\rangle
\end{align*}
where the first step is owing to the product rule of derivative, the second step is based on {\bf Part 4, Part 8, and Part 10} of Lemma~\ref{lem:gradient_x}, and the last step comes from simple algebra.

We have
\begin{align*}
& ~ \frac{\d }{\d x_l} ( \frac{\d }{\d x_i} L_{j_0,i_0} ) \\
= & ~ \frac{\d }{\d x_l} (  c(x,:)_{j_0, i_0} \cdot ( \langle  f(x)_{j_0} \circ \A_{j_0,i}, v \rangle - \langle  f(x)_{j_0} , v \rangle \cdot \langle f(x)_{j_0}, \A_{j_0,i} \rangle ) ) \\
= & ~ ( \langle  f(x)_{j_0} \circ \A_{j_0,i}, v \rangle - \langle  f(x)_{j_0} , v \rangle \cdot \langle f(x)_{j_0}, \A_{j_0,i} \rangle )  \\
& ~ \cdot ( \langle  f(x)_{j_0} \circ \A_{j_0,l}, v \rangle - \langle  f(x)_{j_0} , v \rangle \cdot \langle f(x)_{j_0}, \A_{j_0,l} \rangle ) \\
& ~ + c(x,:)_{j_0, i_0} \cdot \frac{\d}{\d x_l}  ( \langle  f(x)_{j_0} \circ \A_{j_0,i}, v \rangle - \langle  f(x)_{j_0} , v \rangle \cdot \langle f(x)_{j_0}, \A_{j_0,i} \rangle )
\end{align*}

Combining the above two equations, we complete the proof.
\end{proof}

\subsection{A Helpful Lemma}\label{sub:hessian_x:help_lem}

In this section, we present a helpful Lemma.

\begin{lemma}\label{lem:helpful_lemma}
We have
\begin{itemize}
    \item Part 1.
    \begin{align*}
        \langle f(x)_{j_0} \circ \A_{j_0,i} \circ \A_{j_0,l} , v \rangle = \underbrace{ \A_{j_0,i}^\top }_{d^2 \times n} \underbrace{ \diag ( f(x)_{j_0} \circ v ) }_{n \times n} \underbrace{ \A_{j_0,l} }_{n \times d^2}
    \end{align*}
    \item Part 2.
    \begin{align*}
        & ~ \langle f(x)_{j_0} \circ \A_{j_0,i}, v \rangle \cdot \langle f(x)_{j_0}, \A_{j_0,l} \rangle  + \langle f(x)_{j_0} \circ \A_{j_0,l}, v \rangle \cdot \langle f(x)_{j_0}, \A_{j_0,i} \rangle \\
        = & ~ \A_{j_0,i}^\top ( \underbrace{ ( f(x)_{j_0} \circ v ) ( f(x)_{j_0} )^\top + f(x)_{j_0} (f(x)_{j_0} \circ v )^\top }_{\rank-2} ) \A_{j_0,l}
    \end{align*}
    \item Part 3.
    \begin{align*}
        \langle f(x)_{j_0} \circ \A_{j_0,i} ,  v \rangle \cdot \langle f(x)_{j_0} \circ \A_{j_0,l} ,  v \rangle = \A_{j_0,i}^\top \underbrace{ ( f(x)_{j_0} \circ v ) ( f(x)_{j_0} \circ v )^\top }_{\rank-1} \A_{j_0,l}
    \end{align*}
    \item Part 4.
    \begin{align*}
        \langle f(x)_{j_0} , \A_{j_0,i} \rangle \cdot \langle f(x)_{j_0} , \A_{j_0,l}  \rangle = \A_{j_0,i}^\top \underbrace{ ( f(x)_{j_0}  ) ( f(x)_{j_0}  )^\top }_{\rank-1} \A_{j_0,l}
    \end{align*}
\end{itemize}
\end{lemma}
\begin{proof}

{\bf Proof of Part 1.}
We have
\begin{align*}
     \langle f(x)_{j_0} \circ \A_{j_0,i} \circ \A_{j_0,l} , v \rangle 
     = & ~  \A_{j_0,i}^\top \diag ( f(x)_{j_0} \circ v ) \A_{j_0,l} 
\end{align*}
where the first step follows from Fact~\ref{fac:circ_rules}.

{\bf Proof of Part 2.}
We have
\begin{align*}
    & ~\langle f(x)_{j_0} \circ \A_{j_0,i}, v \rangle \cdot \langle f(x)_{j_0}, \A_{j_0,l} \rangle  + \langle f(x)_{j_0} \circ \A_{j_0,l}, v \rangle \cdot \langle f(x)_{j_0}, \A_{j_0,i} \rangle \\ 
    = & ~ \langle f(x)_{j_0} \circ v , \A_{j_0,i} \rangle \cdot f(x)_{j_0}^{\top} \A_{j_0,l} \\
     & ~ + \langle f(x)_{j_0} \circ v, \A_{j_0,l} \rangle \cdot \A_{j_0,i}^{\top} \cdot f(x)_{j_0} \\
    = & ~ \A_{j_0,i}^{\top} \cdot (f(x)_{j_0} \circ v)(f(x)_{j_0})^{\top} \A_{j_0,i} \\
     & ~ +\A_{j_0,i}^{\top} f(x)_{j_0}(f(x)_{j_0} \circ v)^{\top} \A_{j_0,l} \\
    = & ~ \A_{j_0,i}^\top (  ( f(x)_{j_0} \circ v ) ( f(x)_{j_0} )^\top \\
     & ~ + f(x)_{j_0} (f(x)_{j_0} \circ v )^\top  ) \A_{j_0,l}
\end{align*}
where the first step follows from Fact~\ref{fac:circ_rules}, the second step follows from Fact~\ref{fac:circ_rules}, and the last step follows from the simple algebra. 

{\bf Proof of Part 3.}
We have
\begin{align*}
    \langle f(x)_{j_0} \circ \A_{j_0,i} ,  v \rangle \cdot \langle f(x)_{j_0} \circ \A_{j_0,l} ,  v \rangle
    = & ~ \langle f(x)_{j_0} \circ v ,  \A_{j_0,i} \rangle \langle f(x)_{j_0} \circ v ,  \A_{j_0,l} \rangle \\
    = & ~ \A_{j_0,i}^\top  ( f(x)_{j_0} \circ v ) ( f(x)_{j_0} \circ v )^\top  \A_{j_0,l}
\end{align*}
where the first step follows from Fact~\ref{fac:circ_rules}, and the last step follows from Fact~\ref{fac:circ_rules}.

{\bf Proof of Part 4.}
We have
\begin{align*}
    \langle f(x)_{j_0} , \A_{j_0,i} \rangle \cdot \langle f(x)_{j_0} , \A_{j_0,l}  \rangle
    = & ~ \A_{j_0,i}^{\top} f(x)_{j_0}f(x)_{j_0}^{\top} \A_{j_0,l}
\end{align*}
where the first step follows from Fact~\ref{fac:circ_rules}.
\end{proof}

\subsection{Defining \texorpdfstring{$B(x)$}{}}\label{sub:hessian_x:B}

In this section, we formally define $B(x)$.

\begin{definition}\label{def:B(x)}
If the following conditions hold
\begin{itemize}
    \item  Let $\gamma_{j_0}(x) = \langle f(x)_{j_0},v \rangle$
\end{itemize}
We define $B(x) \in \R^{n \times n}$ as follows
\begin{align*}
    B(x) := & B_{\diag}^1 \\
     & ~ + B_{\rank}^1 + B_{\rank}^2 + B_{\rank}^3
\end{align*}
where
\begin{itemize}
    \item $B_{\diag}^1 := (1-\gamma_{j_0}(x)) \cdot c(x,:)_{j_0,i_0} \cdot \diag( f(x)_{j_0} \circ v )$
\end{itemize}
and 
\begin{itemize}
    \item $B_{\rank}^1 := -( 2 \gamma_{j_0}(x) + c(x,:)_{j_0,i_0}) \cdot ( ( f(x)_{j_0} \circ v ) f(x)_{j_0}^\top + f(x)_{j_0} (f(x)_{j_0} \circ v)^\top )$
    \item $B_{\rank}^2 := ( 2 \gamma_{j_0}(x) c(x,:)_{j_0,i_0} + \gamma_{j_0}(x)^2 ) \cdot f(x)_{j_0} f(x)_{j_0}^\top $
    \item $B_{\rank}^3 := (f(x)_{j_0} \circ v) \cdot ( f(x)_{j_0} \circ v )^\top$
\end{itemize}
\end{definition}

\begin{lemma}
Let $B(x)$ be defined as Definition~\ref{def:B(x)}, then we have
\begin{align*}
 \frac{\d^2 L_{j_0,i_0}}{\d x \d x} = \underbrace{ \A_{j_0}^\top }_{d^2 \times n} \underbrace{ B(x) }_{n \times n} \underbrace{ \A_{j_0} }_{n \times d^2}
\end{align*}
\end{lemma}
\begin{proof}
The proof follows by combining Lemma~\ref{lem:hessian_l} and Lemma~\ref{lem:helpful_lemma}.
\end{proof}

%\newpage 
\section{Lipschitz Property of \texorpdfstring{$H_{x,x}$}{}}\label{sec:lips_H_xx}
In Section~\ref{sub:lips_H_xx:main_res}, we present the main results of the Lipschitz property of $H_{x,x}$. In Section~\ref{sub:lips_H_xx:summary}, we summarize the results from following steps 1-9. In Section~\ref{sub:lips_H_xx:upper_bound}, we compute the upper bound of basic functions for the following proof. In Section~\ref{sub:lips_H_xx:basic_lips}, we compute the Lipschitz Property of basic functions for the following proof. In Section~\ref{sub:lips_H_xx:step1}, we analyze the first step of Lipschitz function $c(x,:)_{j_0,i_0} \cdot \diag( f(x)_{j_0} \circ v )$. In Section~\ref{sub:lips_H_xx:step2}, we analyze the second step of Lipschitz function $-\gamma_{j_0}(x) \cdot c(x,:)_{j_0,i_0} \cdot \diag( f(x)_{j_0} \circ v )$. In Section~\ref{sub:lips_H_xx:step3}, we analyze the third step of Lipschitz function $- 2 \gamma_{j_0}(x) \cdot   (f(x)_{j_0} \circ v ) f(x)_{j_0}^\top $. In Section~\ref{sub:lips_H_xx:step4}, we analyze the fourth step of Lipschitz function $-c(x,:)_{j_0,i_0} \cdot   (f(x)_{j_0} \circ v ) f(x)_{j_0}^\top $. In Section~\ref{sub:lips_H_xx:step5}, we analyze the fifth step of Lipschitz function $- 2 \gamma_{j_0}(x)  \cdot  f(x)_{j_0} (f(x)_{j_0} \circ v)^\top$. In Section~\ref{sub:lips_H_xx:step6}, we analyze the sixth step of Lipschitz function $ - c(x,:)_{j_0,i_0}) \cdot  f(x)_{j_0} (f(x)_{j_0} \circ v)^\top$. In Section~\ref{sub:lips_H_xx:step7}, we analyze the seventh step of Lipschitz function $ 2 \gamma_{j_0}(x) c(x,:)_{j_0,i_0}  \cdot f(x)_{j_0} f(x)_{j_0}^\top$. In Section~\ref{sub:lips_H_xx:step8}, we analyze the eighth step of Lipschitz function $  \gamma_{j_0}(x)^2  \cdot f(x)_{j_0} f(x)_{j_0}^\top$. 
In Section~\ref{sub:lips_H_xx:step9}, we analyze the nineth step of Lipschitz function $   (f(x)_{j_0} \circ v) \cdot ( f(x)_{j_0} \circ v )^\top$.

\subsection{Main Result}\label{sub:lips_H_xx:main_res}

In this section, we present the main result of the Lipschitz property.

\begin{lemma}
If the following conditions hold
\begin{itemize}
    \item Let $H_{j_0,i_0} = \frac{\d^2 L_{j_0,i_0}}{\d x \d x} : \R^{d^2} \rightarrow \R^{d^2 \times d^2}$
    \item Let $H = \sum_{j_0=1}^n \sum_{i_0=1}^d H_{j_0,i_0}$ (because $L = \sum_{j_0=1}^n \sum_{i_0=1}^d L_{j_0,i_0} $)
    \item Let $\A \in \R^{n^2 \times d^2}$ and $u(x)_{j_0} \in \R^n$ be defined as Definition~\ref{def:u}
    \item Let $\alpha(x)_{j_0} \in \R$ be defined as Definition~\ref{def:alpha}
    \item Let $f(x)_{j_0} \in \R^n$ be defined as Definition~\ref{def:f}
    \item Let $c(x,:)_{j_0,i_0} \in \R$ be defined as Definition~\ref{def:c}
    \item Let $\gamma(x)_{j_0} = \langle f(x)_{j_0}, v \rangle \in \R$
    \item $\| A_1 \|, \| A_2 \| , \| A_3 \| \leq R$, $\| \A_{j_0} \| \leq R$, $\| x \|_2 \leq R$,$| b_{j_0,i_0} | \leq R$, $\| v \|_2 \leq R^2$
    \item Let $R \geq 4$
   
    \item Let $M:=\exp( O( R^2+ \log (nd ) )  )$
\end{itemize}

Then, we have for all $x, \wt{x} \in \R^{d^2}$
\begin{itemize}
\item Part 1. For each $j_0 \in [n]$, $i_0 \in [d]$
\begin{align*}
    \| H_{j_0,i_0}(x) - H_{j_0,i_0}( \wt{x} ) \| \leq M \cdot \| x - \wt{x} \|_2
\end{align*}
\item Part 2.
\begin{align*}
    \| H(x) - H( \wt{x} ) \| \leq M \cdot \| x - \wt{x} \|_2 
\end{align*}
\end{itemize}
\end{lemma}

\begin{proof}

{\bf Proof of Part 1.}
We have 
\begin{align*}
 \| H_{j_0,i_0}(x) - H_{j_0,i_0}(\wt{x}) \| \leq & ~ \sum_{k=1}^9 \| \A_{j_0}^\top \| \cdot \| G_k(x) - G_k(\wt{x}) \| \cdot \| \A_{j_0} \| \\
 \leq & ~ 9 R^2 \cdot n^{1.5} \exp(20R^2) \\
 \leq & ~ n^{1.5} \exp(30R^2)
\end{align*}
where the first step follows from definition of $H_{j_0,i_0}(x)$,  the second step follows from Lemma~\ref{lem:summary_Gi}, and last step follows from simple algebra.

{\bf Proof of Part 2.}

Then, we have 
\begin{align*}
\| H(x) - H(\wt{x}) \| 
\leq & ~ \sum_{j_0=1}^n \sum_{i_0=1}^d \| H_{j_0,i_0}(x) - H_{j_0,i_0}(\wt{x}) \| \\
\leq & ~ nd \cdot n^{1.5} \exp(30R^2)
\end{align*}
where the first step follows from triangle inequality and $H = \sum_{j_0=1}^n \sum_{i_0=1}^d H_{j_0,i_0}$, and the second step follows from {\bf Part 1.}
\end{proof}

\subsection{Summary of Nine Steps}\label{sub:lips_H_xx:summary}

In this section, we provide a summary of the nine-step calculation of Lipschitz for different matrix functions.

\begin{lemma}\label{lem:summary_Gi}
    If the following conditions hold 
    \begin{itemize}
        \item $G_1(x) =  c(x,:)_{j_0,i_0} \cdot \diag( f(x)_{j_0} \circ v )$
        \item $G_2(x) = -\gamma_{j_0}(x) \cdot c(x,:)_{j_0,i_0} \cdot \diag( f(x)_{j_0} \circ v )$
        \item $G_3(x) =  - 2 \gamma_{j_0}(x) \cdot   (f(x)_{j_0} \circ v ) f(x)_{j_0}^\top  $
        \item $G_4(x) = -c(x,:)_{j_0,i_0} \cdot   (f(x)_{j_0} \circ v ) f(x)_{j_0}^\top  $
        \item $G_5(x) = - 2 \gamma_{j_0}(x)  \cdot  f(x)_{j_0} (f(x)_{j_0} \circ v)^\top$ (The proof of this is identical to $G_3$)
        \item $G_6(x) =  - c(x,:)_{j_0,i_0} \cdot  f(x)_{j_0} (f(x)_{j_0} \circ v)^\top$ (The proof of this is identical to $G_4$)
        \item $G_7(x) =  2 \gamma_{j_0}(x) c(x,:)_{j_0,i_0} \cdot f(x)_{j_0} f(x)_{j_0}^\top$
        \item $G_8(x) = \gamma_{j_0}(x)^2  \cdot f(x)_{j_0} f(x)_{j_0}^\top $
        \item $G_9(x) = (f(x)_{j_0} \circ v) \cdot ( f(x)_{j_0} \circ v )^\top$
    \end{itemize}
    Then, we have
    \begin{align*}
        \max_{k \in [9]} \| G_k(x) - G_k( \wt{x} ) \| \leq n^{1.5} \exp(20R^2).
    \end{align*}
\end{lemma}
\begin{proof}
The proof follows from Lemma~\ref{lem:lipschitz_G1}, Lemma~\ref{lem:lipschitz_G2}, Lemma~\ref{lem:lipschitz_G3}, Lemma~\ref{lem:lipschitz_G4},
Lemma~\ref{lem:lipschitz_G5},
Lemma~\ref{lem:lipschitz_G6},
Lemma~\ref{lem:lipschitz_G7},
Lemma~\ref{lem:lipschitz_G8}, and 
Lemma~\ref{lem:lipschitz_G9}.
\end{proof}

\subsection{A Core Tool: Upper Bound for Several Basic Functions}\label{sub:lips_H_xx:upper_bound}

In this section, we analyze the upper bound of several basic functions.

\begin{lemma}[\cite{dls23,gsx23_incontext}]\label{lem:lower_bound_A:beta}
Provided that the subsequent requirements are satisfied
\begin{itemize}
    \item Let $\A \in \R^{n^2 \times d^2}$ satisfy $\max_{j_0 \in [n]}\| \A_{j_0} \| \leq R$
    \item Let $x \in \R^{d^2}$ satisfy that $\| x \|_2 \leq R $
    \item We define $u(x)$ as Definition~\ref{def:u} 
    \item Let $\beta$ be the greatest lower bound of $\langle u(x)_{j_0} , {\bf 1}_n \rangle$
\end{itemize}
Then we have
\begin{align*}
    \beta \geq \exp(-R^2).
\end{align*}
\end{lemma}

\begin{lemma}[Basic Functions Upper Bound]\label{lem:upper_bound}
If the following conditions hold,
\begin{itemize}
    \item Let $u(x)_{j_0} \in \R^n$ be defined as Definition~\ref{def:u}
    \item Let $\alpha(x)_{j_0} \in \R$ be defined as Definition~\ref{def:alpha}
    \item Let $f(x)_{j_0} \in \R^n$ be defined as Definition~\ref{def:f}
    \item Let $c(x,:)_{j_0,i_0} \in \R$ be defined as Definition~\ref{def:c}
    \item Let $\gamma(x)_{j_0} = \langle f(x)_{j_0}, v \rangle \in \R$
     \item Let $\beta$ be the greatest lower bound of $\langle u(x)_{j_0} , {\bf 1}_n \rangle$
    \item $\| A_1 \|, \| A_2 \| , \| A_3 \| \leq R$
    \item $\| \A_{j_0} \| \leq R$
    \item $\| x \|_2 \leq R$
    \item $| b_{j_0,i_0} | \leq R$  
    \item Let $R \geq 4$
    \item $\| v \|_2 \leq R^2$
\end{itemize}
Then we have: for all $x \in \R^{d^2}$ 
\begin{itemize}
    \item Part 1. $\| u(x)_{j_0} \|_2 \leq \sqrt{n} \cdot \exp(R^2)$
    \item Part 2. $| \alpha(x)_{j_0} | \leq n \exp(R^2)$
    \item Part 3. $| \alpha(x)_{j_0} |^{-1} \leq \exp(R^2)$
    \item Part 4. $\| f(x)_{j_0} \|_2 \leq 1$
    \item Part 5. $| \gamma(x)_{j_0} | \leq R^2$
    \item Part 6. $| c(x,:)_{j_0,i_0} | \leq 2R^2$
\end{itemize}
\end{lemma}
\begin{proof}
We present our proof as follows.

{\bf Proof of Part 1.}
We have
\begin{align*}
    \| u(x)_{j_0} \|_2 = & ~ \| \exp {\A_{j_0} x} \|_2 \\
    \leq & ~ \sqrt{n} \cdot \| \exp(\A_{j_0} x) \|_\infty \\ 
     \leq & ~ \sqrt{n} \cdot  \exp( \| \A_{j_0} x\|_2) \\
    \leq & ~ \sqrt{n} \cdot \exp(R^2)
\end{align*}
where the first step follows from Definition~\ref{def:u}, the second step is based on Fact~\ref{fac:vector_norm}, the third step follows from Fact~\ref{fac:vector_norm}, and the fourth step is because of $\| \A_{j_0} \| \leq R$ and $\| x \|_2 \leq R$ (see from the Lemma statement).

{\bf Proof of Part 2.}
We have  
\begin{align*}
    | \alpha(x)_{j_0} | = & ~ | \langle u(x)_{j_0}, {\bf 1}_n \rangle | \\
    \leq & ~ \sqrt{n} \cdot \| u(x)_{j_0} \|_2 \\
    \leq & ~ \sqrt{n} \cdot \sqrt{n} \cdot \exp(R^2) \\
    = & ~ n \exp(R^2) 
\end{align*}
where the first step is due to Definition~\ref{def:alpha}, the second is based on Fact~\ref{fac:vector_norm}, the third step follows from {\bf Part 1}. and the forth step follows from simple algebra.

{\bf Proof of Part 3.}

We have 
\begin{align*}
   | \alpha^{-1}(x)_{j_0} | = & ~ \frac{1}{\langle u(x)_{j_0} , {\bf 1}_n \rangle} \\
    \leq & ~ \frac{1}{\beta} \\
    \leq & ~ \exp(R^2)
\end{align*}
where the first step is because of Definition~\ref{def:alpha}, the second step follows from the definition of $\beta$ and the third step is due to Lemma~\ref{lem:lower_bound_A:beta}.

{\bf Proof of Part 4.}
We have
\begin{align*}
    \| f(x)_{j_0} \|_2 \leq & ~ \| f(x)_{j_0} \|_1 \\
    = & ~ 1
\end{align*}
where the first step follows from Fact~\ref{fac:vector_norm}, the second step is due to Definition~\ref{def:f} 

{ \bf Proof of Part 5.}
We have
\begin{align*}
    | \gamma(x)_{j_0} | = & ~ | \langle f(x)_{j_0}, v \rangle |\\
    \leq & ~\| f(x)_{j_0} \|_2 \cdot \| v \|_2 \\
    \leq & ~ 1 \cdot R^2 \\
    = & ~ R^2 
\end{align*}
where the first step follows from the definition of $\gamma(x)_{j_0}$ (see from the Lemma statement), the second step follows from Cauchy–Schwarz inequality, the third step follows from {\bf Part 2} and the upper bound for the $\ell_2$ norm of $v$ (from the Lemma statement), and the last step follows from simple algebra.

{\bf Proof of Part 6.}
We have
\begin{align*}
    |c(x,:)_{j_0,i_0}| = & ~ |\langle f(x)_{j_0}, v \rangle - b_{j_0,i_0}| \\
    \leq & ~ |\gamma_{j_0}(x) - b_{j_0,i_0}| \\
    \leq & ~ | \gamma_{j_0}(x) | + | b_{j_0,i_0}| \\
    \leq & ~ R^2 + R \\
    \leq & ~ 2R^2
\end{align*}
where the first step is based on Definition~\ref{def:c}, the second step is because of the definition of $\gamma_{j_0}(x)$, the third step follows from triangle inequality, the fourth step is based on {\bf Part 6} and $|b_{j_0,i_0}| \leq R$ (see from the Lemma statement), and the last step follows from $R \geq 1$.
\end{proof}

\subsection{A Core Tool: Lipschitz Property for Several Basic Functions}\label{sub:lips_H_xx:basic_lips}

In this section, we analyze the Lipschitz property of several basic functions.

\begin{lemma}[Basic Functions Lipschitz Property]\label{lem:basic_lips}
If the following conditions hold,
\begin{itemize}
    \item $\| v \| \leq R^2$
    \item $\| \A_{j_0} \| \leq R$ 
     \item Let $\beta$ be the greatest lower bound of $\langle u(x)_{j_0} , {\bf 1}_n \rangle$
    \item Let $\beta^{-1} \leq \exp(R^2)$
    \item Let $R\geq 4$
\end{itemize}
Then, we have: for all $x, \wt{x} \in \R^{d^2}$
\begin{itemize}
    \item Part 1. $\| u(x)_{j_0} - u( \wt{x} )_{j_0}\|_2 \leq  \sqrt{n}  \exp(2 R^2) \cdot \| x - \wt{x} \|_2$
    \item Part 2. $| \alpha(x)^{-1} - \alpha^{-1}( \wt{x} ) | \leq  n  \exp(4 R^2) \cdot \| x - \wt{x} \|_2$
    \item Part 3. $\| f(x)_{j_0} - f(\wt{x})_{j_0} \|_2 \leq    n^{1.5} R \exp(6 R^2)  \cdot \| x - \wt{x} \|_2$
    \item Part 4. $| \gamma(x)_{j_0} - \gamma (\wt{x})_{j_0} | \leq   n^{1.5} \exp(7 R^2) \cdot \| x - \wt{x} \|_2$
    \item Part 5. $| c(x,:)_{j_0,i_0} - c(\wt{x},:)_{j_0,i_0} | \leq   n^{1.5} \exp(7 R^2) \cdot \| x - \wt{x} \|_2$
\end{itemize}
    
\end{lemma}
\begin{proof}

{\bf Proof of Part 1.}

\begin{align*}
\| u(x)_{j_0} - u(\wt{x})_{j_0} \|_2
= & ~ \| \exp( \A_{j_0} x) - \exp( \A_{j_0} \wt{x}) \|_2 \\ 
\leq & ~ \exp(\| \A_{j_0}x \|_2 )  \cdot 2 \| \A_{j_0} (x-\wt{x}) \|_{\infty} \notag \\
\leq & ~ \sqrt{n} \exp(R^2) \cdot 2 \| \A_{j_0} (x-\wt{x}) \|_2 \notag \\
\leq & ~ \sqrt{n} \exp(R^2)  \cdot 2 \| \A_{j_0} \| \cdot \| x - \wt{x} \|_2 \notag\\
\leq & ~ 2 \sqrt{n} R \exp(R^2) \cdot \|x - \wt{x}\|_2 \\
\leq & ~ \sqrt{n} \exp(2 R^2)\cdot \|x - \wt{x}\|_2 
\end{align*}
where the first step is due to Definition~\ref{def:u}, the second step is because of Fact~\ref{fac:vector_norm}, the third step is based on  Fact~\ref{fac:vector_norm}, the fourth step follows from Fact~\ref{fac:matrix_norm} 
, fifth step is due to $\|\A_{j_0}\| \leq R$.

{\bf Proof of Part 2}
We have
\begin{align*}
    |\alpha(x)^{-1}_{j_0} - \alpha(\wt{x})^{-1}_{j_0}| \leq & ~ \alpha(x)^{-1} \alpha(\wt{x})^{-1} \cdot | \alpha(x) - \alpha(\wt{x})| \\
    \leq & ~ \beta^{-2} \cdot |\alpha(x) - \alpha(\wt{x})| \\
    \leq & ~ \beta^{-2} \cdot |\langle u(x)_{j_0} , {\bf 1}_n \rangle - \langle u(\wt{x})_{j_0}, {\bf 1}_n \rangle  | \\
    \leq & ~ \beta^{-2} \cdot \sqrt{n} \| u(x)_{j_0} - u(\wt{x})_{j_0} \|_2 \\
     \leq & ~  2 \beta^{-2} \cdot n R \exp(R^2)\|x - \wt{x}\|_2 \\
     \leq & ~ n \exp(4R^2) \cdot \|x - \wt{x}\|_2 
\end{align*}
where the first step is due to simple algebra, the second step is due to $\beta \geq \langle u(x)_{j_0}, {\bf 1}_n\rangle$, the third step follows from Definition of $\alpha(x)$ (see Definition~\ref{def:alpha}), the fourth step is based on Fact~\ref{fac:circ_rules} and Fact~\ref{fac:vector_norm}, the fifth step is because of {\bf Part 1}, and the sixth step follows from $R>4$ and $\beta^{-1} \leq \exp(R^2)$.

{\bf Proof of Part 3.}
We have
\begin{align} 
  \| f(x)_{j_0} - f(\wt{x})_{j_0} \|_2 \nonumber = & ~ \| \alpha(x)_{j_0}^{-1} u(x)_{j_0} - \alpha(\wt{x})_{j_0}^{-1} u(\wt{x})_{j_0} \|_2 \nonumber \\
    \leq & ~ \| \alpha(x)_{j_0}^{-1} u(x)_{j_0} - \alpha(\wt{x})_{j_0}^{-1} u(x)_{j_0} \|_2 + \| \alpha(\wt{x})_{j_0}^{-1} u(x)_{j_0} - \alpha(\wt{x})_{j_0}^{-1} u(\wt{x})_{j_0} \|_2 \nonumber \\
    = & ~ |\alpha(x)_{j_0}^{-1} - \alpha(\wt{x})_{j_0}^{-1} | \cdot \| u(x)_{j_0} \|_2 + | \alpha(\wt{x})_{j_0}^{-1} | \cdot \| u(x)_{j_0} -u(\wt{x})_{j_0} \|_2 \notag \\ 
    \leq & ~ n^{1.5} \exp(6R^2) \cdot \| x - \wt{x} \|_2 \notag 
\end{align}
where the first step is due to Definition~\ref{def:f}, the second step is based on triangle inequality, the third step follows from Fact~\ref{fac:vector_norm}, the fourth follows from combination of {\bf Part 1}, {\bf Part 2} and Lemma~\ref{lem:upper_bound}.

{\bf Proof of Part 4.}
We have 
\begin{align*}
    |\gamma_{j_0}(x) - \gamma_{j_0}(\wt{x})|= & ~ | \langle f(x)_{j_0}, v \rangle - \langle f(\wt{x})_{j_0}, v \rangle| \\
    \leq & ~ |\langle f(x)_{j_0} - f(\wt{x})_{j_0},v\rangle| \\
    \leq & ~ \| v \|_2 \cdot \| f(x)_{j_0} - f(\wt{x}) \|_2\\
    \leq & ~ n^{1.5} \exp(7R^2) \cdot \| x - \wt{x} \|_2
\end{align*}
where the first step is based on the definition of $\gamma_{j_0}(x)$, the second is because of Fact~\ref{fac:circ_rules}, the third step is due to Cauchy–Schwarz inequality, and the last step follows from {\bf Part 3}, $\| v \| \leq R^2$ and $R \geq 4$. 

{\bf Proof of Part 5.} We have 
\begin{align*}
    | c(x,:)_{j_0,i_0} - c(\wt{x},:)_{j_0,i_0} | = & ~ |\langle f(x)_{j_0}, v \rangle - \langle f(\wt{x})_{j_0}, v \rangle| \\
    \leq & ~  |\gamma_{j_0}(x) - \gamma_{j_0}(\wt{x})| \\
    \leq & ~ n^{1.5} \exp(7 R^2) \cdot \| x - \wt{x} \|_2 
\end{align*}
where the first step follows from Definition~\ref{def:c}, the second step is based on the definition of $\gamma_{j_0}(x)$ and the last step follows from {\bf Part 4}.
\end{proof}

For convenient, we define
\begin{definition}\label{def:R_0}
We define $R_0$ as follows
\begin{align*}
    R_0 := n^{1.5} \exp(10 R^2).
\end{align*}
\end{definition}

\subsection{Calculation: Step 1 Lipschitz for Matrix Function \texorpdfstring{$c(x,:)_{j_0,i_0} \cdot \diag( f(x)_{j_0} \circ v )$}{}}\label{sub:lips_H_xx:step1}

In this section, we introduce our calculation of Lipschitz for $c(x,:)_{j_0,i_0} \cdot \diag( f(x)_{j_0} \circ v )$.

\begin{lemma}\label{lem:lipschitz_G1}
If the following conditions
\begin{itemize}
    \item Let $G_1(x) = c(x,:)_{j_0,i_0} \cdot \diag( f(x)_{j_0} \circ v )$ 
    \item Let $R_0$ be defined as Definition~\ref{def:R_0}
    \item Let $\A \in \R^{n^2 \times d^2}$ and $u(x)_{j_0} \in \R^n$ be defined as Definition~\ref{def:u}
    \item Let $\alpha(x)_{j_0} \in \R$ be defined as Definition~\ref{def:alpha}
    \item Let $f(x)_{j_0} \in \R^n$ be defined as Definition~\ref{def:f}
    \item Let $c(x,:)_{j_0,i_0} \in \R$ be defined as Definition~\ref{def:c}
    \item Let $\gamma(x)_{j_0} = \langle f(x)_{j_0}, v \rangle \in \R$
    \item $\| A_1 \|, \| A_2 \| , \| A_3 \| \leq R$, $\| \A_{j_0} \| \leq R$, $\| x \|_2 \leq R$,$| b_{j_0,i_0} | \leq R$, $\| v \|_2 \leq R^2$
    \item Let $R \geq 4$
\end{itemize}
Then, we have
\begin{align*}
    \|G_1(x) - G_1( \wt{x} ) \| \leq 10R^4 \cdot R_0 \cdot \|x - \wt{x} \|_2 
\end{align*}
\end{lemma}

\begin{proof}
    We define
    \begin{align*}
        G_{1,1} = & ~  c(x,:)_{j_0,i_0} \cdot \diag( f(x)_{j_0} \circ v ) -  c(\wt{x},:)_{j_0,i_0} \cdot \diag( f(x)_{j_0} \circ v )  \\
        G_{1,2} = & ~ 
         c(\wt{x},:)_{j_0,i_0} \cdot \diag( f(x)_{j_0} \circ v ) - c(\wt{x},:)_{j_0,i_0} \cdot \diag( f(\wt{x})_{j_0} \circ v ) 
    \end{align*}
    we have
    \begin{align*}
         \| G_{1,1} \| = & ~ \| c(x,:)_{j_0,i_0} \cdot \diag( f(x)_{j_0} \circ v ) -  c(\wt{x},:)_{j_0,i_0} \cdot \diag( f(x)_{j_0} \circ v ) \|\\ 
         \leq & ~ | c(x,:)_{j_0,i_0} -  c(\wt{x},:)_{j_0,i_0} | \cdot \| \diag( f(x)_{j_0} \circ v )\| \\
         \leq & ~ R^2 \cdot | c(x,:)_{j_0,i_0} -  c(\wt{x},:)_{j_0,i_0} | \\
         \leq & ~ R^2  R_0 \cdot  \| x - \wt{x} \|_2
    \end{align*}
    where the first step is based on definition~$G_{1,1}$, the second step is due to Fact~\ref{fac:matrix_norm}, the third step follows from Lemma~\ref{lem:upper_bound}, and the fourth step is because of Lemma~\ref{lem:basic_lips}.
    
    Additionally, we have
    \begin{align*}
    \| G_{1,2} \| = & ~ \| c(\wt{x},:)_{j_0,i_0} \cdot \diag( f(x)_{j_0} \circ v ) - c(\wt{x},:)_{j_0,i_0} \cdot \diag( f(\wt{x})_{j_0} \circ v ) \|\\
     \leq & ~ | c(\wt{x}, :)_{j_0,i_0} | \cdot \| v \|_2 \cdot \| \diag(f(x)_{j_0}) - \diag (f(\wt{x})_{j_0})\| \\
     \leq & ~ 2 R^4 \cdot \| f(x)_{j_0} - f( \wt{x})_{j_0} \|_2 \\
     \leq & ~  2 R^4 \cdot R_0  \cdot \| x - \wt{x} \|_2
    \end{align*}
    where the first step is because of definition of $G_{1,2}$, the second step is due to Fact~\ref{fac:matrix_norm}, the third step follows from Lemma~\ref{lem:upper_bound}, and the fourth step is because of Lemma~\ref{lem:basic_lips}.

Combining the above two equations, we complete the proof.
\end{proof}

\subsection{Calculation: Step 2 Lipschitz for Matrix Function \texorpdfstring{$-\gamma_{j_0}(x) \cdot c(x,:)_{j_0,i_0} \cdot \diag( f(x)_{j_0} \circ v )$}{}}\label{sub:lips_H_xx:step2}

In this section, we introduce our calculation of Lipschitz for $-\gamma_{j_0}(x) \cdot c(x,:)_{j_0,i_0} \cdot \diag( f(x)_{j_0} \circ v )$.

\begin{lemma}\label{lem:lipschitz_G2}
If the following conditions hold
\begin{itemize}
    \item Let $G_2(x) = -\gamma_{j_0}(x) \cdot c(x,:)_{j_0,i_0} \cdot \diag( f(x)_{j_0} \circ v )$
    \item Let $\alpha(x)_{j_0} \in \R$ be defined as Definition~\ref{def:alpha}
        \item Let $f(x)_{j_0} \in \R^n$ be defined as Definition~\ref{def:f}
        \item Let $c(x,:)_{j_0,i_0} \in \R$ be defined as Definition~\ref{def:c}
        \item Let $\gamma(x)_{j_0} = \langle f(x)_{j_0}, v \rangle \in \R$
        \item Let $R \geq 4$
\end{itemize}
Then, we have
\begin{align*}
    \|G_2(x) - G_2(\wt{x}) \| \leq 10 R^4 \cdot R_0 \|x -\wt{x} \|_2 
\end{align*}
\end{lemma}

\begin{proof}
    We define
    \begin{align*}
        G_{2,1} = & ~ -\gamma_{j_0}(x) \cdot c(x,:)_{j_0,i_0} \cdot \diag( f(x)_{j_0} \circ v ) - (-\gamma_{j_0}(\wt{x})) \cdot c(x,:)_{j_0,i_0} \cdot \diag( f(x)_{j_0} \circ v )  \\
        G_{2,2} = & ~ 
        -\gamma_{j_0}(\wt{x}) \cdot c(x,:)_{j_0,i_0} \cdot \diag( f(x)_{j_0} \circ v ) - (-\gamma_{j_0}(\wt{x})) \cdot c(\wt{x},:)_{j_0,i_0} \cdot \diag( f(x)_{j_0} \circ v ) \\ 
       G_{2,3} =  & ~
       -\gamma_{j_0}(\wt{x}) \cdot c(\wt{x},:)_{j_0,i_0} \cdot \diag( f(x)_{j_0} \circ v ) - (-\gamma_{j_0}(\wt{x})) \cdot c(\wt{x},:)_{j_0,i_0} \cdot \diag( f(\wt{x})_{j_0} \circ v ) 
    \end{align*}
    We have
    \begin{align*}
        \|G_{2,1}\| = & ~ \|(-\gamma_{j_0}(x)) \cdot c(x,:)_{j_0,i_0} \cdot \diag( f(x)_{j_0} \circ v ) - (-\gamma_{j_0}(\wt{x})) \cdot c(x,:)_{j_0,i_0} \cdot \diag( f(x)_{j_0} \circ v ) \|  \\
        \leq & ~ | \gamma_{j_0}(x) - \gamma_{j_0}(\wt{x}) | \cdot | c(x,:)_{j_0,i_0} | \cdot \| \diag( f(x)_{j_0} \circ v ) \|\\
        \leq & ~ 2 R^4 \cdot \| \gamma_{j_0}(x) - \gamma_{j_0}(\wt{x})\| \\
        \leq & ~ 2 R^4 \cdot R_0 \cdot \| x - \wt{x} \|_2,
    \end{align*}
     where the first step is because of definition of $G_{2,1}$, the second step is due to Fact~\ref{fac:matrix_norm}, the third step follows from Lemma~\ref{lem:upper_bound}, and the fourth step is because of Lemma~\ref{lem:basic_lips}.
     
    Additionally, we have 
    \begin{align*}
    \| G_{2,2} \| = & ~ 
        \| -\gamma_{j_0}(\wt{x}) \cdot c(x,:)_{j_0,i_0} \cdot \diag( f(x)_{j_0} \circ v ) - (-\gamma_{j_0}(\wt{x})) \cdot c(\wt{x},:)_{j_0,i_0} \cdot \diag( f(x)_{j_0} \circ v ) \| \\ 
         \leq & ~ \| \gamma_{j_0}(\wt{x}) \cdot \diag(f(\wt{x})_{j_0} \circ v)\| \cdot \|c(x,:)_{j_0,i_0} - c(\wt{x},:)_{j_0,i_0}  \| \\
         \leq & ~ R^4 \cdot |c(x,:)_{j_0,i_0} - c(\wt{x},:)_{j_0,i_0}  | \\
        \leq & ~ R^4 R_0 \cdot  \| x - \wt{x} \|_2
    \end{align*}
     where the first step is because of definition of $G_{2,2}$, the second step is due to Fact~\ref{fac:matrix_norm}, the third step follows from Lemma~\ref{lem:upper_bound}, and the fourth step is because of Lemma~\ref{lem:basic_lips}.

    Additionally, we have
    \begin{align*}
        \| G_{2,3} \|=  & ~
       \|-\gamma_{j_0}(\wt{x}) \cdot c(\wt{x},:)_{j_0,i_0} \cdot \diag( f(x)_{j_0} \circ v ) - (-\gamma_{j_0}(\wt{x})) \cdot c(\wt{x},:)_{j_0,i_0} \cdot \diag( f(\wt{x})_{j_0} \circ v ) \|\\
        \leq & ~ \| \gamma_{j_0}(\wt{x}) \| \cdot \| c(\wt{x},:)_{j_0,i_0} \| \cdot \| c(x,:)_{j_0,i_0} - c(\wt{x},:)_{j_0,i_0} \| \\
        \leq & ~ 2R^4 \cdot R_0 \cdot \| x - \wt{x} \|_2
    \end{align*}
     where the first step is because of definition of $G_{2,3}$, the second step is due to Fact~\ref{fac:matrix_norm}, the third step follows from Lemma~\ref{lem:upper_bound} and Lemma~\ref{lem:basic_lips}.
     
    Combining all the above equations finish the proof.
\end{proof}

\subsection{Calculation: Step 3  
 Lipschitz for Matrix Function \texorpdfstring{$ - 2 \gamma_{j_0}(x) \cdot   (f(x)_{j_0} \circ v ) f(x)_{j_0}^\top $}{}}\label{sub:lips_H_xx:step3}

In this section, we introduce our calculation of Lipschitz for $- 2 \gamma_{j_0}(x) \cdot   (f(x)_{j_0} \circ v ) f(x)_{j_0}^\top$.

\begin{lemma}\label{lem:lipschitz_G3}
    If the following conditions hold
    \begin{itemize}
        \item Let $G_3(x) = - 2 \gamma_{j_0}(x) \cdot   (f(x)_{j_0} \circ v ) f(x)_{j_0}^\top$.
        \item  Let $R_0$ be defined in Definition~\ref{def:R_0}.
         \item Let $\alpha(x)_{j_0} \in \R$ be defined as Definition~\ref{def:alpha}
        \item Let $f(x)_{j_0} \in \R^n$ be defined as Definition~\ref{def:f}
        \item Let $c(x,:)_{j_0,i_0} \in \R$ be defined as Definition~\ref{def:c}
        \item Let $\gamma(x)_{j_0} = \langle f(x)_{j_0}, v \rangle \in \R$
        \item $\| A_1 \|, \| A_2 \| , \| A_3 \| \leq R$, $\| \A_{j_0} \| \leq R$, $\| x \|_2 \leq R$,$| b_{j_0,i_0} | \leq R$, $\| v \|_2 \leq R^2$
        \item Let $R \geq 4$
    \end{itemize}
    
\end{lemma}

Then, we have
\begin{align*}
    \|G_3(x) - G_3(\wt{x}) \| \leq 10 R^4 \cdot R_0 \|x -\wt{x} \|_2
\end{align*}
\begin{proof}
    We define
    \begin{align*}
        G_{3,1} = & ~ - 2 \gamma_{j_0}(x) \cdot   (f(x)_{j_0} \circ v ) f(x)_{j_0}^\top - (- 2 \gamma_{j_0}(\wt{x}) \cdot   (f(x)_{j_0} \circ v ) f(x)_{j_0}^\top) \\
        G_{3,2} = & ~ - 2 \gamma_{j_0}(\wt{x}) \cdot   (f(x)_{j_0} \circ v ) f(x)_{j_0}^\top - (- 2 \gamma_{j_0}(\wt{x}) \cdot   (f(\wt{x})_{j_0} \circ v ) f(x)_{j_0}^\top) \\
        G_{3,3} = & ~ - 2 \gamma_{j_0}(\wt{x}) \cdot   (f(\wt{x})_{j_0} \circ v ) f(x)_{j_0}^\top - (- 2 \gamma_{j_0}(\wt{x}) \cdot   (f(\wt{x})_{j_0} \circ v ) f(\wt{x})_{j_0}^\top)
    \end{align*}

For $G_{3,1}$, we have
\begin{align*}
    \| G_{3,1} \| 
    \leq & ~ 2 \cdot | \gamma(x)_{j_0} - \gamma( \wt{x} )_{j_0} | \cdot \| f(x)_{j_0} \circ v \|_2 \cdot \| f(x)_{j_0} \|_2 \\
    \leq & ~ 2 R_0 \cdot R^2 \| x - \wt{x} \|_2
\end{align*}
where the first step is based on Fact~\ref{fac:matrix_norm} and the second step is due to Lemma~\ref{lem:upper_bound} and Lemma~\ref{lem:basic_lips}.

Similarly, we have
\begin{align*}
    \| G_{3,2} \| \leq 2 R_0 \cdot R^4 \| x - \wt{x} \|_2
\end{align*}
and
\begin{align*}
    \| G_{3,3} \| \leq 2R_0 \cdot R^4 \| x - \wt{x} \|_2
\end{align*}

\end{proof}

 \subsection{Calculation: Step 4  
 Lipschitz for Matrix Function \texorpdfstring{$ -c(x,:)_{j_0,i_0} \cdot   (f(x)_{j_0} \circ v ) f(x)_{j_0}^\top$}{}}\label{sub:lips_H_xx:step4}

In this section, we introduce our calculation of Lipschitz for $-c(x,:)_{j_0,i_0} \cdot   (f(x)_{j_0} \circ v ) f(x)_{j_0}^\top$.

\begin{lemma}\label{lem:lipschitz_G4}
If the following conditions hold
\begin{itemize}
     \item Let $\alpha(x)_{j_0} \in \R$ be defined as Definition~\ref{def:alpha}
    \item Let $f(x)_{j_0} \in \R^n$ be defined as Definition~\ref{def:f}
    \item Let $c(x,:)_{j_0,i_0} \in \R$ be defined as Definition~\ref{def:c}
    \item Let $\gamma(x)_{j_0} = \langle f(x)_{j_0}, v \rangle \in \R$
    \item $\| A_1 \|, \| A_2 \| , \| A_3 \| \leq R$, $\| \A_{j_0} \| \leq R$, $\| x \|_2 \leq R$,$| b_{j_0,i_0} | \leq R$, $\| v \|_2 \leq R^2$
    \item Let $R \geq 4$
    \item Let $G_{4}(x) = -c(x,:)_{j_0,i_0} \cdot   (f(x)_{j_0} \circ v ) f(x)_{j_0}^\top$
\end{itemize}
    Then, we have 
    \begin{align*}
        \|G_4(x) - G_4(\wt{x}) \| \leq 10 R^4 \cdot R_0 \|x -\wt{x} \|_2
    \end{align*}
\end{lemma}

\begin{proof}
    We define
    \begin{align*}
        G_{4,1} = & ~ 
        -c(x,:)_{j_0,i_0} \cdot   (f(x)_{j_0} \circ v ) f(x)_{j_0}^\top - (-c(\wt{x},:)_{j_0,i_0} \cdot   (f(x)_{j_0} \circ v ) f(x)_{j_0}^\top)\\
        G_{4,2} = & ~ 
        -c(\wt{x},:)_{j_0,i_0} \cdot   (f(x)_{j_0} \circ v ) f(x)_{j_0}^\top - (-c(\wt{x},:)_{j_0,i_0} \cdot   (f(\wt{x})_{j_0} \circ v ) f(x)_{j_0}^\top) \\
        G_{4,3} = & ~ -c(\wt{x},:)_{j_0,i_0} \cdot   (f(\wt{x})_{j_0} \circ v ) f(x)_{j_0}^\top - (-c(\wt{x},:)_{j_0,i_0} \cdot   (f(\wt{x})_{j_0} \circ v ) f(\wt{x})_{j_0}^\top)
    \end{align*}

For $G_{4,1}$, we have
\begin{align*}
    \| G_{4,1} \| \leq R^2 \cdot R_0 \cdot \| x - \wt{x} \|_2
\end{align*}
For $G_{4,2}$, we have
\begin{align*}
    \| G_{4,2} \| \leq 2 R^4 \cdot R_0 \cdot \| x - \wt{x} \|_2
\end{align*}
For $G_{4,3}$, we have
\begin{align*}
    \| G_{4,3} \| \leq 2 R^4 \cdot R_0 \cdot \| x - \wt{x} \|_2
\end{align*}
\end{proof}

  \subsection{Calculation: Step 5  
 Lipschitz for Matrix Function \texorpdfstring{$ - 2 \gamma_{j_0}(x)  \cdot  f(x)_{j_0} (f(x)_{j_0} \circ v)^\top$}{}}\label{sub:lips_H_xx:step5}

In this section, we introduce our calculation of Lipschitz for $- 2 \gamma_{j_0}(x)  \cdot  f(x)_{j_0} (f(x)_{j_0} \circ v)^\top$.

\begin{lemma}\label{lem:lipschitz_G5}
If the following conditions hold
\begin{itemize}
    \item Let $R_0$ be defined as Definition~\ref{def:R_0}
     \item Let $\alpha(x)_{j_0} \in \R$ be defined as Definition~\ref{def:alpha}
    \item Let $f(x)_{j_0} \in \R^n$ be defined as Definition~\ref{def:f}
    \item Let $c(x,:)_{j_0,i_0} \in \R$ be defined as Definition~\ref{def:c}
    \item Let $\gamma(x)_{j_0} = \langle f(x)_{j_0}, v \rangle \in \R$
    \item $\| A_1 \|, \| A_2 \| , \| A_3 \| \leq R$, $\| \A_{j_0} \| \leq R$, $\| x \|_2 \leq R$,$| b_{j_0,i_0} | \leq R$, $\| v \|_2 \leq R^2$
    \item Let $R \geq 4$
    \item Let $G_{5}(x) = - 2 \gamma_{j_0}(x)  \cdot  f(x)_{j_0} (f(x)_{j_0} \circ v)^\top$
\end{itemize}
    Then, we have
    \begin{align*}
        \|G_5(x) - G_5(\wt{x}) \| \leq 10R^4 \cdot R_0 \|x -\wt{x} \|_2
    \end{align*}
\end{lemma}
\begin{proof}
    This proof is similar to the proof of Lemma~\ref{lem:lipschitz_G3}, so we omit it here.
\end{proof}

  \subsection{Calculation: Step 6  
 Lipschitz for Matrix Function \texorpdfstring{$ - c(x,:)_{j_0,i_0} \cdot  f(x)_{j_0} (f(x)_{j_0} \circ v)^\top$}{}}\label{sub:lips_H_xx:step6}

In this section, we introduce our calculation of Lipschitz for $ - c(x,:)_{j_0,i_0} \cdot  f(x)_{j_0} (f(x)_{j_0} \circ v)^\top$.

\begin{lemma}\label{lem:lipschitz_G6}
If the following conditions hold
\begin{itemize}
     \item Let $\alpha(x)_{j_0} \in \R$ be defined as Definition~\ref{def:alpha}
    \item Let $f(x)_{j_0} \in \R^n$ be defined as Definition~\ref{def:f}
    \item Let $c(x,:)_{j_0,i_0} \in \R$ be defined as Definition~\ref{def:c}
    \item Let $\gamma(x)_{j_0} = \langle f(x)_{j_0}, v \rangle \in \R$
    \item $\| A_1 \|, \| A_2 \| , \| A_3 \| \leq R$, $\| \A_{j_0} \| \leq R$, $\| x \|_2 \leq R$,$| b_{j_0,i_0} | \leq R$, $\| v \|_2 \leq R^2$
    \item Let $R \geq 4$
    \item Let $G_{6}(x) = - c(x,:)_{j_0,i_0} \cdot  f(x)_{j_0} (f(x)_{j_0} \circ v)^\top$
\end{itemize}

    Then, we have
    \begin{align*}
        \|G_5(x) - G_5(\wt{x}) \| \leq 10 R^4 \cdot R_0  \|x -\wt{x} \|_2
    \end{align*}
\end{lemma}
\begin{proof}
    This proof is similar to the proof of Lemma~\ref{lem:lipschitz_G4}, so we omit it here.
\end{proof}

   \subsection{Calculation: Step 7  
 Lipschitz for Matrix Function \texorpdfstring{$ 2 \gamma_{j_0}(x) c(x,:)_{j_0,i_0}  \cdot f(x)_{j_0} f(x)_{j_0}^\top$}{}}\label{sub:lips_H_xx:step7}

In this section, we introduce our calculation of Lipschitz for $2 \gamma_{j_0}(x) c(x,:)_{j_0,i_0}  \cdot f(x)_{j_0} f(x)_{j_0}^\top$.

 \begin{lemma}\label{lem:lipschitz_G7}
 If the following conditions hold
 \begin{itemize}
     \item Let $\alpha(x)_{j_0} \in \R$ be defined as Definition~\ref{def:alpha}
    \item Let $f(x)_{j_0} \in \R^n$ be defined as Definition~\ref{def:f}
    \item Let $c(x,:)_{j_0,i_0} \in \R$ be defined as Definition~\ref{def:c}
    \item Let $\gamma(x)_{j_0} = \langle f(x)_{j_0}, v \rangle \in \R$
    \item $\| A_1 \|, \| A_2 \| , \| A_3 \| \leq R$, $\| \A_{j_0} \| \leq R$, $\| x \|_2 \leq R$,$| b_{j_0,i_0} | \leq R$, $\| v \|_2 \leq R^2$
    \item Let $R \geq 4$
    \item  Let $G_{7}(x) = 2 \gamma_{j_0}(x) c(x,:)_{j_0,i_0}  \cdot f(x)_{j_0} f(x)_{j_0}^\top$
 \end{itemize}

     Then, we have 
     \begin{align*}
         \|G_7(x) - G_7(\wt{x}) \| \leq 10 R^4 R_0 \|x -\wt{x} \|_2
     \end{align*}
 \end{lemma}
 \begin{proof}
     We define
     \begin{align*}
         G_{7,1} = & ~ 2 \gamma_{j_0}(x) c(x,:)_{j_0,i_0}  \cdot f(x)_{j_0} f(x)_{j_0}^\top - 2 \gamma_{j_0}(\wt{x}) c(x,:)_{j_0,i_0}  \cdot f(x)_{j_0} f(x)_{j_0}^\top \\
        G_{7,2} = & ~ 2 \gamma_{j_0}(\wt{x}) c(x,:)_{j_0,i_0}  \cdot f(x)_{j_0} f(x)_{j_0}^\top - 2 \gamma_{j_0}(\wt{x}) c(\wt{x},:)_{j_0,i_0}  \cdot f(x)_{j_0} f(x)_{j_0}^\top \\
        G_{7,3} = & ~ 2 \gamma_{j_0}(\wt{x}) c(\wt{x},:)_{j_0,i_0}  \cdot f(x)_{j_0} f(x)_{j_0}^\top - 2 \gamma_{j_0}(\wt{x}) c(\wt{x},:)_{j_0,i_0}  \cdot f(\wt{x})_{j_0} f(x)_{j_0}^\top \\
        G_{7,4} = & ~ 2 \gamma_{j_0}(\wt{x}) c(\wt{x},:)_{j_0,i_0}  \cdot f(\wt{x})_{j_0} f(x)_{j_0}^\top - 2 \gamma_{j_0}(\wt{x}) c(\wt{x},:)_{j_0,i_0}  \cdot f(\wt{x})_{j_0} f(\wt{x})_{j_0}^\top 
     \end{align*}
    For $G_{7,1}$, we have
    \begin{align*}
        \| G_{7,1} \| =  & ~ \|2 \gamma_{j_0}(x) c(x,:)_{j_0,i_0}  \cdot f(x)_{j_0} f(x)_{j_0}^\top - 2 \gamma_{j_0}(\wt{x}) c(x,:)_{j_0,i_0}  \cdot f(x)_{j_0} f(x)_{j_0}^\top \|\\
        \leq & ~ 2|\gamma_{j_0}(x) - \gamma_{j_0}(\wt{x})| \| c(x,:)_{j_0,i_0}  \cdot f(x)_{j_0} f(x)_{j_0}^\top \| \\
        \leq & ~2 R_0 \cdot|c(x,:)_{j_0,i_0} | \cdot \|f(x)_{j_0} \| \cdot \| f(x)_{j_0}^\top \| \\
        \leq & ~
        2 R_0 \cdot 2R^2 \cdot \| x - \wt{x} \|_2
    \end{align*}
    where the first step is due to the definition of $G_{7,1}$, the second step is because of Fact~\ref{fac:matrix_norm}, the third step is based on {\bf Part 4} of Lemma~\ref{lem:basic_lips} and Fact~\ref{fac:matrix_norm}, and the last step comes from {\bf Part 4 and Part 6} of Lemma~\ref{lem:upper_bound}.
    
    Similarly, for $G_{7,2}$, we have
    \begin{align*}
        \| G_{7,2} \| \leq 2 R_0 \cdot R^2 \cdot \| x - \wt{x} \|_2
    \end{align*}
    
   For $G_{7,3}$, we have
    \begin{align*}
        \| G_{7,3} \| \leq 2 R_0 \cdot 2R^4 \cdot \| x - \wt{x} \|_2
    \end{align*}
    
    For $G_{7,4}$, we have
    \begin{align*}
        \| G_{7,4} \| \leq 2 R_0 \cdot 2R^4 \cdot \| x - \wt{x} \|_2
    \end{align*}
 \end{proof}

\subsection{Calculation: Step 8  
 Lipschitz for Matrix Function \texorpdfstring{$ \gamma_{j_0}(x)^2  \cdot f(x)_{j_0} f(x)_{j_0}^\top$}{}}\label{sub:lips_H_xx:step8}

In this section, we introduce our calculation of Lipschitz for $\gamma_{j_0}(x)^2  \cdot f(x)_{j_0} f(x)_{j_0}^\top$.

\begin{lemma}\label{lem:lipschitz_G8}

 If the following conditions hold
 \begin{itemize}
     \item Let $\alpha(x)_{j_0} \in \R$ be defined as Definition~\ref{def:alpha}
    \item Let $f(x)_{j_0} \in \R^n$ be defined as Definition~\ref{def:f}
    \item Let $c(x,:)_{j_0,i_0} \in \R$ be defined as Definition~\ref{def:c}
    \item Let $\gamma(x)_{j_0} = \langle f(x)_{j_0}, v \rangle \in \R$
    \item $\| A_1 \|, \| A_2 \| , \| A_3 \| \leq R$, $\| \A_{j_0} \| \leq R$, $\| x \|_2 \leq R$,$| b_{j_0,i_0} | \leq R$, $\| v \|_2 \leq R^2$
    \item Let $R \geq 4$
    \item Let $G_{8,1} =\gamma_{j_0}(x)^2  \cdot f(x)_{j_0} f(x)_{j_0}^\top $
 \end{itemize}

    Then, we have 
    \begin{align*}
         \|G_8(x) - G_8(\wt{x}) \| \leq 10 R^4 R_0 \|x -\wt{x} \|_2
    \end{align*}
\end{lemma}
\begin{proof}
    We define
    \begin{align*}
        G_{8,1} = & ~ \gamma_{j_0}(x) \gamma_{j_0}(x)  \cdot f(x)_{j_0} f(x)_{j_0}^\top - \gamma_{j_0}(\wt{x}) \gamma_{j_0}(x) \cdot f(x)_{j_0} f(x)_{j_0}^\top \\
         G_{8,2} = & ~ \gamma_{j_0}(\wt{x}) \gamma_{j_0}(x)  \cdot f(x)_{j_0} f(x)_{j_0}^\top - \gamma_{j_0}(\wt{x})^2  \cdot f(x)_{j_0} f(x)_{j_0}^\top \\
        G_{8,3} = & ~ \gamma_{j_0}(\wt{x})^2  \cdot f(x)_{j_0} f(x)_{j_0}^\top - \gamma_{j_0}(\wt{x})^2  \cdot f(\wt{x})_{j_0} f(x)_{j_0}^\top \\
        G_{8,4} = & ~ \gamma_{j_0}(\wt{x})^2  \cdot f(\wt{x})_{j_0} f(x)_{j_0}^\top - \gamma_{j_0}(\wt{x})^2  \cdot f(\wt{x})_{j_0} f(\wt{x})_{j_0}^\top 
    \end{align*}
    We can show that
    \begin{align*}
        \max_{i\in [4] }\| G_{8,i} \| \leq R^4 \cdot R_0 \cdot \| x - \wt{x} \|_2
    \end{align*}
\end{proof}
 
\subsection{Calculation: Step 9  
 Lipschitz for Matrix Function \texorpdfstring{$ (f(x)_{j_0} \circ v) \cdot ( f(x)_{j_0} \circ v )^\top$}{}}\label{sub:lips_H_xx:step9}

In this section, we introduce our calculation of Lipschitz for $(f(x)_{j_0} \circ v) \cdot ( f(x)_{j_0} \circ v )^\top$.

\begin{lemma}\label{lem:lipschitz_G9}
 If the following conditions hold
 \begin{itemize}
     \item Let $\alpha(x)_{j_0} \in \R$ be defined as Definition~\ref{def:alpha}
    \item Let $f(x)_{j_0} \in \R^n$ be defined as Definition~\ref{def:f}
    \item Let $c(x,:)_{j_0,i_0} \in \R$ be defined as Definition~\ref{def:c}
    \item Let $\gamma(x)_{j_0} = \langle f(x)_{j_0}, v \rangle \in \R$
    \item $\| A_1 \|, \| A_2 \| , \| A_3 \| \leq R$, $\| \A_{j_0} \| \leq R$, $\| x \|_2 \leq R$,$| b_{j_0,i_0} | \leq R$, $\| v \|_2 \leq R^2$
    \item Let $R \geq 4$
    \item  Let $G_{9}(x) = (f(x)_{j_0} \circ v) \cdot ( f(x)_{j_0} \circ v )^\top$
 \end{itemize}

    Then, we have 
    \begin{align*}
        \|G_9(x) - G_9(\wt{x}) \| \leq 10 R^4 R_0 \|x -\wt{x} \|_2
    \end{align*}
\end{lemma}
\begin{proof}
    We define 
    \begin{align*}
        G_{9,1} = & ~ (f(x)_{j_0} \circ v) \cdot ( f(x)_{j_0} \circ v )^\top  - (f(\wt{x})_{j_0} \circ v) \cdot ( f(x)_{j_0} \circ v )^\top \\
        G_{9,2} = & ~ (f(\wt{x})_{j_0} \circ v) \cdot ( f(x)_{j_0} \circ v )^\top  - (f(\wt{x})_{j_0} \circ v) \cdot ( f(\wt{x})_{j_0} \circ v )^\top
    \end{align*}
    We can show that
    \begin{align*}
        \max_{i \in [2]} \| G_{9,i} \| \leq R^4 \cdot R_0 \cdot \| x - \wt{x} \|_2
    \end{align*}
\end{proof}

%\newpage
\section{Hessian for \texorpdfstring{$X$}{} Is PSD}\label{sec:psd_H_xx}

In Section~\ref{sub:psd_H_xx:main_res}, we present the main result of PSD bound for Hessian. In Section~\ref{sub:psd_H_xx:psd}, we show the PSD bound for $B(x)$. In this section, our focus will be on establishing the PSD bound for $H_{x,x}$. Throughout this section, we will use the symbol $H$ to represent $H_{x,x}$ for the sake of simplicity.
\subsection{Main Result}\label{sub:psd_H_xx:main_res}

In this section, we introduce the main result of the PSD bound for Hessian.

\begin{lemma}\label{lem:hessian_property:x}
If the following conditions hold
\begin{itemize}
    \item Let $j_0 \in [n]$
    \item Let $i_0 \in [d]$
    \item Let $H_{j_0,i_0} = \frac{\d^2 L_{j_0,i_0}}{\d x \d x} \in \R^{d^2 \times d^2}$
    \item Let $B_{j_0,i_0}(x) \in \R^{n \times n}$ be defined as Definition~\ref{def:B(x)}.
    \begin{itemize}
        \item Therefore, $H_{j_0,i_0} = \A_{j_0}^\top B_{j_0,i_0}(x) \A_{j_0} \in \R^{d^2 \times d^2}$
    \end{itemize}
    \item Let $\max_{j_0 \in [n]} \| \A_{j_0} \| \leq R$
    \item Let $\sigma_{\min}$ be the smallest singular value. We define $\sigma_{\min}(\A_{\min}) := \min_{j_0 \in [n]} \sigma_{\min} (\A_{j_0})$.
    \item Let $H = \sum_{j_0=1}^n \sum_{i_0=1}^d H_{j_0,i_0}$ 
    \item Let $H_{\reg,j_0,i_0} =  \A_{j_0}^\top ( B_{j_0,i_0}(x) + W^2) \A_{j_0}$ where $W \in \R^{n \times n}$ is a positive diagonal matrix.
    \item Let $H_{\reg} = \sum_{j_0=1}^n \sum_{i_0=1}^d H_{\reg,j_0,i_0}$
    \item Let $C_0: = 30 R^8$ (be a local parameter in this lemma)
    \item Let $l > 0$ (denote the strongly convex parameter for hessian)
\end{itemize}
Then, we have
\begin{itemize}
\item {\bf Part 1.} For each $j_0 \in [n]$, for each $i_0 \in [d]$
\begin{align*}
    - C_0 I_n \preceq B_{j_0,i_0}(x) \preceq C_0 I_n
\end{align*}
\item {\bf Part 2.} For each $j_0 \in [n]$, for each $i_0 \in [d]$
\begin{align*}
    \| H_{j_0,i_0}(x) \| \leq C_0 R^2. 
\end{align*}
\item {\bf Part 3.} For each $j_0 \in [n]$, $i_0 \in [d]$, if $\min_{j_1 \in [n]} w_{j_1,j_1} \geq \frac{ l }{ \sigma_{\min}(\A_{j_0})^{2} } + C_0 $, then we have
\begin{align*}
    H_{\reg,j_0,i_0}(x) \succeq l \cdot I_{d^2}
\end{align*}
\item {\bf Part 4.} For each $j_0 \in [n]$, $i_0 \in [d]$, if $\min_{j_1 \in [n]} w_{j_1,j_1} \geq \frac{ l }{ \sigma_{\min}(\A_{j_0})^2 } + 100 \cdot C_0 $, then we have
\begin{align*}
 1.1 \cdot ( B(x)_{j_0,i_0} + W^2 ) \succeq W^2 \succeq 0.9 \cdot ( B(x)_{j_0,i_0} + W^2 ) 
\end{align*}
and 
\begin{align*}
   1.1 H_{j_0,i_0} \succeq H_{\reg,j_0,i_0} \succeq 0.9 H_{j_0,i_0}
\end{align*}
\item {\bf Part 5.}
For each $j_0 \in [n]$, $i_0 \in [d]$, if $\min_{j_1 \in [n]} w_{j_1,j_1} \geq \frac{ l }{ nd \sigma_{\min}(\A_{\min})^{2} } + C_0 $, then we have
\begin{align*}
    H_{\reg}(x) \succeq l \cdot I_{d^2}
\end{align*}
\item {\bf Part 6.} For each $j_0 \in [n]$, $i_0 \in [d]$, if $\min_{j_1 \in [n]} w_{j_1,j_1} \geq \frac{ l }{ nd \sigma_{\min}(\A_{\min})^{2} } + 100 \cdot C_0 $, then we have
\begin{align*}
1.1 H \succeq H_{\reg} \succeq 0.9  H
\end{align*}
\end{itemize}
\end{lemma}
\begin{proof}
{\bf Proof of Part 1.}

It directly follows from Lemma~\ref{lem:hessian_X_psd_tool}.

{\bf Proof of Part 2.}
We have
\begin{align*}
   \| H_{j_0,i_0} \| 
   = & ~ \| \A_{j_0}^\top B_{j_0,i_0}(x) \A_{j_0} \| \\
   \leq & ~ \| \A_{j_0} \|^2 \cdot \| B_{j_0,i_0}(x) \| \\
   \leq & ~ R^2 \cdot \| B_{j_0,i_0}(x) \| \\
   \leq & ~ 30 R^{10} 
\end{align*}
where the first step follows from the $H_{j_0,i_0} = \A_{j_0}^\top B_{j_0,i_0}(x) \A_{j_0}$, the second step follows from Fact~\ref{fac:matrix_norm}, the third step follows from $\max_{j_0 \in [n]} \| \A_{j_0} \| \leq R$, and the last step follow from {\bf Part 1}.

{\bf Proof of Part 3.}

The proof is similar to \cite{dls23}.

{\bf Proof of Part 4.}

The proof is similar to \cite{dls23}.

{\bf Proof of Part 5 and Part 6.}
It is because we can write $H$ as summation of $nd$ terms $H_{j_0,i_0}$ for all $j_0 \in [d]$, $i_0 \in [d]$.
\end{proof}

\subsection{PSD Bound}\label{sub:psd_H_xx:psd}

In this section, we analyze the PSD bound for each of the $B_{\rank}$ and $B_{\diag}$.

\begin{lemma}\label{lem:hessian_X_psd_tool}
If the following condition holds
\begin{itemize}
    \item $B_{\diag}^1 := (1-\gamma_{j_0}(x)) \cdot c(x,:)_{j_0,i_0} \cdot \diag( f(x)_{j_0} \circ v )$
    \item $B_{\rank}^1 := -( 2 \gamma_{j_0}(x) + c(x,:)_{j_0,i_0}) \cdot ( ( f(x)_{j_0} \circ v ) f(x)_{j_0}^\top + f(x)_{j_0} (f(x)_{j_0} \circ v)^\top )$
    \item $B_{\rank}^2 := ( 2 \gamma_{j_0}(x) c(x,:)_{j_0,i_0} + \gamma_{j_0}(x)^2 ) \cdot f(x)_{j_0} f(x)_{j_0}^\top $
    \item $B_{\rank}^3 := (f(x)_{j_0} \circ v) \cdot ( f(x)_{j_0} \circ v )^\top$
    \item $|\gamma(x)_{j_0}| \leq R^2$
    \item $|c(x,:)_{j_0,i_0}| \leq 2R^2$
    \item $\| v \|_2 \leq R^2$
\end{itemize}
Then, we have
\begin{itemize}
    \item Part 1.
    \begin{align*}
      -8R^6 \cdot I_n \preceq B_{\diag}^1 \preceq 8 R^6 \cdot I_n
    \end{align*}   
    \item Part 2. 
    \begin{align*}
        -16 R^8 \cdot I_n \preceq B_{\rank}^1 \preceq 16 R^8 \cdot I_n
    \end{align*}   
    \item Part 3.
    \begin{align*}
        -8 R^4 \cdot I_n \preceq B_{\rank}^2 \preceq 8 R^4 \cdot I_n
    \end{align*} 
    \item Part 4.
    \begin{align*}
        0 \cdot I_n \preceq B_{\rank}^3 \preceq 8 R^4 \cdot I_n
    \end{align*}
\end{itemize}
\end{lemma}
\begin{proof}
{\bf Proof of Part 1.}
\begin{align*}
     B_{\diag}^1 = & ~(1-\gamma_{j_0}(x)) \cdot c(x,:)_{j_0,i_0} \cdot \diag( f(x)_{j_0} \circ v ) \\
     \preceq & ~ |1-\gamma_{j_0}(x) ||c(x,:)_{j_0,i_0}|\|f(x)_{j_0}\|_2 
    \| v \|_2 \\
    \preceq & ~ 8R^2 \cdot I_n
\end{align*}
where the first step follows from the definition of $B_{\diag}^1$, the second step follows from Fact~\ref{fac:psd_rule}, and the last step follows from Lemma~\ref{lem:upper_bound}, $|\gamma(x)_{j_0}| \leq R^2,~|c(x,:)_{j_0,i_0}| \leq 2R^2$, and $\| v \|_2 \leq R^2$. 

{\bf Proof of Part 2.}
\begin{align*}
    B_{\rank}^1 
    = & ~ -( 2 \gamma_{j_0}(x) + c(x,:)_{j_0,i_0}) \cdot ( ( f(x)_{j_0} \circ v ) f(x)_{j_0}^\top + f(x)_{j_0} (f(x)_{j_0} \circ v)^\top ) \\
    \succeq & ~ - |2 \gamma_{j_0}(x) + c(x,:)_{j_0,i_0} | \cdot( f(x)_{j_0} \circ v )\cdot  f(x)_{j_0} \circ v )^{\top} + f(x)_{j_0}f(x)_{j_0}^\top) \\
    \succeq & ~ 4R^2 \cdot \|f(x)_{j_0} \circ v\|_2^2 \cdot \| f(x)_{j_0}\|_2^2 \\
    \succeq & ~ 4R^2\| f(x)_{j_0}\|_2^2 \| v\|_2^2 \cdot \| f(x)_{j_0}\|_2^2 \\
    \succeq & ~ 16R^8 \cdot I_n
\end{align*}
where the first step follows from the definition of $B_{\rank}^1$, the second step follows from Fact~\ref{fac:psd_rule}, the third step follows from $|\gamma(x)_{j_0}| \leq R^2,~
    |c(x,:)_{j_0,i_0}| \leq 2R^2$ and Fact~\ref{fac:psd_rule}, the fourth step follows from Fact~\ref{fac:circ_rules}, and last step follows from Lemma~\ref{lem:upper_bound} and $\| v\|_2 \leq R^2$. 

{\bf Proof of Part 3.}
\begin{align*}
    B_{\rank}^2 
    = & ~( 2 \gamma_{j_0}(x) c(x,:)_{j_0,i_0} + \gamma_{j_0}(x)^2 ) \cdot f(x)_{j_0} f(x)_{j_0}^\top \\
    \preceq & ~ |2 \gamma_{j_0}(x) c(x,:)_{j_0,i_0} + \gamma_{j_0}(x)^2 | \|f(x)_{j_0} \|_2^2 \\
    \preceq & ~ 8R^4 \cdot I_n
\end{align*}
where the first step follows from definition of $B_{\rank}^2$, the second step follows from Fact~\ref{fac:psd_rule}, and the last step follows from $|\gamma(x)_{j_0}| \leq R^2,~
    |c(x,:)_{j_0,i_0}| \leq 2R^2$ and Lemma~\ref{lem:upper_bound}.

{\bf Proof of Part 4.}
\begin{align*}
    B_{\rank}^3 
    = & ~ (f(x)_{j_0} \circ v) \cdot ( f(x)_{j_0} \circ v )^\top \\
    \preceq & ~ \|f(x)_{j_0} \circ v\|_2^2 \\
    \preceq & ~ \|f(x)_{j_0} \|_2^2 \|  v\|_2^2 \\
    \preceq & ~ 8R^4 \cdot I_n
\end{align*}
where the first step follows from definition of $B_{\rank}^3$, the second step follows from Fact~\ref{fac:psd_rule}, the third step follows from Fact~\ref{fac:circ_rules}, and the last step follows from $\| v\|_2 \leq R^2$ and Lemma~\ref{lem:upper_bound}.

\end{proof}

%\newpage
\section{Hessian for \texorpdfstring{$Y$}{}}\label{sec:hessian_Y}
In Section~\ref{sub:hessian_Y:hessian}, we present the hessian property with respect to $Y$. In Section~\ref{sub:hessian_Y:one_j0i0}, we compute the Hessian matrix with respect to $Y$ for one $j_0,i_0$.
\subsection{Hessian Property}\label{sub:hessian_Y:hessian}

In this section, we analyze the Hessian properties. 

\begin{lemma}\label{lem:hessian_property:y}
If the following conditions hold
\begin{itemize}
    \item Let $B_{j_0}(x) = f(x)_{j_0} f(x)_{j_0}^\top \in \R^{n \times n}$ (because of Lemma~\ref{lem:hessian_y_j_0_i_0})
    \item Let $B(x) = \sum_{j_0=1}^n B_{j_0}(x)$
    \item Let $H_{j_0,i_0} = \frac{\d^2 L_{j_0,i_0}}{ \d y_{i_0} \d y_{i_0}} = A_3^\top B_{j_0}(x) A_3 \in \R^{d \times d}$
    \item Let $H_{i_0} \in \R^{d \times d}$ be $H_{i_0} = \frac{\d^2 L}{\d y_{i_0} \d y_{i_0}} = \sum_{j_0=1}^d H_{j_0,i_0} $
    \item Let $H_{\reg,i_0} = A_3^\top ( B(x) + W^2 ) A_3$ where $W \in \R^{n \times n}$ is a positive diagonal matrix
    \item Let $H(y) \in \R^{d^2 \times d^2} $ be $ H(y)= \begin{bmatrix}
    H_{1} & 0 & \cdots & 0 \\
    0 & H_2& \cdots & 0 \\
    \vdots & \vdots & \ddots & \vdots \\
    0 & 0 & \cdots & H_d
    \end{bmatrix}$
\end{itemize}
Then, we have
\begin{itemize}
    \item {\bf Part 1.}
    \begin{align*}
        0 \preceq B_{j_0}(x) \preceq I_n
    \end{align*}
    \item {\bf Part 2.}
    \begin{align*}
        0 \preceq B(x) \preceq n \cdot I_n
    \end{align*}
    \item {\bf Part 3.} If $\min_{j_1 \in [n]} w_{j_1,j_1}^2 \geq \frac{l}{\sigma_{\min}(A_3)^2} $
    \begin{align*}
        H_{\reg, i_0} \succeq l \cdot I_{d}, ~~~ H(y) \succeq l\cdot I_{d^2}
    \end{align*}
    \item {\bf Part 4.} If $\min_{j_1 \in [n]} w_{j_1,j_1}^2 \geq \frac{l}{\sigma_{\min}(A_3)^2} + 100 n$
    \begin{align*}
        0.9  (W^2+B(x)) \preceq    W^2 \preceq 1.1 (W^2+B(x))
    \end{align*}
    \item {\bf Part 5.} Lipschitz, Due to $H(y)$ is independent of $y$, then
    \begin{align*}
        \| H(y) - H(\wt{y}) \| \leq \| y - \wt{y} \|_2
    \end{align*}
\end{itemize}
\end{lemma}
\begin{proof}
For hessian closed-form, we can obtain them from Lemma~\ref{lem:hessian_y_j_0_i_0}.

The proofs are straightforward, so we omit the details here.
\end{proof}

\subsection{Hessian for One \texorpdfstring{$j_0,i_0$}{}}\label{sub:hessian_Y:one_j0i0}

In this section, we analyze the Hessian for the matrix $Y$ with one $j_0,i_0$.

\begin{lemma}\label{lem:hessian_y_j_0_i_0}
If the following conditions hold
\begin{itemize}
    \item We define a temporary notation here $v := f(x)_{j_0}$ (for simplicity we drop the index $j_0$ in the statement. Note that $v$ could have different meaning in other sections.)  
    \item Let $f(x)_{j_0}$ be defined as Definition~\ref{def:f}.
    \item Let $c(x,:)_{j_0,i_0}$ be defined as Definition~\ref{def:c}.
    \item Let $h(y)_{i_0}$ be defined as Definition~\ref{def:h}.
    \item Let $L_{j_0,i_0}$ be defined as Definition~\ref{def:f}.
\end{itemize}
Then, we have
\begin{itemize}
    \item {\bf Part 1.} For $i_1 = i_2$, the diagonal case
    \begin{align*}
        \frac{\d^2 L_{j_0,i_0}}{\d y_{i_0,i_1} \d y_{i_0,i_1}} = A_{3,*,i_1}^\top v v^\top A_{3,*,i_1}
    \end{align*}
    \item {\bf Part 2.} For $i_1 \neq i_2$, the off-diagonal case
    \begin{align*}
        \frac{\d^2 L_{j_0,i_0} }{ \d y_{i_0,i_1} \d y_{i_0,i_2} } = A_{3,*,i_1}^\top v v^\top A_{3,*,i_2}
    \end{align*}
    \item {\bf Part 3.} The $\frac{\d^2 L_{j_0,i_0}}{\d y_{i_0} \d y_{i_0}} \in \R^{d \times d}$
    \begin{align*}
        \frac{\d^2 L_{j_0,i_0}}{\d y_{i_0} \d y_{i_0}} = A_{3}^\top vv^\top A_3
    \end{align*}
\end{itemize}
\end{lemma}
\begin{proof}

{\bf Proof of Part 1.}

\begin{align*}
\frac{\d^2 L_{j_0,i_0}}{\d y_{i_0,i_1} \d y_{i_0,i_1}} 
= & ~ \frac{\d }{\d y_{i_0,i_1}} ( \frac{\d }{\d y_{i_0,i_1}} L_{j_0,i_0} ) \\
= & ~ \frac{\d }{\d y_{i_0,i_1}} ( c(:,y)_{j_0,i_0} \langle v, A_{3,*,i_1} \rangle ) \\
= & ~ \langle v, A_{3,*,i_1} \rangle \cdot \langle v, A_{3,*,i_1} \rangle \\
= & ~ A_{3,*,i_1}^\top v v^\top A_{3,*,i_1}
\end{align*}
where the first step follows from simple algebra, the second step follows from Lemma~\ref{lem:gradient_y}, the third step follows from Lemma~\ref{lem:gradient_y}, and the last step follows from Fact~\ref{fac:circ_rules}.

{\bf Proof of Part 2.}

\begin{align*}
\frac{\d^2 L_{j_0,i_0}}{\d y_{i_0,i_2} \d y_{i_0,i_1}} 
= & ~ \frac{\d }{\d y_{i_0,i_2}} ( \frac{\d }{\d y_{i_0,i_1}} L_{j_0,i_0} ) \\
= & ~ \frac{\d }{\d y_{i_0,i_2}} ( c(:,y)_{j_0,i_0} \langle v, A_{3,*,i_1} \rangle ) \\
= & ~ \langle v, A_{3,*,i_2} \rangle \cdot \langle v, A_{3,*,i_1} \rangle \\
= & ~ A_{3,*,i_1}^\top v v^\top A_{3,*,i_2}
\end{align*}
where the first step follows from simple algebra, the second step follows from Lemma~\ref{lem:gradient_y}, the third step follows from Lemma~\ref{lem:gradient_y}, and the last step follows from Fact~\ref{fac:circ_rules}.

{\bf Proof of Part 3.}

It follows by combining above two parts directly.
\end{proof}

%\newpage
\section{Hessian for \texorpdfstring{$X$}{} and \texorpdfstring{$Y$}{}}\label{sec:hessian_XY}
In Section~\ref{sub:hessian_XY:hessian}, we compute the Hessian matrix with respect to both $X$ and $Y$. In Section~\ref{sub:hessian_XY:help_lem}, we present several helpful lemmas for the following proof. In Section~\ref{sub:hessian_XY:help_lem}, we create $B(x)$ for the further analysis.

\subsection{Computing Hessian}\label{sub:hessian_XY:hessian}

In this section, we compute the Hessian matrix for $X$ and $Y$.

\begin{lemma}\label{lem:hessian_xy}
If the following conditions hold
\begin{itemize}
    \item Let $f(x)_{j_0}$ be defined as Definition~\ref{def:f}.
    \item Let $c(x,y)_{j_0,i_0}$ be defined as Definition~\ref{def:c}.
    \item Let $h(y)_{i_0}$ be defined as Definition~\ref{def:h}.
    \item Let $L_{j_0,i_0}$ be defined as Definition~\ref{def:L}.
    
\end{itemize}
Then, we have
\begin{itemize}
    \item Part 1.
    \begin{align*}
        \frac{\d }{\d y_{i_0,i_1} } ( \frac{\d }{\d x_i} L_{j_0,i_0} )  
        = & ~ \langle f(x)_{j_0}, A_{3,*,i_1} \rangle \cdot \langle f(x)_{j_0} \circ \A_{j_0,i} , h(y)_{i_0} \rangle \\
& ~ - \langle f(x)_{j_0}, A_{3,*,i_1} \rangle \langle f(x)_{j_0}, h(y)_{i_0} \rangle \cdot \langle f(x)_{j_0}, \A_{j_0,i} \rangle \\
& ~ + c(x,y)_{j_0, i_0} \cdot ( \langle  f(x)_{j_0} \circ \A_{j_0,i}, A_{3,*,i_1} \rangle - \langle  f(x)_{j_0} , A_{3,*,i_1} \rangle \cdot \langle f(x)_{j_0}, \A_{j_0,i} \rangle )
    \end{align*}
\end{itemize}
\end{lemma}
\begin{proof}

We can show
\begin{align*}
 & ~ \frac{\d }{\d y_{i_0,i_1} } ( \frac{\d }{\d x_i} L_{j_0,i_0} ) \\
= & ~ \frac{\d }{\d y_{i_0,i_1}} (  c(x,y)_{j_0, i_0} \cdot ( \langle  f(x)_{j_0} \circ \A_{j_0,i}, h(y)_{i_0} \rangle - \langle  f(x)_{j_0} , h(y)_{i_0} \rangle \cdot \langle f(x)_{j_0}, \A_{j_0,i} \rangle ) ) \\
= & \frac{\d }{\d y_{i_0,i_1}} (  c(x,y)_{j_0, i_0} ) \cdot ( \langle  f(x)_{j_0} \circ \A_{j_0,i}, h(y)_{i_0} \rangle - \langle  f(x)_{j_0} , h(y)_{i_0} \rangle \cdot \langle f(x)_{j_0}, \A_{j_0,i} \rangle )  \\
& ~ +(  c(x,y)_{j_0, i_0}) \cdot  \frac{\d }{\d y_{i_0,i_1}}  ( \langle  f(x)_{j_0} \circ \A_{j_0,i}, h(y)_{i_0} \rangle - \langle  f(x)_{j_0} , h(y)_{i_0} \rangle \cdot \langle f(x)_{j_0}, \A_{j_0,i} \rangle )  \\
= & ~ \langle f(x)_{j_0}, A_{3,*,i_1} \rangle \cdot ( \langle  f(x)_{j_0} \circ \A_{j_0,i}, h(y)_{i_0} \rangle - \langle  f(x)_{j_0} , h(y)_{i_0} \rangle \cdot \langle f(x)_{j_0}, \A_{j_0,i} \rangle )   \\
& ~ + c(x,y)_{j_0, i_0} \cdot ( \langle  f(x)_{j_0} \circ \A_{j_0,i}, A_{3,*,i_1} \rangle - \langle  f(x)_{j_0} , A_{3,*,i_1} \rangle \cdot \langle f(x)_{j_0}, \A_{j_0,i} \rangle ) \\
= & ~ \langle f(x)_{j_0}, A_{3,*,i_1} \rangle \cdot \langle f(x)_{j_0} \circ \A_{j_0,i} , h(y)_{i_0} \rangle \\
& ~ - \langle f(x)_{j_0}, A_{3,*,i_1} \rangle \langle f(x)_{j_0}, h(y)_{i_0} \rangle \cdot \langle f(x)_{j_0}, \A_{j_0,i} \rangle \\
& ~ + c(x,y)_{j_0, i_0} \cdot ( \langle  f(x)_{j_0} \circ \A_{j_0,i}, A_{3,*,i_1} \rangle - \langle  f(x)_{j_0} , A_{3,*,i_1} \rangle \cdot \langle f(x)_{j_0}, \A_{j_0,i} \rangle )
\end{align*}
where the first step is due to {\bf Part 6} of Lemma~\ref{lem:gradient_x}, the second step comes from the product rule of derivative, the third step is based on Lemma~\ref{lem:hessian_y_j_0_i_0}, and the last step follows from simple algebra.

Thus, we complete the proof.
\end{proof}

\subsection{A Helpful Lemma}\label{sub:hessian_XY:help_lem}

In this section, we provide a helpful Lemma.

\begin{lemma}\label{lem:hessian_xy_help_lem}
If the following conditions hold
\begin{itemize}
    \item Let $f(x)_{j_0}$ be defined in Definition~\ref{def:f}.
    \item Let $\A \in \R^{n^2 \times d^2}$ be defined in Definition~\ref{def:u}.
    \item Let $c(x,y)_{j_0,i_0}$ be defined as Definition~\ref{def:c}.
    \item Let $h(y)_{i_0}$ be defined as Definition~\ref{def:h}.
    \item Let $L_{j_0,i_0}$ be defined as Definition~\ref{def:L}.
\end{itemize}
Then, we have
\begin{itemize}
    \item {\bf Part 1.} 
    \begin{align*}
        \langle f(x)_{j_0}, A_{3,*,i_1} \rangle \cdot \langle f(x)_{j_0} \circ \A_{j_0,i} , h(y)_{i_0} \rangle = \A_{j_0,i}^\top (f(x)_{j_0} \circ h(y)_{i_0}) f(x)_{j_0}^\top A_{3,*,i_1} 
    \end{align*}
    \item {\bf Part 2.}
    \begin{align*}
         \langle f(x)_{j_0}, A_{3,*,i_1} \rangle \cdot \langle f(x)_{j_0}, h(y)_{i_0} \rangle \cdot \langle f(x)_{j_0}, \A_{j_0,i} \rangle  = \langle f(x)_{j_0}, h(y)_{i_0} \rangle \cdot \A_{j_0,i}^\top f(x)_{j_0} f(x)_{j_0}^\top A_{3,*,i_1}
    \end{align*}
    \item {\bf Part 3.}
    \begin{align*}
         \langle f(x)_{j_0} \circ \A_{j_0,i}^{\top}, A_{3,*,i_1} \rangle =  \A_{j_0,i} ^{\top}\diag(f(x)_{j_0}) A_{3,*,i_1}
    \end{align*}
    \item {\bf Part 4.}
    \begin{align*}
         \langle  f(x)_{j_0} , A_{3,*,i_1} \rangle \cdot \langle f(x)_{j_0}, \A_{j_0,i} \rangle = \A_{j_0,i}^{\top} f(x)_{j_0} f(x)_{j_0}^\top A_{3,*,i_1}
    \end{align*}
\end{itemize}
\end{lemma}
\begin{proof}
{\bf Proof of Part 1.}
\begin{align*}
     \langle f(x)_{j_0}, A_{3,*,i_1} \rangle \cdot \langle f(x)_{j_0} \circ \A_{j_0,i} , h(y)_{i_0} \rangle 
     = & ~ \langle f(x)_{j_0} \circ  h(y)_{i_0} , \A_{j_0,i} \rangle f(x)_{j_0}^{\top}  A_{3,*,i_1} \\ 
     = & ~ \A_{j_0,i}^{\top}(f(x)_{j_0} \circ  h(y)_{i_0})  f(x)_{j_0}^{\top}  A_{3,*,i_1}
\end{align*}
where the first step follows from Fact~\ref{fac:circ_rules}, and the second step follows from Fact~\ref{fac:circ_rules}.

{\bf Proof of Part 2.}
\begin{align*}
     \langle f(x)_{j_0}, A_{3,*,i_1} \rangle \cdot \langle f(x)_{j_0}, h(y)_{i_0} \rangle \cdot \langle f(x)_{j_0}, \A_{j_0,i} \rangle  
     = & ~  \langle f(x)_{j_0}, h(y)_{i_0} \rangle \A_{j_0,i}^{\top} f(x)_{j_0} f(x)_{j_0}^{\top} A_{3,*,i_1}
\end{align*}
where the first step follows from Fact~\ref{fac:circ_rules}.

{\bf Proof of Part 3.}
\begin{align*}
    \langle f(x)_{j_0} \circ \A_{j_0,i}, A_{3,*,i_1} \rangle = & ~  (f(x)_{j_0} \circ \A_{j_0,i})^{\top} A_{3,*,i_1} \\
    = & ~ (\diag(f(x)_{j_0}) \A_{j_0,i})^{\top} A_{3,*,i_1}\\
    = & ~ \A_{j_0,i}^{\top} \diag(f(x)_{j_0}) A_{3,*,i_1}
\end{align*}
where the first, second, and last step follows from Fact~\ref{fac:circ_rules}.

{\bf Proof of Part 4.}
\begin{align*}
    \langle  f(x)_{j_0} , A_{3,*,i_1} \rangle \cdot \langle f(x)_{j_0}, \A_{j_0,i} \rangle = \A_{j_0,i}^{\top} f(x)_{j_0} f(x)_{j_0}^\top A_{3,*,i_1}
\end{align*}
where the first step follows from Fact~\ref{fac:circ_rules}.

\end{proof}

\subsection{Creating \texorpdfstring{$B(x,y)$}{}}\label{sub:hessian_XY:B}

In this section, we give a formal definition of $B(x,y)$.

\begin{definition}\label{def:B(x,y)}
We define $B(x,y)$ 
\begin{align*}
    B(x,y) = B_{\diag}^1 + B_{\rank}^1 + B_{\rank}^2 + B_{\rank}^1
\end{align*}
where
\begin{itemize}
    \item $B_{\rank}^1 (x,y) = ( f(x)_{j_0} \circ h(y)_{i_0} ) f(x)_{j_0}^\top$
    \item $B_{\rank}^2(x,y) = - \langle f(x)_{j_0}, h(y)_{i_0} \rangle f(x)_{j_0} f(x)_{j_0}^\top$
    \item $B_{\diag}^1(x,y) = - c(x,y)_{j_0,i_0} \diag( f(x)_{j_0} ) $
    \item $B_{\rank}^3(x,y) = c(x,y)_{j_0,i_0} f(x)_{j_0} f(x)_{j_0}^\top$
\end{itemize}

\end{definition}

\begin{lemma}
If the following conditions 
\begin{itemize}
    \item Let $B(x,y)$ be defined as Definition~\ref{def:B(x,y)}.
\end{itemize}
Then, we have 
\begin{itemize}
\item {\bf Part 1.} 
\begin{align*}
   \frac{\d^2 L_{j_0,i_0}}{ \d y_{i_0} \d x} = \A_{j_0}^\top B(x,y) A_3 \in \R^{d^2 \times d}
\end{align*}
\item {\bf Part 2.} $i_1 \neq i_0$
\begin{align*}
   \frac{\d^2 L_{j_0,i_0}}{ \d y_{i_1} \d x} = \A_{j_0}^\top {\bf 0}_{n \times n} A_3 \in \R^{d^2 \times d} = {\bf 0}_{n \times n}
\end{align*}
\end{itemize}
\end{lemma}
\begin{proof}
{\bf Proof of Part 1.}
We have
\begin{align*}
    \frac{\d^2 L_{j_0,i_0}}{ \d y_{i_0,i_2} \d x_i} 
    = & ~ \A^{\top}_{j_0,i} B(x,y) A_{3,*,i_2} \\
\end{align*}
where the first step follows from combining Lemma~\ref{lem:hessian_xy} and Lemma~\ref{lem:hessian_xy_help_lem}.

Then, we can have
\begin{align*}
    \frac{\d^2 L_{j_0,i_0}}{ \d y_{i_0} \d x} = \A_{j_0}^\top B(x,y) A_3 
\end{align*}

{\bf Proof of Part 2.}
We have
\begin{align*}
    \frac{\d^2 L_{j_0,i_0}}{ \d y_{i_1,i_2} \d x_i} 
    = & ~ \A^{\top}_{j_0,i} {\bf 0}_{n \times n} A_{3,*,i_2} = {\bf 0}_{n \times n} \\
\end{align*}
where the first step follows from combining Lemma~\ref{lem:hessian_xy} and Lemma~\ref{lem:hessian_xy_help_lem}.

Then, we can have
\begin{align*}
    \frac{\d^2 L_{j_0,i_0}}{ \d y_{i_1} \d x} = \A_{j_0}^\top {\bf 0}_{n \times n} A_3 = {\bf 0}_{n \times n}
\end{align*}
\end{proof}

%\newpage
\section{Lipschitz for Hessian of \texorpdfstring{$x,y$}{}}\label{sec:lips_H_xy}
In Section~\ref{sub:lips_H_xy:main_res}, we present the main results of the Lipschitz property of $H_{x,y}$. In Section~\ref{sub:lips_H_xy:summary}, we summarize the results from the following steps 1-4. In Section~\ref{sub:lips_H_xy:upper_bound}, we compute the upper bound of basic functions for the following proof. In Section~\ref{sub:lips_H_xy:basic_lips}, we compute the Lipschitz Property of basic functions for the following proof. In Section~\ref{sub:lips_H_xy:step1}, we analyze the first step of Lipschitz function $( f(x)_{j_0} \circ h(y)_{i_0} ) f(x)_{j_0}^\top$. In Section~\ref{sub:lips_H_xy:step2}, we analyze the second step of Lipschitz function $- \langle f(x)_{j_0}, h(y)_{i_0} \rangle f(x)_{j_0} f(x)_{j_0}^\top$. In Section~\ref{sub:lips_H_xy:step3}, we analyze the third step of Lipschitz function $- c(x,y)_{j_0,i_0} \diag( f(x)_{j_0} ) $. In Section~\ref{sub:lips_H_xy:step4}, we analyze the fourth step of Lipschitz function $c(x,y)_{j_0,i_0} f(x)_{j_0} f(x)_{j_0}^\top $. In Section~\ref{sub:lips_H_xy:psd}, we compute the PSD upper bound for the Hessian matrix. In Section~\ref{sub:lips_H_xy:summary_psd}, we summarize PSD upper bound of $G(x,y)$.

\subsection{Main Results}\label{sub:lips_H_xy:main_res}

In this section, we present the main result of Section~\ref{sec:lips_H_xy}.

\begin{lemma} \label{lem:lips_H_xy}
If the following conditions hold
 \begin{itemize}
    \item $\max_{j_0 \in [n]} \| \A_{j_0}  \| \leq R$
    \item Let $H(x,y)_{j_0,i_0} \in \R^{d^2 \times d}$ denote $\frac{\d^2 L_{j_0,i_0}}{ \d x \d y_{i_0} }$
    \item $\frac{\d^2 L_{j_0,i_0}}{ \d x \d y_{i_1}} = {\bf 0}_{d^2 \times d}$
    \item Let $H(x,y) \in \R^{d^2 \times d^2}$ be 
    \begin{align*}
    H(x,y) : =
    \begin{bmatrix}
        \sum_{j_0=1}^n H_{j_0,1}(x,y) & \sum_{j_0=1}^n H_{j_0,2}(x,y) & \cdots & \sum_{j_0=1}^n H_{j_0,d}(x,y)
    \end{bmatrix}
    \end{align*}
\end{itemize} 

Then we have
\begin{itemize}
    \item Part 1. For $j_0 \in [d], i_0 \in [n]$ 
    \begin{align*}
    \|H(x,y)_{j_0,i_0} - H(\wt{x},\wt{y})_{j_0,i_0} \| \leq n^{1.5} \exp(20R^2) \cdot ( \|x -\wt{x} \|_2 +   \| y -\wt{y} \|_2 )
\end{align*}
    \item Part 2. \begin{align*}
    \|H(x,y) - H(\wt{x},\wt{y}) \| \leq n^{2.5} d(  \|x -\wt{x} \|_2 + \| y -\wt{y} \|_2)
\end{align*}
\end{itemize}
\end{lemma}
\begin{proof}
    {\bf Proof of Part 1.}
    It follows from Lemma~\ref{lem:summary_Gi_xy}.

    {\bf Proof of Part 2}.
    We can show that
    \begin{align*}
        \| H(x,y) - H(\wt{x},\wt{y}) \| \leq nd \cdot n^{1.5} \exp(20R^2) ( \| x - \wt{x} \|_2 + \| y - \wt{y} \|_2 )
    \end{align*}
    where the first step follows from that  we can write $H$ as summation of $nd$ terms $H_{j_0,i_0}$ for all $j_0 \in [d]$, $i_0 \in [d]$.
\end{proof}

\subsection{Summary of Four Steps on Lipschitz for Matrix Functions}\label{sub:lips_H_xy:summary}

In this section, we summarize the four steps for analyzing the Lipschitz for different matrix functions.

\begin{lemma}\label{lem:summary_Gi_xy}
    If the following conditions hold 
    \begin{itemize}
        \item $G_{1}(x,y) = ( f(x)_{j_0} \circ h(y)_{i_0} ) f(x)_{j_0}^\top
        $
        \item $G_{2}(x,y) = - \langle f(x)_{j_0}, h(y)_{i_0} \rangle f(x)_{j_0} f(x)_{j_0}^\top$
        \item $G_3(x,y) = - c(x,y)_{j_0,i_0} \diag( f(x)_{j_0} ) $
        \item $G_4(x,y) = c(x,y)_{j_0,i_0} f(x)_{j_0} f(x)_{j_0}^\top$
        
    \end{itemize}

    Then, we have
    \begin{align*}
        \sum_{k=1}^4 \|G_{k }(x,y) - G_{k}(\wt{x}, \wt{y})\| \leq n^{1.5} \exp(20R^2) ( \|x -\wt{x} \|_2 +  \| y -\wt{y} \|_2 )
    \end{align*}
\end{lemma}
\begin{proof}
    The proof follows from Lemma~\ref{lem:lipschitz_xy_G1}, Lemma~\ref{lem:lipschitz_xy_G2}, Lemma~\ref{lem:lipschitz_xy_G3}, and Lemma~\ref{lem:lipschitz_xy_G4}.
\end{proof}

\subsection{A Core Tool: Upper Bound for Several Basic Functions}\label{sub:lips_H_xy:upper_bound}

In this section, we give an upper bound for each of the basic functions.

\begin{lemma}
    If the following conditions hold
    \begin{itemize}
     \item Let $f(y)_{j_0} \in \R^n $ be defined as Definition~\ref{def:f}.
    \item Let $h(y)_{i_0} \in \R^n $ be defined as Definition~\ref{def:h}.
    \item Let $c(x,y)_{j_0,i_0} \in \R $ be defined as Definition~\ref{def:c}.
    \item Let $ R \geq 4$
    \item $\|A_3 \| \leq R$
    \item $\|y_{i_0} \| \leq R$ 
    \item $\|b_{j_0,i_0} \|_2 \leq   R$
    
    \end{itemize}

    Then, we have
    \begin{itemize}
        \item Part 1. $\| h(y)_{i_0}\|_2 \leq R^2$
        \item Part 2. $| c(x,y)_{j_0,i_0}| \leq 2R^2$
    \end{itemize}
\end{lemma}
\begin{proof}
    {\bf Proof of Part 1.}
    \begin{align*}
        \|h(y)_{i_0} \|_2 = & ~ \|A_3 y_{i_0} \|_2 \\
        \leq & ~ \|A_3 \| \|y_{i_0} \|_2 \\
        \leq & ~ R^2 
    \end{align*}
    where the first step is due to Definition~\ref{def:h}, the second step is based on Fact~\ref{fac:matrix_norm} and the third step is because of Lemma~\ref{lem:upper_bound}.
    
    {\bf Proof of Part 2.}
    \begin{align*}
        |c(x,y)_{j_0,i_0}| = & ~ |\langle f(x)_{j_0}, h(y)_{i_0}\rangle  - b_{j_0,i_0}| \\
        \leq  & ~ \| f(x)_{j_0}\|_2 \|h(y)_{i_0} \|_2 + |b_{j_0,i_0} | \\
        \leq & ~ R^2 + R\\
        \leq & ~ 2R^2
    \end{align*}
    where the first step is because of Definition~\ref{def:c}, the second step is based on triangle inequality and Cauchy–Schwarz inequality, the third step is due to Lemma~\ref{lem:upper_bound}, and the last step follows from $R \geq 4$. 
\end{proof}
\subsection{A Core Tool: Lipschitz Property for Several Basic Functions}\label{sub:lips_H_xy:basic_lips}

In this section, we introduce the Lipschitz property for several basic functions.

\begin{lemma}\label{lem:upper_bound:y}
    If the following conditions hold
    \begin{itemize}
         \item Let $f(y)_{j_0} \in \R^n $ be defined as Definition~\ref{def:f}.
    \item Let $h(y)_{i_0} \in \R^n $ be defined as Definition~\ref{def:h}.
    \item Let $c(x,y)_{j_0,i_0} \in \R $ be defined as Definition~\ref{def:c}.
    \item Let $ R \geq 4$
    \item $\|A_3 \| \leq R$
    \item $\|y_{i_0} \| \leq R$ 
    \item $\|b_{j_0,i_0} \|_2 \leq   R$
    \item Let $R_0$ be defined as Definition~\ref{def:R_0}.
    \end{itemize}

    Then, we have
    \begin{itemize}
        \item Part 1. $\|h(y)_{i_0} -  h(\wt{y})_{i_0}\|_2 \leq R\|y - \wt{y} \|_2$
        \item Part 2.
         $| c(x,y)_{j_0,i_0} - c(\wt{x},y_{j_0,i_0})| \leq R^2 \cdot R_0 \|x -\wt{x} \|$
         \item Part 3.
         $|c(x,y)_{j_0,i_0} -c(x,\wt{y})_{j_0,i_0})| \leq R \| y - \wt{y} \|_2 $
    \end{itemize}
\end{lemma}
\begin{proof}
    {\bf Proof of Part 1.}
    \begin{align*}
        \|h(y)_{i_0} -  h(\wt{y})_{i_0}\|_2 = & ~ \|A_3 y_{i_0} - A_3 \wt{y}_{i_0} \|_2 \\
        \leq & ~\|A_3 \| \|y_{i_0} - \wt{y}_{i_0}\|_2 \\
        \leq & ~ R\|y - \wt{y} \|_2
    \end{align*}
    where the first step follows from Definition~\ref{def:h}, the second step is based on Fact~\ref{fac:matrix_norm}, and the third step is due to Lemma~\ref{lem:upper_bound}.
    
    {\bf Proof of Part 2.}
    \begin{align*}
        | c(x,y)_{j_0,i_0} - c(\wt{x},y_{j_0,i_0})| = & ~ |\langle f(x)_{j_0}, h(y)_{i_0} \rangle - b_{j_0,i_0} - (\langle f(\wt{x})_{j_0}, h(y)_{i_0} \rangle - b_{j_0,i_0}) |\\
        \leq & ~ \|f(x)_{j_0} - f(\wt{x})_{j_0} \|_2 \| h(y)_{i_0} \|_2\\
        \leq & ~ R^2 \cdot R_0 \|x-\wt{x}\|_2
    \end{align*}
    where the first step is due to Definition~\ref{def:c}, the second step follows from Cauchy–Schwarz inequality, and the third step is because of {\bf Part 1} of Lemma~\ref{sub:lips_H_xy:upper_bound} and {\bf Part 3} of Lemma~\ref{lem:basic_lips}.
    
    {\bf Proof of Part 3.}
    \begin{align*}
        |c(x,y)_{j_0,i_0} -c(x,\wt{y})_{j_0,i_0})| = & ~
        |\langle f(x)_{j_0}, h(y)_{i_0} \rangle - b_{j_0,i_0} - (\langle f(x)_{j_0}, h(\wt{y})_{i_0} \rangle - b_{j_0,i_0}) | \\
        \leq & ~ \|f(x)_{j_0} \|_2 \cdot \|h(y)_{i_0} - h(\wt{y})_{i_0} \|_2 \\
        \leq & ~ R \|y -\wt{y} \|_2
    \end{align*}
    where the first step follows from Definition~\ref{def:c}, the second step is due to Cauchy–Schwarz inequality and the third step is because of {\bf Part 4} of Lemma~\ref{lem:upper_bound} and {\bf Part 1} of this Lemma.
\end{proof}

\subsection{Calculation: Step 1 Lipschitz for Matrix Function \texorpdfstring{$( f(x)_{j_0} \circ h(y)_{i_0} ) f(x)_{j_0}^\top$}{}}\label{sub:lips_H_xy:step1}

In this section, we calculate the Lipschitz for $( f(x)_{j_0} \circ h(y)_{i_0} ) f(x)_{j_0}^\top$.

\begin{lemma}\label{lem:lipschitz_xy_G1}
If the following conditions
\begin{itemize}
    \item Let $G_1(x,y) = ( f(x)_{j_0} \circ h(y)_{i_0} ) f(x)_{j_0}^\top$ 
    \item Let $R_0$ be defined in Definition~\ref{def:R_0}.
    \item Let $\alpha(x)_{j_0} \in \R$ be defined as Definition~\ref{def:alpha}
    \item Let $f(x)_{j_0} \in \R^n$ be defined as Definition~\ref{def:f}
    \item Let $c(x,y)_{j_0,i_0} \in \R$ be defined as Definition~\ref{def:c}
    \item Let $\gamma(x)_{j_0} = \langle f(x)_{j_0}, v \rangle \in \R$
    \item $\| A_1 \|, \| A_2 \| , \| A_3 \| \leq R$, $\| \A_{j_0} \| \leq R$, $\| x \|_2 \leq R$,$| b_{j_0,i_0} | \leq R$, $\| v \|_2 \leq R^2$
    \item Let $R \geq 4$
\end{itemize}
Then, we have 
\begin{align*}
    \|G_1(x,y) - G_1( \wt{x}, \wt{y} ) \| \leq 2R^2 \cdot R_0 (\|x -\wt{x} \|_2 +  \| y -\wt{y} \|_2)
\end{align*}
\end{lemma}
\begin{proof}
We define
\begin{align*}
    G_{1,1} = & ~ ( f(x)_{j_0} \circ h(y)_{i_0} ) f(x)_{j_0}^\top - ( f(\wt{x})_{j_0} \circ h(y)_{i_0} ) f(x)_{j_0}^\top \\
    G_{1,2} = & ~ ( f(\wt{x})_{j_0} \circ h(y)_{i_0} ) f(x)_{j_0}^\top - ( f(\wt{x})_{j_0} \circ h(\wt{y})_{i_0} ) f(x)_{j_0}^\top \\
    G_{1,3} = & ~ ( f(\wt{x})_{j_0} \circ h(\wt{y})_{i_0} ) f(x)_{j_0}^\top - ( f(\wt{x})_{j_0} \circ h(\wt{y})_{i_0} ) f(\wt{x})_{j_0}^\top 
\end{align*}
where the first step follows from definition of $G_{1,1}$, the second step is based on Fact~\ref{fac:vector_norm} and the third step is due to Lemma~\ref{lem:upper_bound}.

We have
\begin{align*}
    \|G_{1,1}\| = & ~ \| ( f(x)_{j_0} \circ h(y)_{i_0} ) f(x)_{j_0}^\top - ( f(\wt{x})_{j_0} \circ h(y)_{i_0} ) f(x)_{j_0}^\top \| \\
    \leq & ~ \|f(x)_{j_0} -f(\wt{x})_{j_0}\|_{\infty} \cdot \| h(y)_{i_0} \|_2 \cdot \|f(x)_{j_0} \|_2 \\
    \leq & ~ R^2 \cdot R_0 \|x - \wt{x} \|_2
\end{align*}
where the first step follows from definition of $G_{1,1}$, the second step is due to Fact~\ref{fac:matrix_norm}, and the third step is based on combining Lemma~\ref{lem:upper_bound}, Lemma~\ref{lem:basic_lips}, and Lemma~\ref{sub:lips_H_xy:upper_bound}.

Also, we have 
\begin{align*}
    \|G_{1,2} \| = & ~ 
    \|( f(\wt{x})_{j_0} \circ h(y)_{i_0} ) f(x)_{j_0}^\top - ( f(\wt{x})_{j_0} \circ h(\wt{y})_{i_0} ) f(x)_{j_0}^\top\| \\
    \leq & ~ \|f(\wt{x})_{j_0} \|_2 \cdot \|h(y)_{i_0} - h(\wt{y})_{i_0} \|_2 \cdot \|f(x)_{j_0} \|_2 \\
    \leq & ~ R\|y -\wt{y} \|_2
\end{align*}
where the first step is based on definition of $G_{1,2}$, the second step is because of Fact~\ref{fac:matrix_norm}, and the third step follows from Lemma~\ref{lem:upper_bound:y}.

Additionally, 
\begin{align*}
    \|G_{1,3} \| = & ~ \|( f(\wt{x})_{j_0} \circ h(\wt{y})_{i_0} ) f(x)_{j_0}^\top - ( f(\wt{x})_{j_0} \circ h(\wt{y})_{i_0} ) f(\wt{x})_{j_0}^\top\| \\
    \leq & ~  \|f(\wt{x})_{j_0} \|_2 \cdot \|h(\wt{y})_{i_0} \|_2 \cdot \|f(x)_{j_0} - f(\wt{x})_{j_0} \|_2 \\
    \leq & ~ R^2 \cdot R_0\|x - \wt{x}\|_2
\end{align*}
where the first step follows from the definition of $G_{1,3}$, the second step follows from Fact~\ref{fac:matrix_norm}, and the third step is because of Lemma~\ref{lem:basic_lips}.

Combining all the above equations we complete the proof.
\end{proof}

\subsection{Calculation: Step 2 Lipschitz for Matrix Function \texorpdfstring{$- \langle f(x)_{j_0}, h(y)_{i_0} \rangle f(x)_{j_0} f(x)_{j_0}^\top$}{}}\label{sub:lips_H_xy:step2}

In this section, we calculate the Lipschitz for $- \langle f(x)_{j_0}, h(y)_{i_0} \rangle f(x)_{j_0} f(x)_{j_0}^\top$.

\begin{lemma}\label{lem:lipschitz_xy_G2}
If the following conditions
\begin{itemize}
    \item Let $\alpha(x)_{j_0} \in \R$ be defined as Definition~\ref{def:alpha}
    \item Let $f(x)_{j_0} \in \R^n$ be defined as Definition~\ref{def:f}
    \item Let $c(x,y)_{j_0,i_0} \in \R$ be defined as Definition~\ref{def:c}
    \item Let $\gamma(x)_{j_0} = \langle f(x)_{j_0}, v \rangle \in \R$
    \item $\| A_1 \|, \| A_2 \| , \| A_3 \| \leq R$, $\| \A_{j_0} \| \leq R$, $\| x \|_2 \leq R$,$| b_{j_0,i_0} | \leq R$, $\| v \|_2 \leq R^2$
    \item Let $R \geq 4$
    \item Let $G_2(x,y) = - \langle f(x)_{j_0}, h(y)_{i_0} \rangle f(x)_{j_0} f(x)_{j_0}^\top$ 
\end{itemize}
Then, we have
\begin{align*}
    \|G_2(x,y) - G_2( \wt{x}, \wt{y} ) \| \leq 3R^2 R_0 (\|x -\wt{x} \|_2 + \| y -\wt{y} \|_2)
\end{align*}
\end{lemma}
\begin{proof}
    We define
    \begin{align*}
        G_{2,1} = & ~ - \langle f(x)_{j_0}, h(y)_{i_0} \rangle f(x)_{j_0} f(x)_{j_0}^\top - (- \langle f(\wt{x})_{j_0}, h(y)_{i_0} \rangle f(x)_{j_0} f(x)_{j_0}^\top) \\
        G_{2,2} = & ~ - \langle f(\wt{x})_{j_0}, h(y)_{i_0} \rangle f(x)_{j_0} f(x)_{j_0}^\top - (- \langle f(\wt{x})_{j_0}, h(\wt{y})_{i_0} \rangle f(x)_{j_0} f(x)_{j_0}^\top)\\
        G_{2,3} = & ~ 
        - \langle f(\wt{x})_{j_0}, h(\wt{y})_{i_0} \rangle f(x)_{j_0} f(x)_{j_0}^\top - (- \langle f(\wt{x})_{j_0}, h(\wt{y})_{i_0} \rangle f(\wt{x})_{j_0} f(x)_{j_0}^\top) \\
        G_{2,4} = & ~ 
        - \langle f(\wt{x})_{j_0}, h(\wt{y})_{i_0} \rangle f(\wt{x})_{j_0} f(x)_{j_0}^\top - (- \langle f(\wt{x})_{j_0}, h(\wt{y})_{i_0} \rangle f(\wt{x})_{j_0} f(\wt{x})_{j_0}^\top)
    \end{align*}

We have 
\begin{align*}
    \|G_{2,1} \| = & ~ \|- \langle f(x)_{j_0}, h(y)_{i_0} \rangle f(x)_{j_0} f(x)_{j_0}^\top - (- \langle f(\wt{x})_{j_0}, h(y)_{i_0} \rangle f(x)_{j_0} f(x)_{j_0}^\top)\|\\
    \leq & ~ \|f(x)_{j_0} - f(\wt{x})_{j_0}  \|_2 \cdot \|h(y)_{i_0} \|_2 \cdot \|f(x)_{j_0} \|_2 \cdot 
    \|f(x)_{j_0} \|_2\\
    \leq & ~ R^2 \cdot R_0 \| x- \wt{x} \|_2 
\end{align*}
where the first step is based on the definition of $G_{2,1}$, the second step follows from Fact~\ref{fac:circ_rules}, and the third step is because of Lemma~\ref{lem:upper_bound}.

and 
\begin{align*}
    \|G_{2,2} \| = & ~ \|- \langle f(\wt{x})_{j_0}, h(y)_{i_0} \rangle f(x)_{j_0} f(x)_{j_0}^\top - (- \langle f(\wt{x})_{j_0}, h(\wt{y})_{i_0} \rangle f(x)_{j_0} f(x)_{j_0}^\top)\| \\
    \leq & ~ \| f(\wt{x})_{j_0}  \|_2 \cdot 
 \|h(y)_{i_0} -h(\wt{y})_{i_0} \| \cdot \|f(x)_{j_0} \|_2 \cdot 
    \|f(x)_{j_0} \|_2 \\
    \leq & ~ R \|y -\wt{y} \|_2
\end{align*}
where the first step is due to the definition of $G_{2,1}$, the second step is based on Fact~\ref{fac:circ_rules}, and the third step follows from Lemma~\ref{lem:upper_bound:y}.

Similarly, we have
\begin{align*}
    \|G_{2,3} \| \leq & ~ R^2 \cdot R_0 \| x- \wt{x} \|_2 \\
    \|G_{2,4} \| \leq & ~ R^2 \cdot R_0 \| x- \wt{x} \|_2 
\end{align*}

Combining all the above equations we complete the proof.
\end{proof}

\subsection{Calculation: Step 3 Lipschitz for Matrix Function \texorpdfstring{$- c(x,y)_{j_0,i_0} \diag( f(x)_{j_0} )$}{}}\label{sub:lips_H_xy:step3}

In this section, we calculate the Lipschitz for $- c(x,y)_{j_0,i_0} \diag( f(x)_{j_0} )$.

\begin{lemma}\label{lem:lipschitz_xy_G3}
If the following conditions
\begin{itemize}
    \item Let $\alpha(x)_{j_0} \in \R$ be defined as Definition~\ref{def:alpha}
    \item Let $f(x)_{j_0} \in \R^n$ be defined as Definition~\ref{def:f}
    \item Let $c(x,y)_{j_0,i_0} \in \R$ be defined as Definition~\ref{def:c}
    \item Let $\gamma(x)_{j_0} = \langle f(x)_{j_0}, v \rangle \in \R$
    \item $\| A_1 \|, \| A_2 \| , \| A_3 \| \leq R$, $\| \A_{j_0} \| \leq R$, $\| x \|_2 \leq R$,$| b_{j_0,i_0} | \leq R$, $\| v \|_2 \leq R^2$
    \item Let $R \geq 4$
    \item Let $R_0$ be defined as Definition~\ref{def:R_0}.
    \item Let $G_3(x,y) = - c(x,y)_{j_0,i_0} \diag( f(x)_{j_0} )$ 
\end{itemize}
Then, we have
\begin{align*}
    \|G_3(x,y) - G_3( \wt{x}, \wt{y} ) \| \leq 3R^2 \cdot R_0  (\|x -\wt{x} \|_2 +  \| y -\wt{y} \|_2)
\end{align*}
\end{lemma}
\begin{proof}
    We define
    \begin{align*}
        G_{3,1} = & ~
        - c(x,y)_{j_0,i_0} \diag( f(x)_{j_0} ) - (- c(\wt{x},y)_{j_0,i_0} \diag( f(x)_{j_0} ))\\
        G_{3,2} = & ~
        - c(\wt{x},y)_{j_0,i_0} \diag( f(x)_{j_0} ) - (- c(\wt{x},\wt{y})_{j_0,i_0} \diag( f(x)_{j_0} )) \\
        G_{3,3} = & ~
        - c(\wt{x},\wt{y})_{j_0,i_0} \diag( f(x)_{j_0} ) - (- c(\wt{x},\wt{y})_{j_0,i_0} \diag( f(\wt{x})_{j_0} ))
    \end{align*}

For $G_{3,1}$, we have
\begin{align*}
    \|G_{3,1} \| = & ~  \|- c(x,y)_{j_0,i_0} \diag( f(x)_{j_0} ) - (- c(\wt{x},y)_{j_0,i_0} \diag( f(x)_{j_0} )) \|\\
    \leq & ~ |c(x,y)_{j_0,i_0} - c(\wt{x},y)_{j_0,i_0} | \cdot \|f(x)_{j_0} \|_2 \\
    \leq & ~ R^2 \cdot R_0\|x - \wt{x} \|_2
\end{align*}
where the first step follows from definition of $G_{3,1}$, the second step is based on Fact~\ref{fac:vector_norm} and the third step is because of Lemma~\ref{lem:upper_bound:y}.

Similarly, we have 
\begin{align*}
    \|G_{3,2} \| \leq & ~  R \|y - \wt{y} \|_2 \\
    \|G_{3,3} \| \leq & ~ 2R^2\cdot R_0 \|x -\wt{x} \|_2
\end{align*}

Combining all the above equations we complete the proof.
\end{proof}

\subsection{Calculation: Step 4 Lipschitz for Matrix Function \texorpdfstring{$c(x,y)_{j_0,i_0} f(x)_{j_0} f(x)_{j_0}^\top$}{}}\label{sub:lips_H_xy:step4}

In this section, we calculate the Lipschitz for $c(x,y)_{j_0,i_0} f(x)_{j_0} f(x)_{j_0}^\top$.

\begin{lemma}\label{lem:lipschitz_xy_G4}
If the following conditions
\begin{itemize}
    \item Let $\alpha(x)_{j_0} \in \R$ be defined as Definition~\ref{def:alpha}
    \item Let $f(x)_{j_0} \in \R^n$ be defined as Definition~\ref{def:f}
    \item Let $c(x,y)_{j_0,i_0} \in \R$ be defined as Definition~\ref{def:c}
    \item Let $\gamma(x)_{j_0} = \langle f(x)_{j_0}, v \rangle \in \R$
    \item $\| A_1 \|, \| A_2 \| , \| A_3 \| \leq R$, $\| \A_{j_0} \| \leq R$, $\| x \|_2 \leq R$,$| b_{j_0,i_0} | \leq R$, $\| v \|_2 \leq R^2$
    \item Let $R \geq 4$
    \item Let $R_0$ be defined in Definition~\ref{def:R_0}.
    \item Let $G_4(x,y) = c(x,y)_{j_0,i_0} f(x)_{j_0} f(x)_{j_0}^\top$ 
\end{itemize}
Then, we have
\begin{align*}
    \|G_4(x,y) - G_4( \wt{x}, \wt{y} ) \| \leq 5R^2 \cdot R_0 (\|x -\wt{x} \|_2 + \| y -\wt{y} \|_2)
\end{align*}
\end{lemma}
\begin{proof}
    We define
    \begin{align*}
        G_{4,1} = & ~ c(x,y)_{j_0,i_0} f(x)_{j_0} f(x)_{j_0}^\top - c(\wt{x},y)_{j_0,i_0} f(x)_{j_0} f(x)_{j_0}^\top \\
        G_{4,2} = & ~ c(\wt{x},y)_{j_0,i_0} f(x)_{j_0} f(x)_{j_0}^\top - c(\wt{x},\wt{y})_{j_0,i_0} f(x)_{j_0} f(x)_{j_0}^\top \\
        G_{4,3} = & ~ c(\wt{x},\wt{y})_{j_0,i_0} f(x)_{j_0} f(x)_{j_0}^\top - c(\wt{x},\wt{y})_{j_0,i_0} f(\wt{x})_{j_0} f(x)_{j_0}^\top \\
        G_{4,4} = & ~ c(\wt{x},\wt{y})_{j_0,i_0} f(\wt{x})_{j_0} f(x)_{j_0}^\top - c(\wt{x},\wt{y})_{j_0,i_0} f(\wt{x})_{j_0} f(\wt{x})_{j_0}^\top
    \end{align*}
For $G_{4,1}$, we have
\begin{align*}
    \|G_{4,1}\| = & ~ \|c(x,y)_{j_0,i_0} f(x)_{j_0} f(x)_{j_0}^\top - c(\wt{x},y)_{j_0,i_0} f(x)_{j_0} f(x)_{j_0}^\top \|\\
    \leq & ~ |c(x,y)_{j_0,i_0} -c(\wt{x},y)_{j_0,i_0}| \cdot  \|f(x)_{j_0} \|_2 \cdot \|f(x)_{j_0} \|_2 \\
    \leq & ~ R^2 \cdot R_0 \| x- \wt{x} \|_2
\end{align*}
where the first step is due to definition of $G_{4,1}$, the second step is because of Fact~\ref{fac:vector_norm} and the third step follows from Lemma~\ref{lem:upper_bound} and Lemma~\ref{lem:basic_lips}. 

Similarly, we have 
\begin{align*}
    \|G_{4,2} \| \leq & ~R \|y - \wt{y} \|_2 \\
    \|G_{4,3} \| \leq & ~ 2R^2 \cdot R_0 \| x -\wt{x}\|_2 \\
    \|G_{4,4} \| \leq & ~ 2R^2 \cdot R_0 \| x -\wt{x}\|_2
\end{align*}
Combining all the above equations we complete the proof.
\end{proof}

%\newpage
\subsection{PSD Upper Bound for Hessian \texorpdfstring{$x,y$}{}}\label{sub:lips_H_xy:psd}

In this section, we analyze the PSD upper bound for Hessian. 

\begin{lemma}

If the following conditions hold
 \begin{itemize}
    \item $\max_{j_0 \in [n]} \| \A_{j_0}  \| \leq R$
    \item Let $H(x,y)_{j_0,i_0} \in \R^{d^2 \times d}$ denote $\frac{\d^2 L_{j_0,i_0}}{ \d x \d y_{i_0} }$
    \item $\frac{\d^2 L_{j_0,i_0}}{ \d x \d y_{i_1}} = {\bf 0}_{d^2 \times d}$
    \item Let $H(x,y) \in \R^{d^2 \times d^2}$ be 
    \begin{align*}
    H(x,y) : =
    \begin{bmatrix}
        \sum_{j_0=1}^n H_{j_0,1}(x,y) & \sum_{j_0=1}^n H_{j_0,2}(x,y) & \cdots & \sum_{j_0=1}^n H_{j_0,d}(x,y)
    \end{bmatrix}
    \end{align*}
\end{itemize} 

Then we have
\begin{itemize}
    \item Part 1. For $j_0 \in [d], i_0 \in [n]$ \begin{align*}
    \|H(x,y)_{j_0,i_0}   \| \leq 10R^2
\end{align*}
    \item Part 2. \begin{align*}
    \|H(x,y)  \| \leq nd \cdot 10 R^2
\end{align*}
\end{itemize}
\end{lemma}
\begin{proof}
    {\bf Proof of Part 1.}
    It follows from Lemma~\ref{lem:summary_Gi_xy_psd}.

    {\bf Proof of Part 2}.
    We can show that
    \begin{align*}
        \| H(x,y) \| = & ~ \sum_{j_0}^d \sum_{i_0}^n \| H(x,y)_{j_0,i_0} \| \\
        \leq & ~ nd \cdot 10 R^2
    \end{align*}
where the first step is due to the assumption of $H(x,y)$, and the second step comes from {\bf Part 1.}
\end{proof}

\subsection{Upper Bound on Hessian Spectral Norms}\label{sub:lips_H_xy:summary_psd}

In this section, we find the upper bound for the Hessian spectral norms.

\begin{lemma}\label{lem:summary_Gi_xy_psd}
    If the following conditions hold 
    \begin{itemize}
        \item $G_{1}(x,y) = ( f(x)_{j_0} \circ h(y)_{i_0} ) f(x)_{j_0}^\top
        $
        \item $G_{2}(x,y) = - \langle f(x)_{j_0}, h(y)_{i_0} \rangle f(x)_{j_0} f(x)_{j_0}^\top$
        \item $G_3(x,y) = - c(x,y)_{j_0,i_0} \diag( f(x)_{j_0} ) $
        \item $G_4(x,y) = c(x,y)_{j_0,i_0} f(x)_{j_0} f(x)_{j_0}^\top$
        
    \end{itemize}

    Then, we have
    \begin{itemize}
    \item Part 1. $\| G_1(x,y) \| \leq R^2$
    \item Part 2. $\| G_2(x,y) \| \leq R^2$
    \item Part 3. $\| G_3(x,y) \| \leq 2R^2$
    \item Part 4. $\| G_4(x,y) \| \leq 2R^2$
    \item Part 5.
    \begin{align*}
        \sum_{k=1}^4 \|G_{k }(x,y) \| \leq 10R^2
    \end{align*}
    \end{itemize}
\end{lemma}
\begin{proof}
    The proof is straightforward by using upper bound on each term
\end{proof}

%\newpage
\section{Generating a Spectral Sparsifier via TensorSketch}
\label{sec:tensorsketch}

Tensor type sketching has been widely used in problems \cite{swz19_tensor,dssw18,djs+19,akk+20,swyz21,szz21,sxz22,z22,syz23_sdp}. Section~\ref{sec:tensorsketch:ose} presents the definition of oblivious subspace embedding. In Section~\ref{sub:tensorsketch:TensorSRHT}, we give an overview of $\mathsf{TensorSRHT}$ and introduce its basic property. In Section~\ref{sub:tensorsketch:TensorSparse}, we present the definition of the property of $\mathsf{TensorSparse}$. In Section~\ref{sub:tensorsketch:Sketching}, we introduce the fast approximation for hessian via sketching.

\subsection{Oblivious Subspace Embedding}
\label{sec:tensorsketch:ose}

We define oblivious subspace embedding,
\begin{definition}[Oblivious subspace embedding, \cite{s06}]
We define $(\epsilon,\delta,d,n)$-Oblivious subspace embedding (\textsf{OSE}) as follows: Suppose $\Pi$ is a distribution on $m \times n$ matrices $S$, where $m$ is a function of $n, d, \epsilon$, and $\delta$. Suppose that with probability at least $1 - \delta$, for
any fixed $n \times d$ orthonormal basis $U$, a matrix $S$ drawn from the distribution $\Pi$ has the property that the singular values of $SU$ lie in the range $[1-\epsilon,1+\epsilon]$.
\end{definition}

\subsection{TensorSRHT}
\label{sub:tensorsketch:TensorSRHT}

We define a well-known sketching matrix family called {\sf TensorSRHT} \cite{ldfu13,akk+20}. It has been used in many optimization literature \cite{swyz21,szz21,sxz22}.
\begin{definition}[Tensor subsampled randomized Hadamard transform (\textsf{TensorSRHT}) \cite{akk+20,swyz21}]\label{def:tensor_srht}
The $\mathsf{TensorSRHT}$ $S: \R^n \times \R^n \to \R^m$ is defined as 
\begin{align*}
S := \frac{1}{\sqrt{m}} P \cdot (HD_1 \otimes HD_2),
\end{align*}
where each row of $P \in \{0, 1\}^{m \times n^2}$ contains only one $1$ at a random coordinate and one can view $P$ as a sampling matrix. $H$ is a $n \times n$ Hadamard matrix, and $D_1$, $D_2$ are two $n \times n$ independent diagonal matrices with diagonals that are each independently set to be a Rademacher random variable (uniform in $\{-1, 1\}$).  
\end{definition}

It is known \cite{akk+20} that {\sf TensorSRHT} matrices imply the {\sf OSE}.
\begin{lemma}[\cite{akk+20,swyz21} , see for example, Lemma 2.12 in~\cite{swyz21}]\label{lem:tensor_srht}
    Let $S$ be a {\sf TensorSRHT} matrix defined in Definition~\ref{def:tensor_srht}. If 
    \begin{align*}
        m=O(\epsilon^{-2}d^2 \log^3(nd / \epsilon\delta) ),
    \end{align*}
    then $S$ is an $(\epsilon, \delta, d^2, n^2)$-{\sf OSE} for degree-$2$ tensors. 

    Further for matrices $A_1,A_2 \in \R^{n \times d}$, $S(A_1 \otimes A_2)$ can be computed in $\wt{O}(nd + md^2)$ time.
\end{lemma}

\subsection{TensorSparse}
\label{sub:tensorsketch:TensorSparse}

\cite{sxz22} define {\sf TensorSparse} by compose Sparse embedding \cite{nn13,c16} with tensor operation \cite{p13}.
\begin{definition}[{\sf TensorSparse}, see Definition~7.6 in \cite{sxz22}]\label{def:tensor_sparse}
Let $h_1,h_2:[n] \times [s]\rightarrow [m/s]$ be $O(\log 1/\delta)$-wise independent hash functions and let $\sigma_1,\sigma_2:[n ]\times [s]\rightarrow \{\pm 1\}$ be $O(\log 1/\delta)$-wise independent random sign functions. Then, the degree two tensor sparse transform, $S:\R^n \times \R^n \rightarrow \R^m$ is given as: 
\begin{align*}
    R_{r,(i,j)} = & ~ \exists k\in [s]: \sigma_1(i,k)\sigma_2(j,k)/\sqrt{s}\cdot {\bf 1}[ ((h_1(i,k)+h_2(j,k))~\text{mod~}m/s)+(k-1)m/s=r]
\end{align*}
\end{definition}

\begin{lemma}[Theorem 7.10 in \cite{sxz22}]\label{lem:tensor_sparse}
Let $\epsilon\in (0,1)$ be precision parameter and $\delta\in (0,1)$ be success probability. Let $S \in \R^{m \times n^2}$ be a ${\sf TensorSparse}$ matrix (Def.~\ref{def:tensor_sparse}). Suppose $m=\Omega(\epsilon^{-2} d^2 \log(n/\delta))$ and $s=\epsilon^{-1} \log(n/\delta)$, then {\sf TensorSparse} provides $(\epsilon,\delta,d^2,n^2)$-{\sf OSE}.

Further for matrices $A_1,A_2 \in \R^{n \times d}$, $S(A_1 \otimes A_2)$ can be computed in $O( (\nnz(A_1) + \nnz(A_2) ) s + m d^2)$ time
\end{lemma}

\subsection{Fast Approximation for Hessian via Sketching}
\label{sub:tensorsketch:Sketching}

In this section, we present the fast approximation for hessian via sketching.

\begin{lemma}\label{lem:compute_hessian_approximate}
If the following conditions hold
\begin{itemize}
    \item Let $A_1 \in \R^{n \times d}$, let $A_2 \in \R^{n \times d}$
    \item Let $\A = (A_1 \otimes A_2) \in \R^{n^2 \times d^2}$
    \item Let $W\in \R^{n \times n}$ denote a positive diagonal matrix 
    \item Let $\ov{A}_1 = W A_1$
    \item Let $\ov{\A} = (\ov{A}_1 \otimes A_2) \in \R^{n^2 \times d^2}$
\end{itemize}
Then, we have
\begin{itemize}
    \item {\bf Part 1.} 
    \begin{align*}
    \A^\top (W^2 \otimes I_n) \A = \ov{\A}^\top \ov{\A}
    \end{align*}
    \item {\bf Part 2.} For any constant $\epsilon \in (0,0.1)$, there is an algorithm runs in $\wt{O}(nd + d^4)$ time to compute $S \ov{\A}$ such that
    \begin{align*}
        (1-\epsilon) \cdot \ov{\A}^\top \ov{\A} \preceq \ov{\A}^\top S^\top S \ov{\A} \preceq (1+\epsilon) \cdot \ov{\A}^\top \ov{\A}
    \end{align*}
    holds with probability $1-\delta$.
    \item {\bf Part 3.} For any $\epsilon \in (0,0.1)$, there is an algorithm runs in $\wt{O}(\nnz(A_1) + \nnz(A_2) + d^4)$ time to compute $S \ov{\A}$ such that
    \begin{align*}
        (1-\epsilon) \cdot \ov{\A}^\top \ov{\A} \preceq \ov{\A}^\top S^\top S \ov{\A} \preceq (1+\epsilon) \cdot \ov{\A}^\top \ov{\A}
    \end{align*}
    holds with probability $1-\delta$.
\end{itemize}
\end{lemma}
\begin{proof}

{\bf Proof of Part 1.}

We can show
\begin{align*}
    \A^\top (W^2 \otimes I_n) \A 
    = & ~ \A^\top (W \otimes I_n) \cdot (W \otimes I_n) \A \\
    = & ~  ( (A_1 \otimes A_2) (W \otimes I_n) )^\top  \cdot ( (A_1 \otimes A_2) (W \otimes I_n) ) \\
    = & ~ (\ov{A}_1 \otimes A_2 )^\top ( \ov{A}_1 \otimes A_2 ) \\
    = & ~ \ov{\A}^\top \ov{\A}
\end{align*}
where the first step follows from $(W^2 \otimes I) = (W \otimes I_n) \cdot (W \otimes I_n)$ (where $\otimes$ operation and $W$ is a diagonal matrix), the second step follows from the definition of $\A$, the third step follows from the definition of $\ov{A}_1$, and the last step follows from the definition of $\ov{\A}$. 

{\bf Proof of Part 2.}

It follows from using Lemma~\ref{lem:tensor_srht}.

{\bf Proof of Part 3.}

It follows from using Lemma~\ref{lem:tensor_sparse}.

\end{proof}

%\newpage
\section{Analysis Of Algorithm~\ref{alg:main_result}}\label{sec:newton}

We introduce the concept of a $(l,M)$-good function in Section~\ref{sub:newton:good} and discuss the notion of a well-initialized point. Subsequently, we will present our approximation and update rule methods in Section~\ref{sub:newton:approximation}. In light of the optimization problem introduced in Definition~\ref{def:attention}, we put forward Algorithm~\ref{alg:main_result}, and in this section, we establish the correctness and convergence of the algorithm.

\subsection{\texorpdfstring{$(l,M)$}{}-Good Loss Function}
\label{sub:newton:good}

We will now introduce the definition of a $(l, M)$-Good Loss Function. Next, let's revisit the optimization problem defined in Definition~\ref{def:L} as follows:
\begin{align*}
   L(X,Y) := 0.5 \cdot \| \underbrace{ D(X)^{-1} }_{n \times n} \underbrace{ \exp(A_1 X A_2^\top) }_{n \times n} \underbrace{ A_3 }_{n \times d} \underbrace{Y}_{d \times d} - \underbrace{ B  }_{n \times d} \|_F^2
\end{align*}
We will now demonstrate that our optimization function possesses the following properties.

\begin{definition}[$(l,M)$-good Loss function]\label{def:assumptions}
For a function $L : \R^d \rightarrow \R$, if the following conditions hold,
\begin{itemize}
    \item {\bf Hessian is $M$-Lipschitz.} If there exists a positive scalar $M>0$ such that
    \begin{align*}
       \| \nabla^2 L(x,y) - \nabla^2 L(\wt{x},\wt{y}) \| \leq M\cdot ( \| x - \wt{x} \|_2 + \| y - \wt{y} \|_2 )
    \end{align*}
    \item {\bf $l$-local Minimum.}  
    Given $l >0$ as a positive scalar. If there exists a vector $x^* \in \R^{d^2}$ and $y^* \in \R^{d^2}$ such that the following holds
    \begin{itemize}
        \item $\nabla L(x^*, y^*) = {\bf 0}_d$.
        \item $\nabla^2 L(x^*, y^*) \succeq l \cdot I_{2 d^2}$.
    \end{itemize}
    \item {\bf Good Initialization Point.} Let $x_0$ and $y_0$ denote the initialization point. If $r_0:= (\| x_0 -x_*\|_2 + \| y_0 - y_*\|_2 )$ satisfies
    \begin{align*}
        r_0 M \leq 0.1 l.
    \end{align*}    
\end{itemize}
 we say $L$ is $(l,M)$-good 
\end{definition}

Drawing upon Lemma~\ref{lem:hessian_lower_bound} and Lemma~\ref{lem:lips_H_xy}, we can establish that our loss function (See Definition~\ref{def:L}) satisfies the aforementioned assumption.

\subsection{Convergence}\label{sub:newton:approximation}
After introducing the approximation method 'Sparsifier via TensorSketch' in Section~\ref{sec:tensorsketch}, we will now proceed to introduce the update method employed in Algorithm~\ref{alg:main_result}.
In this section, we demonstrate the concept of approximate update and present an induction hypothesis. 

\begin{definition}[Approximate Update]\label{def:update_x_t}

The following process is considered by us
\begin{align*}
    \begin{bmatrix} x(t+1) \\ y(t+1) \end{bmatrix} \gets \begin{bmatrix} x(t) \\ y(t) \end{bmatrix} -  \begin{bmatrix} g( x(t)) \\ g(y(t)) \end{bmatrix} \wt{H}^{-1}
\end{align*}
\end{definition}

A tool from previous work is presented by us now.
\begin{lemma}[Iterative shrinking, a variation of Lemma 6.9 on page 32 of \cite{lsz23}]\label{lem:one_step_shrinking}
If the following conditions hold
\begin{itemize}
    \item Loss Function $L$ is $(l,M)$-good (see  Definition~\ref{def:assumptions}). 
    \item Let $\epsilon \in (0,0.1)$ (see Lemma~\ref{lem:compute_hessian_approximate}). 
    \item Let $x^*, y^*$ be defined in Definition~\ref{def:assumptions} and $x_t,y_t$ be defined in Definition~\ref{def:update_x_t}.
    \item Let $r_t:= \| x_t - x^* \|_2 + \| y_t - y^*\|_2$.
    \item Let $\ov{r}_t: = M \cdot r_t$
\end{itemize}
It follows that  
\begin{align*}
r_{t+1} \leq 2 \cdot (\epsilon_0 + \ov{r}_t/( l - \ov{r}_t ) ) \cdot r_t.
\end{align*} 
\end{lemma}

In this context, where $T$ denotes the total number of iterations in the algorithm, we require the following lemma based on the induction hypothesis to apply Lemma~\ref{lem:one_step_shrinking}. This lemma is a well-established concept in the literature, and for further details, you can refer to \cite{lsz23}.
\begin{lemma}[Induction hypothesis, Lemma 6.10 on page 34 of \cite{lsz23}]\label{lem:newton_induction}
If the following condition hold
\begin{itemize}
    \item $\epsilon = 0.01$ (see Lemma~\ref{lem:compute_hessian_approximate})
    \item Let $x^*, y^*$ be defined in Definition~\ref{def:assumptions} and $x_t, y_t$ be defined in Definition~\ref{def:update_x_t}.
    \item Let $r_t:= \| x_t - x^* \|_2 + \| y_t - y^*\|_2$.
    \item  For each $i \in [T]$, $r_{i} \leq 0.4 \cdot r_{i-1}$, for all $i \in [t]$
    \item Let $l$ and $M$ be Defined in Definition~\ref{def:assumptions}
    \item $M \cdot r_i \leq 0.1 l$, for all $i \in [t]$.
\end{itemize}
It follows that
\begin{itemize}
    \item $r_{t+1} \leq 0.4 r_t$
    \item $M \cdot r_{t+1} \leq 0.1 l$
\end{itemize}
\end{lemma}

\ifdefined\isarxiv
%\section*{Acknowledgments}
\bibliographystyle{alpha}
\bibliography{ref}
\else
\bibliography{ref}
\bibliographystyle{alpha}

\fi

\newpage
\onecolumn
\appendix

%%%% Cut-line between first 10 pages and appendix

%%% some writing rules

%% Writing rule for creating tags.
%% Tags :
%% Theorem    \ref{thm:bla_bla}
%% Lemma      \ref{lem:bla_bla}
%% Claim      \ref{cla:bla_bla}
%% Corollary  \ref{cor:bla_bla}
%% Fact       \ref{fac:bla_bla}
%% Definition \ref{def:bla_bla}
%% Section    \ref{sec:bla_bla}
%% Subsection \ref{sub:bla_bla}
%% Equation   \ref{eq:bla_bla}

\end{document}